\documentclass[journal=jacsat,manuscript=article]{achemso}
\usepackage[version=3]{mhchem} 
\usepackage[T1]{fontenc}
\usepackage{graphicx}
\usepackage{subcaption}
\usepackage{braket}
\usepackage[colorlinks, linkcolor = blue, citecolor = blue, filecolor = blue,
urlcolor = blue]{hyperref}
\usepackage{url}
\usepackage{amsmath}
\usepackage{geometry}

\usepackage{algorithm}
\usepackage{algorithmic}

\author{Dongyu Lyu}
\affiliation[Constructor University]
{School of Science, Constructor University, Bremen 28759, Germany}
\author{Matthias Holzenkamp}
\affiliation[University of Wuppertal]
{School of Mathematics and Natural Sciences, University of Wuppertal, Wuppertal 42119, Germany}
\author{Vivin Vinod}
\affiliation[University of Wuppertal]
{School of Mathematics and Natural Sciences, University of Wuppertal, Wuppertal 42119, Germany}
\author{Yannick Marcel Holtkamp}
\affiliation[Constructor University]
{School of Science, Constructor University, Bremen 28759, Germany}
\author{Sayan Maity}
\affiliation[Constructor University]
{School of Science, Constructor University, Bremen 28759, Germany}
\author{Carlos R. Salazar}
\affiliation[Constructor University]
{School of Science, Constructor University, Bremen 28759, Germany}
\altaffiliation{Department of Physics, Chemistry and Biology (IFM), Linköping University, SE-581 83 Linköping, Sweden}
\author{Ulrich Kleinekath\"ofer}
\affiliation[Constructor University]
{School of Science, Constructor University, Bremen 28759, Germany}
\email{ukleinekathoefer@constructor.university}
\author{Peter Zaspel}
\affiliation[University of Wuppertal]
{School of Mathematics and Natural Sciences, University of Wuppertal, Wuppertal 42119, Germany}
\email{zaspel@uni-wuppertal.de}

\title{Excitation Energy Transfer between Porphyrin Dyes on a Clay Surface: A study employing Multifidelity Machine Learning}

\begin{document}

\begin{abstract}
Natural light-harvesting antenna complexes efficiently capture solar energy mostly using chlorophyll molecules, i.e., magnesium porphyrin pigments, embedded in a protein matrix. Inspired by this natural configuration, artificial clay-porphyrin antenna structures have experimentally been synthesized and  demonstrated to exhibit remarkable excitation energy transfer properties. The current study presents a computational design and simulations of a synthetic light-harvesting system that emulates natural mechanisms by arranging cationic free-base porphyrin molecules on an anionic clay surface. We investigated the transfer of excitation energy  among the porphyrin dyes using a multiscale quantum mechanics/molecular mechanics (QM/MM) approach based on the semi-empirical density functional-based tight-binding theory for the ground state dynamics. To improve the accuracy of our results, we incorporated an innovative multifidelity machine learning  approach, which allows the prediction of excitation energies at the numerically demanding time-dependent density functional theory level together with the def2-SVP basis set.  This approach was applied to an extensive dataset of 640K geometries for the 90-atom porphyrin structures, facilitating a thorough analysis of the excitation energy diffusion among the porphyrin molecules adsorbed to the clay surface.  The insights gained from this study, inspired by natural light-harvesting complexes, demonstrate the potential of porphyrin-clay systems as effective energy transfer systems.
\end{abstract}

\section{Introduction}
In biological systems, sunlight is collected efficiently by pigment molecules embedded in a protein matrix and subsequently transferred to a reaction center.  For man-made systems,  remarkable progress has been made in recent years in dye chemistry developing  functional materials
\cite{bial21a}. To identify next-generation energy devices, researchers are exploring both experimental and theoretical biohybrid approaches inspired by biological systems  \cite{rich17b}. Of particular interest are
efforts to translate biological principles directly into synthetic energy systems \cite{alm20a}. To this end, photochemical systems with a light-harvesting function on inorganic nanosheets are being developed
\cite{ishi11b,ohta17a,tsuk18a,nish20a}. In particular, porphyrins and other tetrapyrrole macrocycles have been shown to have an impressive variety of functional properties that have  been and can be exploited in natural and artificial systems \cite{auwa15a}. 

Experimental results indicate that clay-porphyrin complexes have the potential to be used in the development of highly efficient artificial light-harvesting systems \cite{ishi11b,hass24a}.
The assembly of porphyrin molecules on clay surfaces can  be
effectively controlled using electrostatic  interactions \cite{ishi11b,ohta17a,fuji18a}. 
The authors reported an almost 100 \% efficiency of the
energy-transfer reaction \cite{ishi11b}. 
In terms of adsorption, the “size-matching effect”, which involves matching the anionic charge distribution with the dimensions of the adsorbing molecules, has been proposed as a means of achieving enhanced adsorption. Furthermore, multiple-step energy transfer reactions have been observed between three adsorbed dyes on the clay surface \cite{ohta17a}. In addition, 
The combination of a metalloporphyrin as a photocatalyst and a subporphyrin as a photosensitizer in a self-assembling manner can enable the photochemical conversion of cyclohexene using a wider range of visible light without any energy loss due to the suppression of unexpected deactivation processes \cite{tsuk18a}. Other porphyrin-clay arrangements include, e.g., the intercalation of a double-decker porphyrin metal complex into clay nanosheets \cite{yama24a}. All these examples demonstrate potential applications of porphyrin-clay systems.  In the present study, we examine a clay-pophyrin system that emulates
the experimental setup described in Ref.~\citenum{ishi11b}. To this end, we have developed a molecular multiscale approach coupled to multi-fidelity machine learning in order to simulate the excitation energy transfer in that system.

In several of the experimental studies, the clay saponite is utilized. 
While the atomistic structure has yet to be resolved, numerous individual features have been identified \cite{chan18a}. Given the lack of full atomic-level details for saponite, this study will utilize montmorillonite nanosheets for simulations, as has been done in other studies on clays, including the adsorption of esters and polymers \cite{will19a,sun20b,will22a,wang24a}. Very recently, ClayCode, a software facilitating the modeling of clay systems closely resembling experimentally determined
structures, has been made public \cite{poll24a}. In the future, this will facilitate the simulation of an even greater number of clay types. 

\begin{figure}[b!]
    \centering
    \begin{subfigure}[b]{0.4\textwidth}
        \centering
        \includegraphics[width=\textwidth]{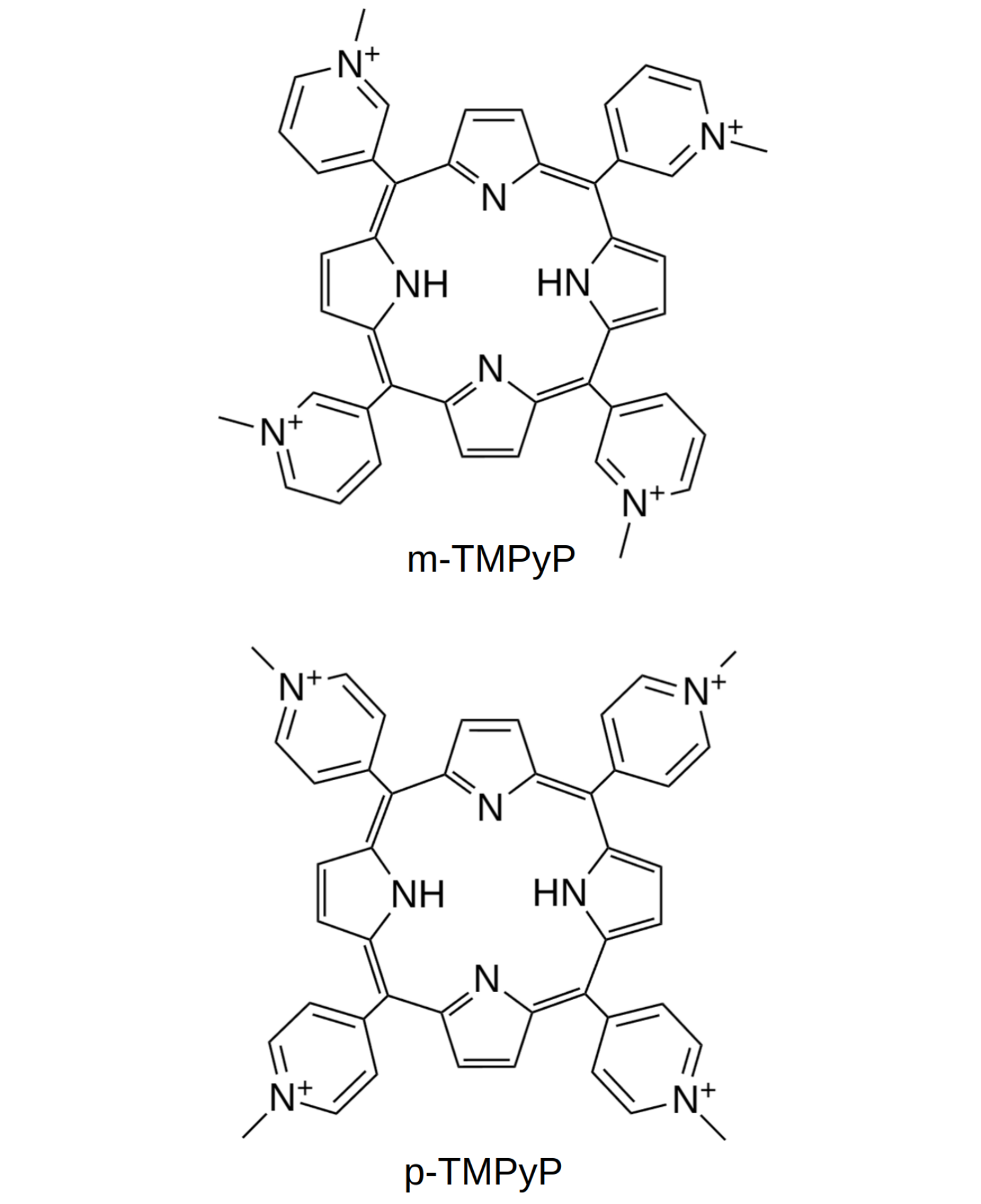}
        \caption{}
    \end{subfigure}
    \begin{subfigure}[b]{0.52\textwidth}
        \centering
        \includegraphics[width=\textwidth]{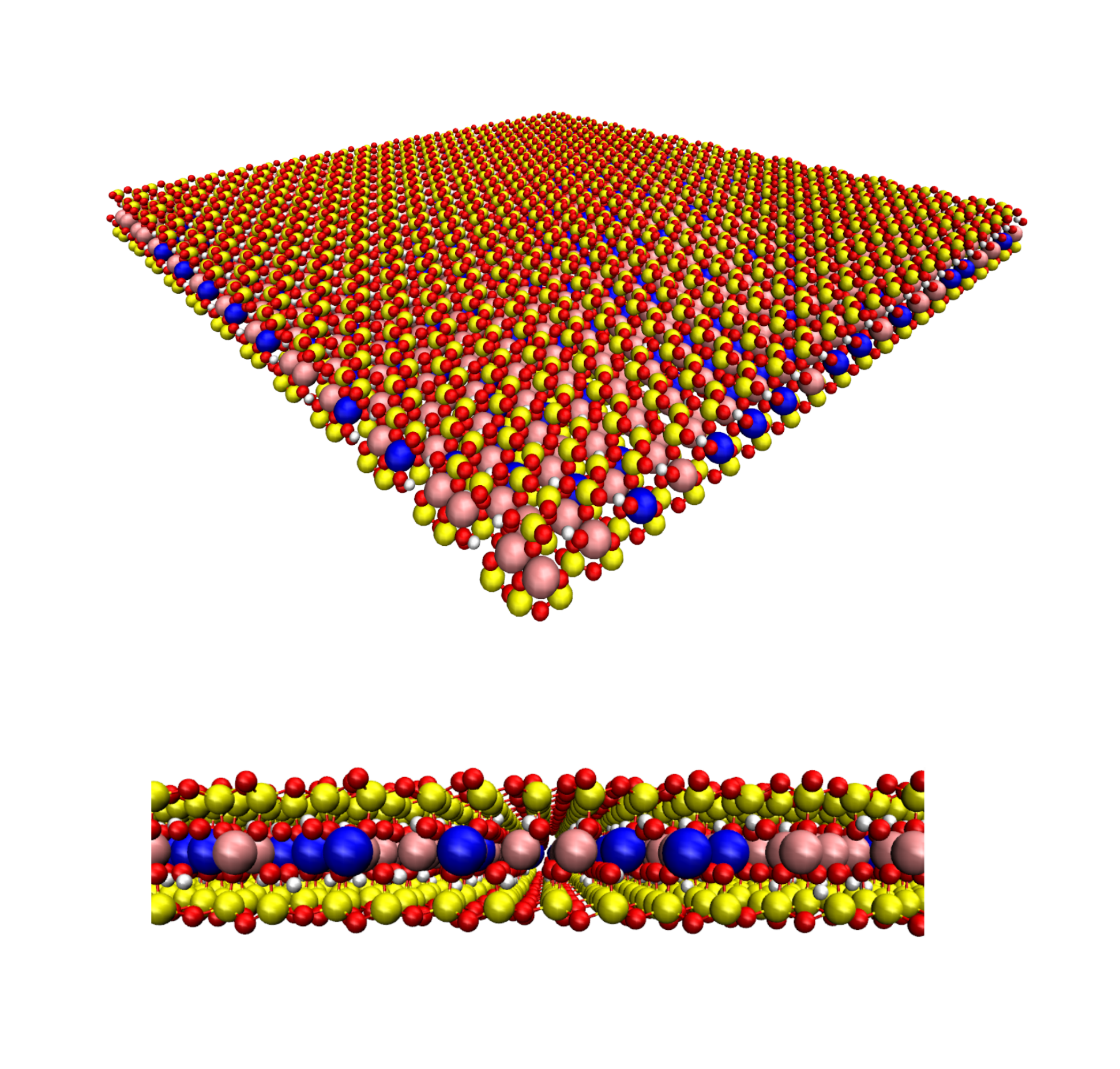}
        \caption{}
        \label{fig:montmorillonite surface}
    \end{subfigure}    
    \caption{(a) The chemical structures of the two porphyrin types used in experiment \cite{ishi11b} and in the present study. (b) The montmorillonite surface used in the present simulations. Oxygen, silicon, and hydrogen atoms are shown in red, yellow, and white, respectively, while the aluminum and magnesium atoms are colored in pink and blue, respectively. Visible in the structure is the non-regular arrangement of the magnesium defects  located in the middle layer.}
    \label{fig:surface}
\end{figure}

During photosynthesis in plants, algae, and some bacteria, the pigment molecules responsible for the absorption of sunlight are mostly (bacterio)chlrophyll molecules. These molecules play a dual role in the process, serving both as light absorbers and as mediators of excitation energy transfer to the reaction centers.
In the experimental system, which will be mimic in this study and which is depicted in Figure \ref{fig:surface}, 
tetrakis(1-methylpyridinium-3-yl)  (m-TMPyP) and tetrakis(1-methylpyridinium-4-yl)   (p-TMPyP) porphyrins adsorbed on a clay surface were selected for the same purpose
\cite{ishi11b}. Therefore, the present simulation study will focus on the same porphyrin types, with the understanding that the results may similarly also be applicable to other porphyrin molecules. 
 
Regarding modeling of the excitation energy transfer in the porphyrin-clay system, we will largely follow a procedure that has been successfully employed to model biological light-harvesting systems \cite{mait20a,mait21a,mait21b,sarn22a,mait23a,sarn24a}. 
In summary, the initial step is to conduct a molecular dynamics (MD) simulation of the entire system. The equilibrated structure is then used as the starting point for QM/MM (quantum mechanics/molecular mechanics) simulations simulating the pigment molecules. These employ the  efficient density functional tight binding (DFTB) approach \cite{hour20a} to improve the description of the vibrational motion of the pigment molecules. Subsequently, the excited states are determined using a machine-learning approach as detailed below.
At the same time, the excitonic couplings are calculated using the popular TrESP (Transition charges from Electrostatic potential) approach  \cite{reng09a,bold20a}. Based on this information, a time-dependent Hamiltonian can be constructed, which in turn is used to estimate the exciton dynamics employing the NISE  (Numerical Integration of the
Schrödinger equation) formalism \cite{jans06a},  also known as Ehrenfest scheme without back-reaction \cite{aght12a}. 

Concerning the calculation of the excitation energies along the QM/MM trajectories, we aim at improving the accuracy of the excitation energies by employing the multi-fidelity machine learning (MFML) to the present problem \cite{vino23a,vino24a}. Starting from a training set of (descriptors for) the molecular geometries and the corresponding excitation energies for the first excited state, 
we can construct regression models in machine learning (ML) to be used for the determination of the excitation energies along trajectories. These models are cheap to evaluate and ideally have low prediction error, relative to unseen molecular geometries that are picked from a neighborhood of the conformation space spanned up by the training data \cite{west20a}. We use ML models to predict excitation energies at the level of time-dependent density functional  theory using the CAM-B3LYP functional (TD-DFT/CAM-B3LYP)  with the basis set def2-SVP. To get a strong model with low prediction error, we would need a large amount of such training samples, which would however come at a prohibitive computational cost. Instead, we use MFML \cite{zasp19a,vino23a,vino24a,vinod2024_gamma_curve_error_contours} models, which are specifically constructed from training data at different levels of fidelity, here different basis set sizes (STO-3G, 3-21G, 6-31G, dev2-SVP) for the TD-DFT/CAM-B3LYP calculations with a strongly decreasing amount of samples with growing level of the hierarchy \cite{vinod2024_gamma_curve_error_contours}. Thereby, as we will show, we clearly reduce the total amount of computational effort over building “classical” single-fidelity ML models from standard training data. In addition, we investigate active learning strategies \cite{rupp2014machine,uteva2018active,zaverkin2022exploring,wilson2022batch} to reduce the required amount of training samples, even on a single level. Active learning aims at choosing or creating new training samples, which are maximally informative to the model and therefore reduce redundant information in the training data. In our setup, the selection process is done on a large amount of candidate molecular configurations (without calculated energies) from the MD runs via uncertainty sampling \cite{uncertainty_sampling}. 

The following section outlines the atomistic setup of the porphyrin-clay system, which will be simulated at the classical MD and QM/MM levels. The MD simulations allow us to assess the stability and movement of porphyrin molecules on the insulating clay surface. Furthermore, we examine the nearest-neighbor distances of select porphyrin molecules and the resulting excitonic couplings between them. Subsequently, we will address the topic of active learning and multi-fidelity machine learning of excitation energies. Based on these results, we can determine the spectral densities, which describe the interaction of the primary modes and those treated as a thermal bath.
In an alternative ansatz, the excitation energies and excitonic couplings are used in the NISE approach to determine the exciton dynamics and the diffusion of excitation energy along the clay surface. The contribution concludes with a brief overview of the key findings and future directions, while a Methods section gives more insights into technical details of the used approaches.  
 

\section{MD and QM/MM simulations}\label{Sec:MDSimulation} 

In a first step, the clay surface has to be modelled. Here we decided to model a sheet of montmorillonite since it is widely used in experiments and structural data is available. The chemical formula of montmorillonite is
$(K,Na)_x[Si_4O_8][Al_{(2-x)}Mg_xO_2(OH)_2]$. In the present study, we use Na  ions as counterions dissolved in the surrounding water.
The parameter $x$, which can vary between 0 and 0.95, specifies the percentage  of aluminum atoms being replaced by magnesium atoms in the system and can be used to adjust the charge of the surface. 
Four different values of $x$, i.e., 0.13, 0.2, 0.45 and 0.94, were tested together with four m-TMPyP and twelve p-TMPyP molecules (see Methods section for details of the MD and QM/MM setups).
Larger values of $x$ make the clay surface more anionic, though the cationic porphyrin molecules have to compete with sodium ion concerning the binding to the surface.  It has to be noted that in the present cationic montmorillonite model, the negative nature of the surface is due to atom exchange in the so-called
octahedral layer while in the experimental setup with the saponite system the ions were exchanged  in the tetrahedral layer \cite{ishi11b} (see Methods section for details). The latter layer is
closer to the surface of the material. Therefore, the negative charges in the present modeling setup are slightly more screened at the surface, leading to a somewhat reduced binding of the molecules to the surface compared to experiment. The montmorillonite surface with a $x$ value of 0.45 was chosen for the following calculations, since all porphyrin molecules stayed on the surface after the MD equilibrium for 30~ns. In simulations with  the other values of $x$, at least one dye molecule detached from the surface during the simulations.

\begin{figure}[bt!]
    \centering
    \begin{subfigure}[b]{0.48\textwidth}
        \centering
        \includegraphics[width=\textwidth]{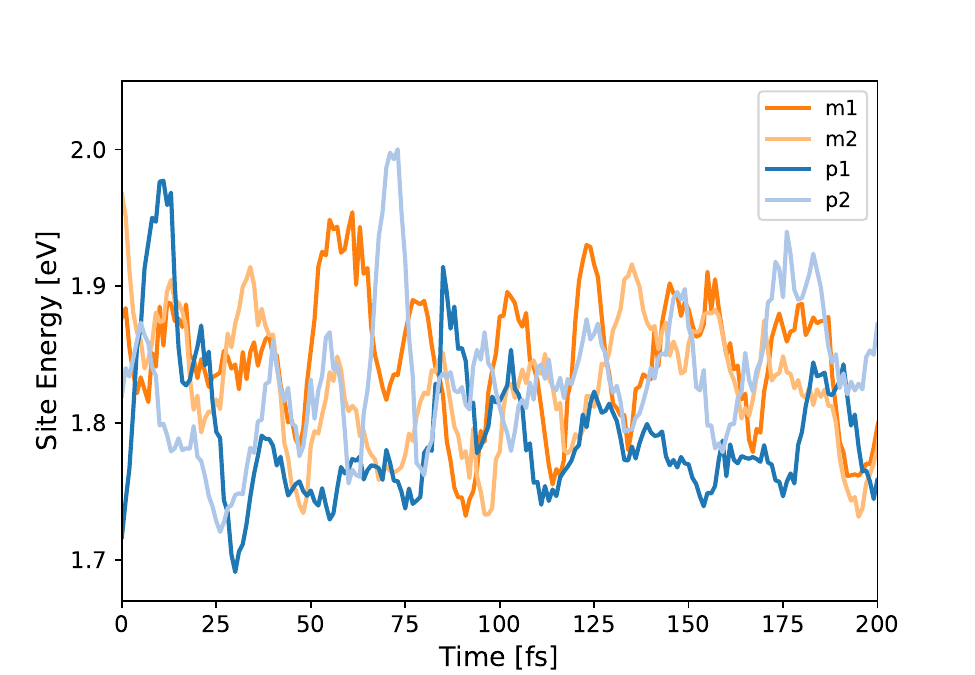}
        \caption{}
        \label{fig:fluctuation}
    \end{subfigure} 
    \begin{subfigure}[b]{0.48\textwidth}
        \centering
        \includegraphics[width=\textwidth]{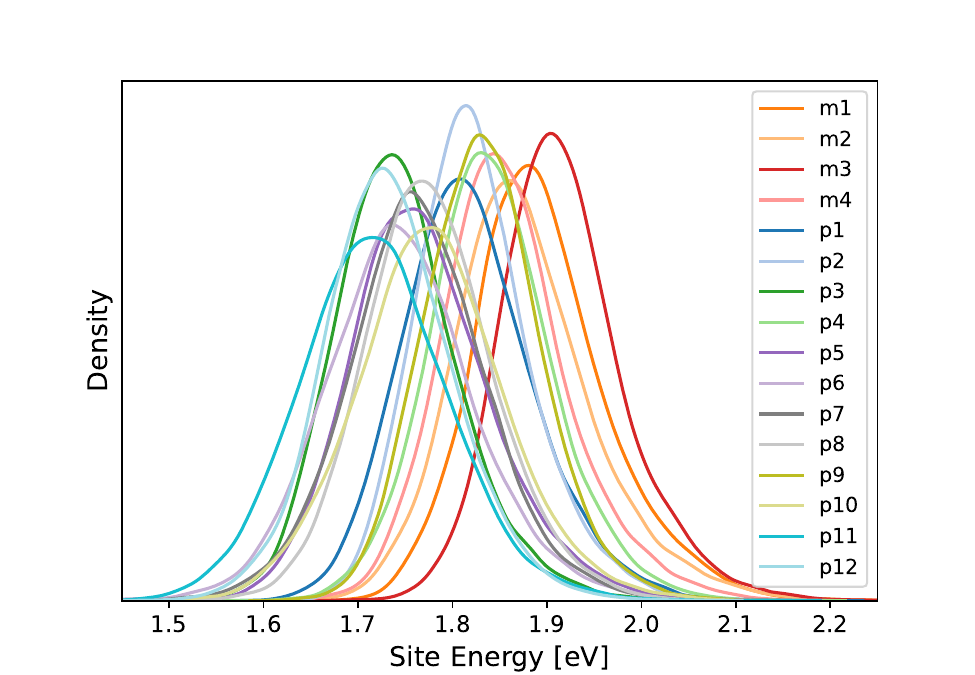}
        \caption{}
        \label{fig:dos_DFTB}
    \end{subfigure}
        \begin{subfigure}[b]{0.48\textwidth}
        \centering
        \includegraphics[width=\textwidth]{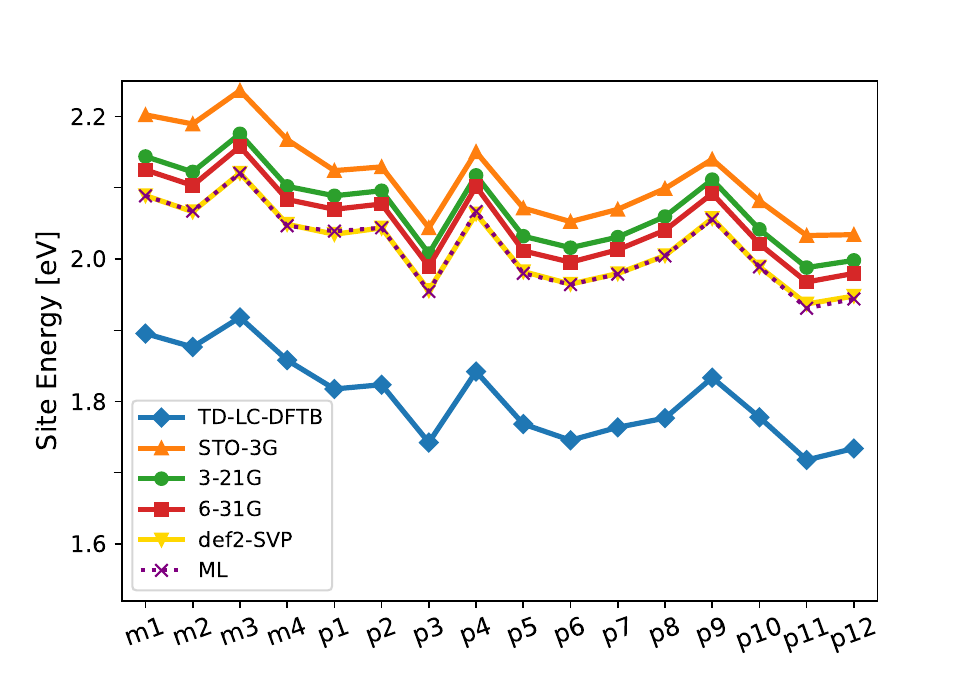}
        \caption{}
        \label{fig:average_energy}
    \end{subfigure}
    \caption{ (a) An example fragment of the energy fluctuations at TD-LC-DFTB level of theory for four porphyrins on the surface. (b) Distribution of the excited state energies using TD-LC-DFTB, which are approximately Gaussian distributed. (c) The average site energies using different levels of quantum chemistry and the predictions of the  MFML model. The numbering of the different m-type (m1-m4) and p-type (p1-p12) porphyrin dyes on the clay surface are discussed below. }
    \label{fig:dos_methods}
\end{figure}
Starting from the last frame of the MD simulation,
quantum mechanics/molecular mechanics dynamics (QM/MM) ground state simulations are performed. For each porphyrin molecule, 40,000 conformations including all surrounding point charges were saved with a stride of 1~fs. Subsequently, these conformations were used as the basis of excited state calculations using TD-LC-DFTB \cite{kran17a,bold20a}. In addition, TD-DFT/CAM-B3LYP calculations in combination with  four different basis sets, i.e., STO-3G, 3-21G, 6-31G and def2-SVP, were also used to calculate the excited states of these conformations with a stride of 8, 16, 32 and 64 fs, respectively. 
Unlike the protein environment in natural light-harvesting systems, the clay surface is basically rigid and has a high degree of periodicity if one neglects the non-periodic charge distribution due to the Mg doping inside the material. The conformations of the porphyrin molecules, however, are flexible and show fluctuations. These changes in the conformations lead to energy fluctuation. A short piece of such energy fluctuations is depicted in Figure \ref{fig:fluctuation}.
The excitation energies, also termed site energies below, lead to distributions as shown in Figure \ref{fig:dos_DFTB}. It becomes clear that the m-type porphyrins have site energies which are, on average, higher in energy than those of the p-type porphyrin dyes. The average site energies obtained by different levels of theory show offsets with respect to each other, but an overall consistent trend as shown in Figure S1 and the average excitation energies are shown in Figure \ref{fig:average_energy}.  The same type of pigments  have somewhat different site energies due to their different placements on the surface and the interactions with other porphyrins on the surface. 
These energy values from different levels of quantum chemistry are then used as input for the MFML model, as described in the next section.

\begin{figure}[b!]
    \centering
    \begin{subfigure}[b]{0.32\textwidth}
        \centering
        \includegraphics[width=\textwidth]{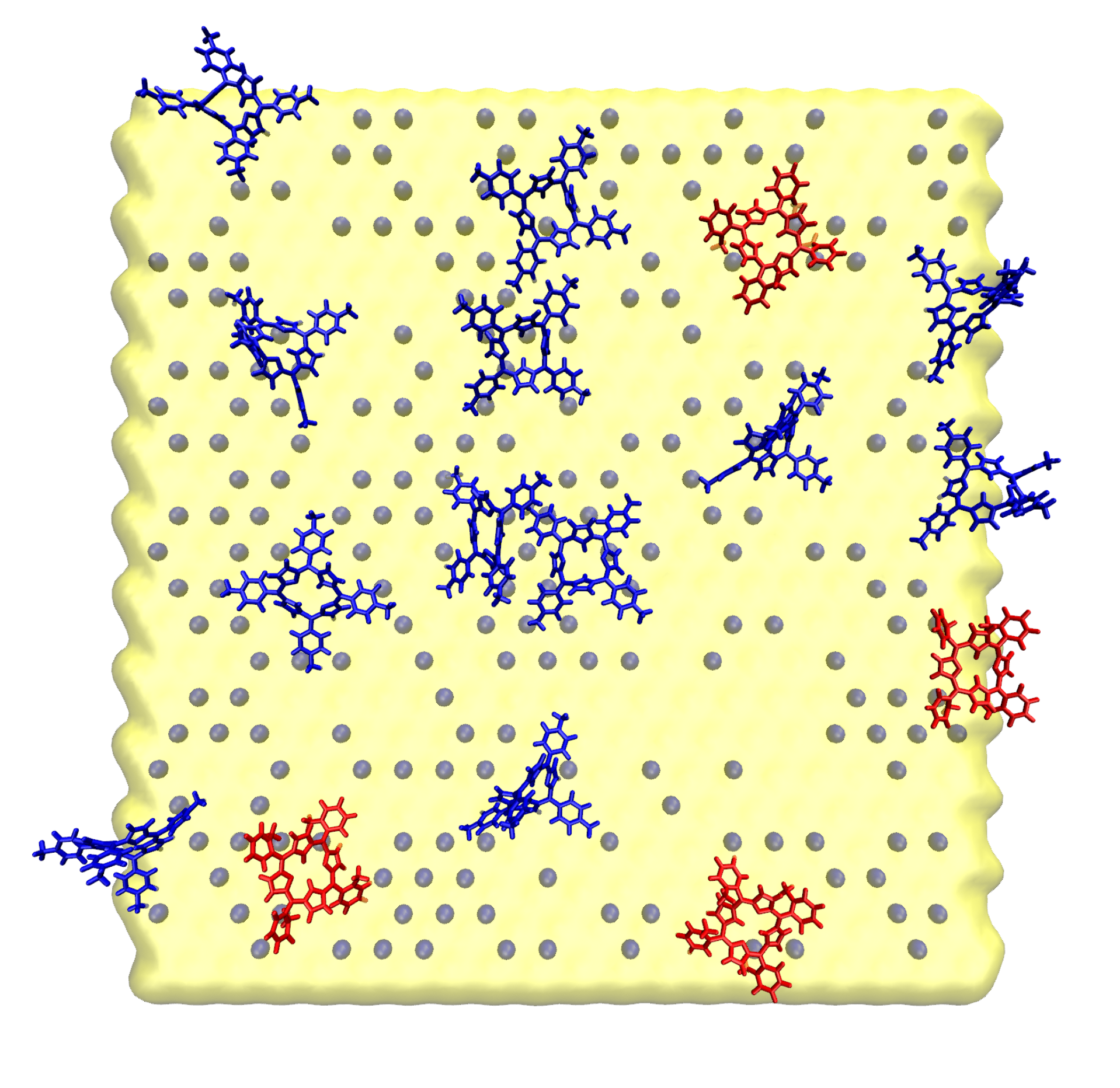}
        \caption{0ns}
    \end{subfigure}
    \hfill
    \begin{subfigure}[b]{0.32\textwidth}
        \centering
        \includegraphics[width=\textwidth]{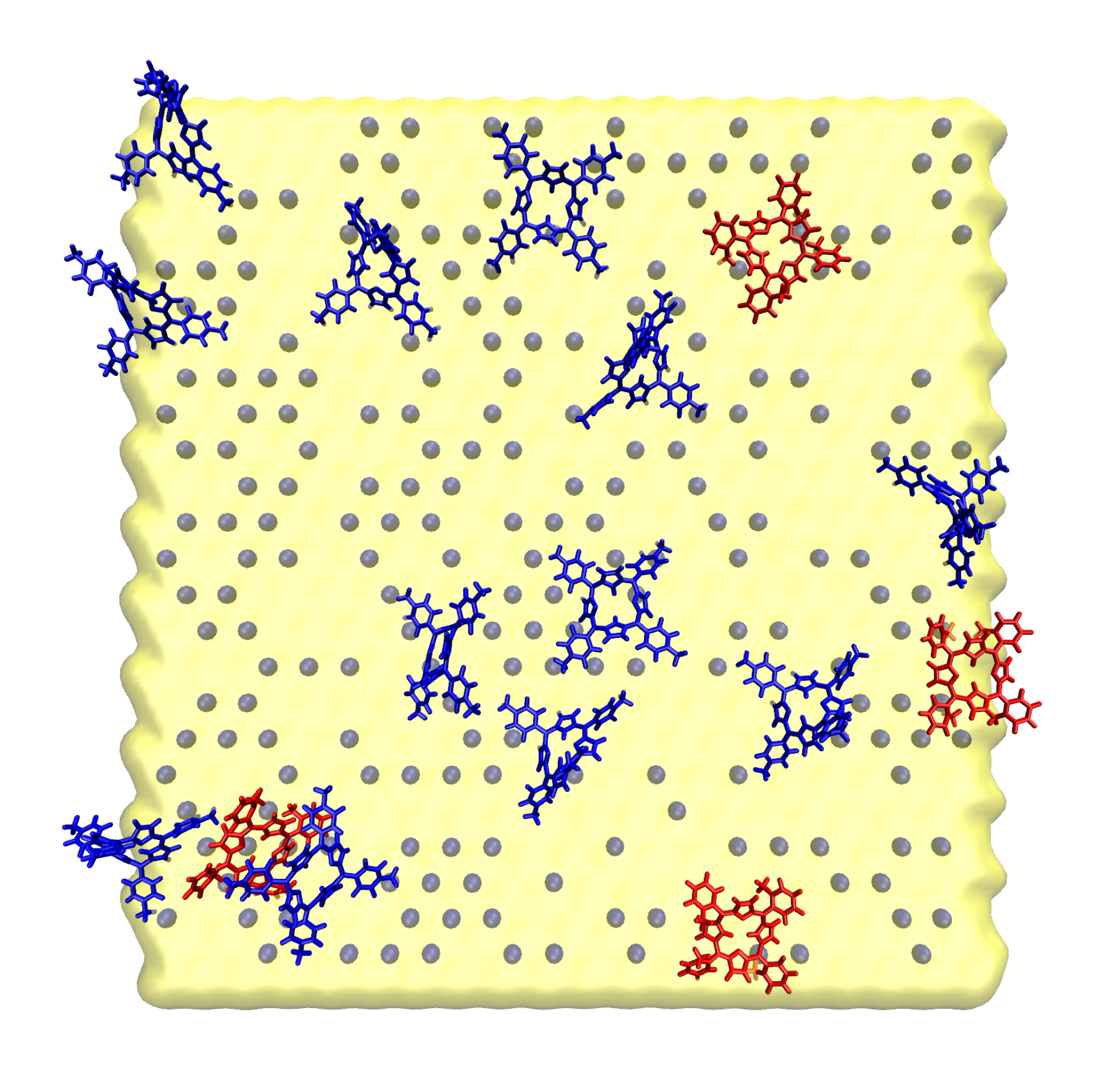}
        \caption{50ns}
    \end{subfigure}
    \hfill
    \begin{subfigure}[b]{0.32\textwidth}
        \centering
        \includegraphics[width=\textwidth]{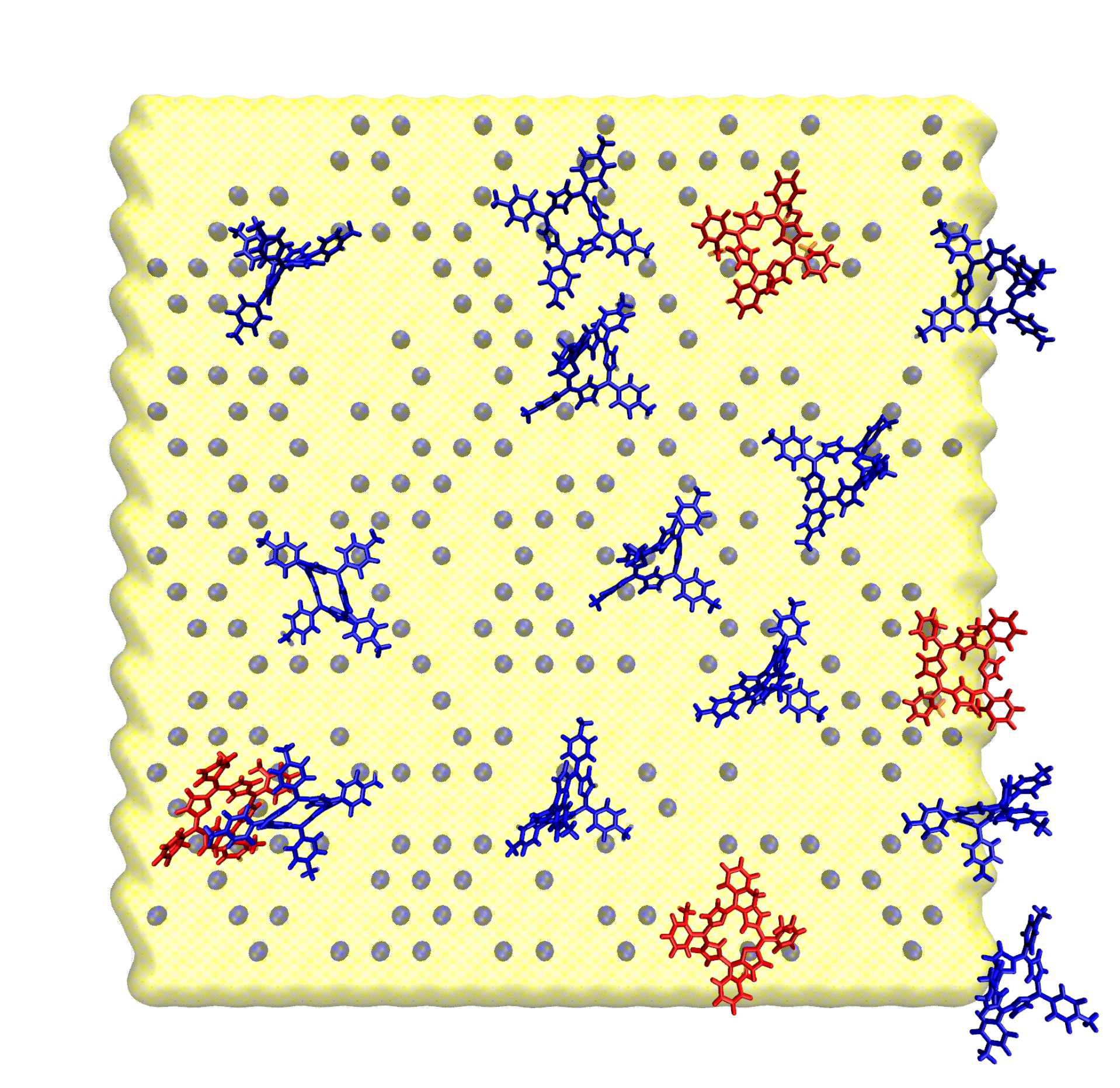}
        \caption{100ns}
    \end{subfigure}
    \caption{Several snapshots of the classical MD trajectory. The montmorillonite surface is shown in yellow in the background, with gray beads representing the magnesium defects inside the clay material. The m-TMPyP and p-TMPyP molecules on the surface are colored in red and blue, respectively.}
    \label{fig:md}
\end{figure}
\begin{figure}[b!]
    \centering
    \begin{subfigure}[b]{0.48\textwidth}
        \centering
        \includegraphics[width=\textwidth]{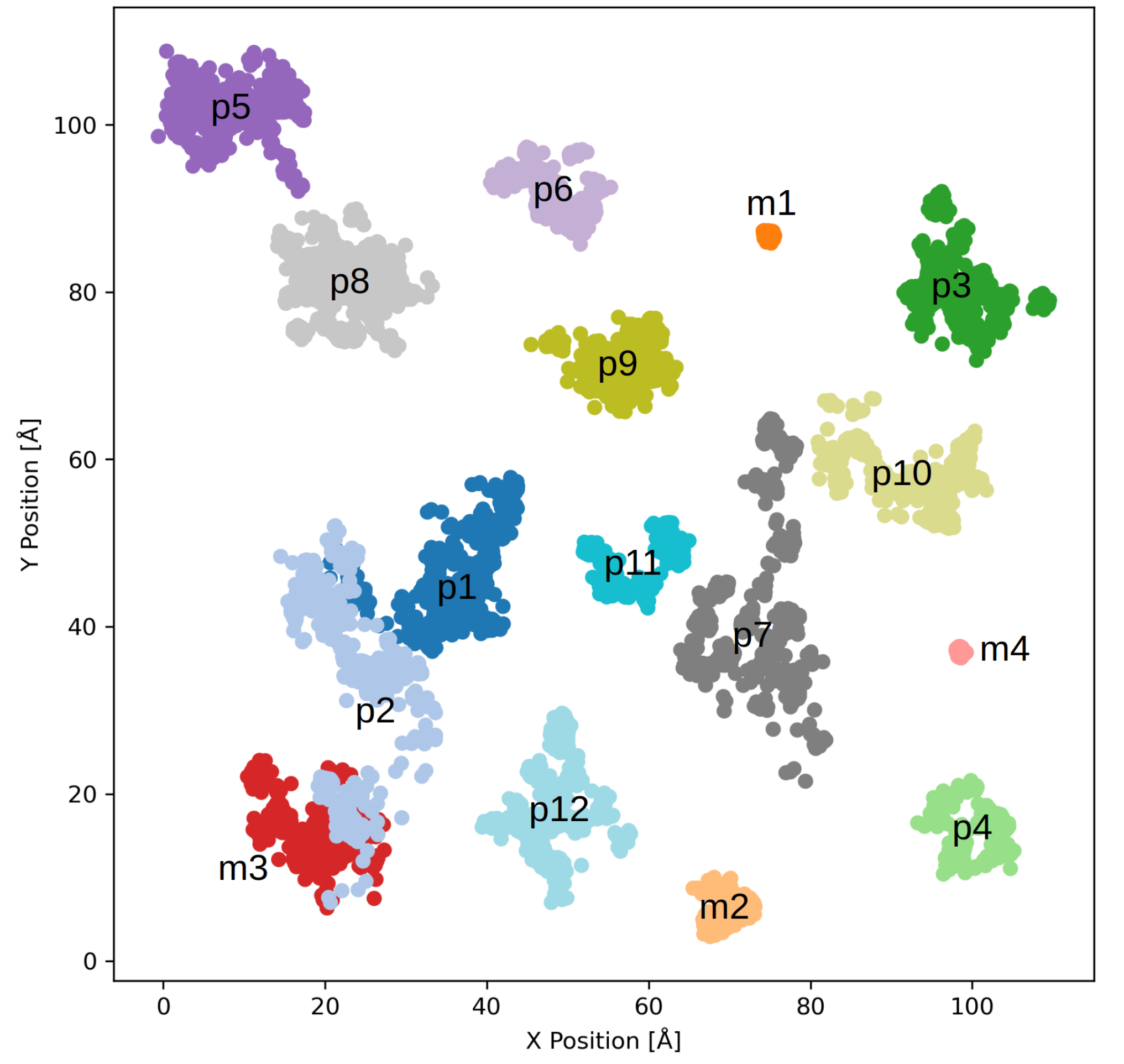}
        \caption{Single Copy}
        \label{fig:cm_single}
    \end{subfigure}
    \hfill
    \begin{subfigure}[b]{0.48\textwidth}
        \centering
        \includegraphics[width=\textwidth]{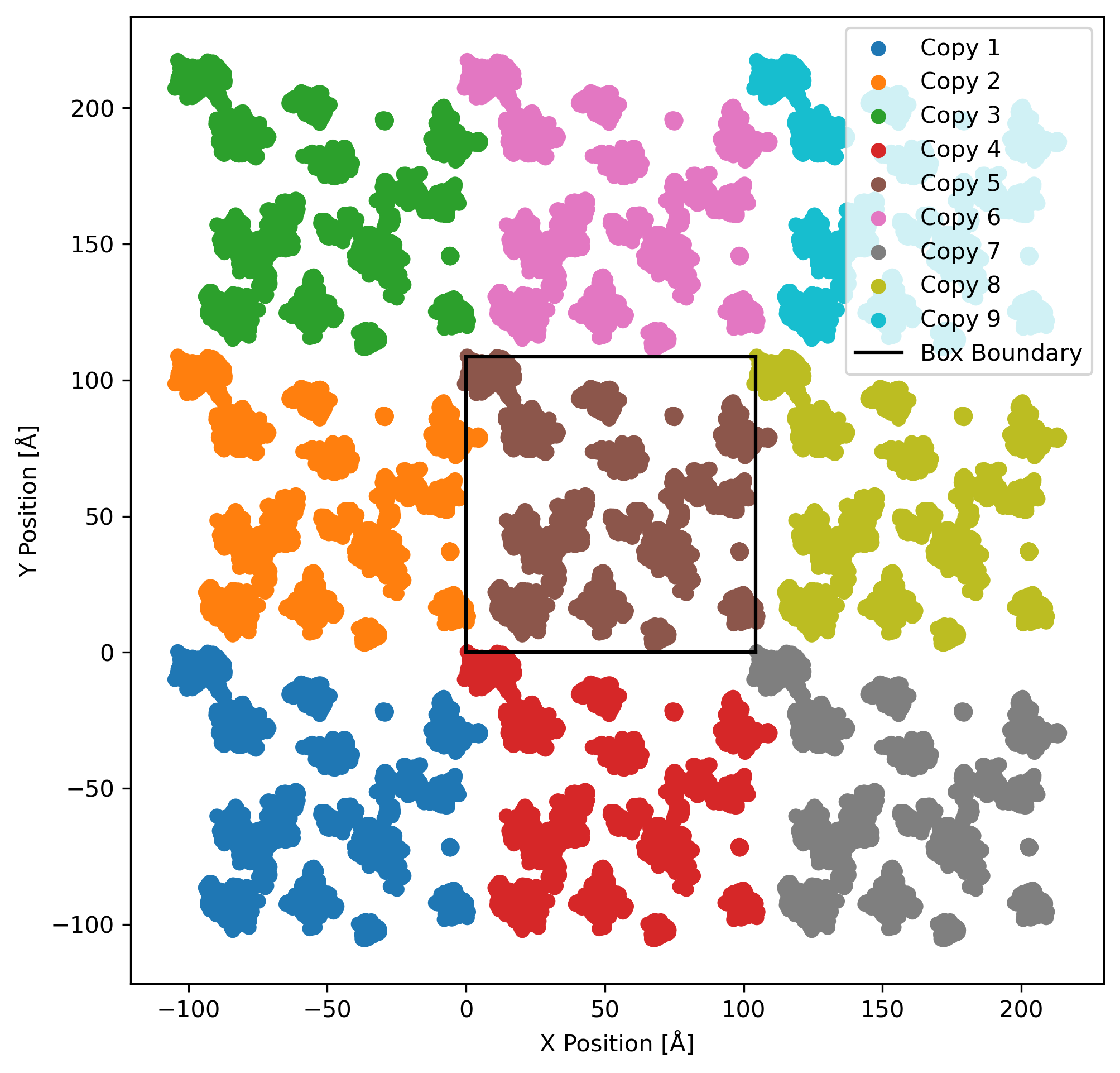}
        \caption{All Copies}
        \label{fig:cm_9copy}
    \end{subfigure}
    \hfill
    \caption{Scatter plot of the center of masses of the dyes reported every 100~ps along the 100~ns trajectory in (a) a single unit cell and (b) in a plot also showing the nearest periodic images to all sides in the plane of the clay surface.}
    \label{fig:cm_scatter}
\end{figure}
Furthermore, an additional 100~ns-long classical MD simulation was performed producing 10,000 snapshots at a stride of 10~ps  in order to be able to better evaluate the positioning  and movement of the porphyrin dyes  on the clay surface and to determine the excitonic couplings between them.
In Figure \ref{fig:md},  three snapshots at 0, 50 and 100 ns of this additional trajectory give an impression of the positions and conformational difference of the porphyrin molecules along the trajectory. 
The m-TMPyP molecules (red) are tightly adsorbed, with all four side chain groups interacting closely with the clay surface for most of the simulation time. It should be noted that the adsorption positions of the side chains are quite close to the positions of the magnesium defects in the montmorillonite. In Refs.~\citenum{eguc04a} and  \citenum{ishi11b} this has been termed the “size-matching rule”. 
On the contrary, only two to three side chains of the p-TMPyPs molecules (blue) are adsorbed on the surface, and the rest are “wandering” in the water environment, making the molecules “stand” with a certain angle, and move more easily on the clay surface.
In order to better visualize the range of movements of the porphyrins along the trajectory, the center of mass of the pigments along the trajectory was extracted and plotted in Figure \ref{fig:cm_scatter}. Compared to the m-type porphyrins, which are approximately fixed on the surface, there are distinct movements of the p-type ones.  These movements also significantly modify   
the now time-dependent distances between porphyrins. In addition, note in particular that although Figure \ref{fig:cm_single} suggests a large range of motion for the porphyrin m3, it only starts moving after molecule p2 approaches.
\begin{figure}[b!]
    \centering
    \begin{subfigure}[b]{0.32\textwidth}
        \centering
        \includegraphics[width=\textwidth]{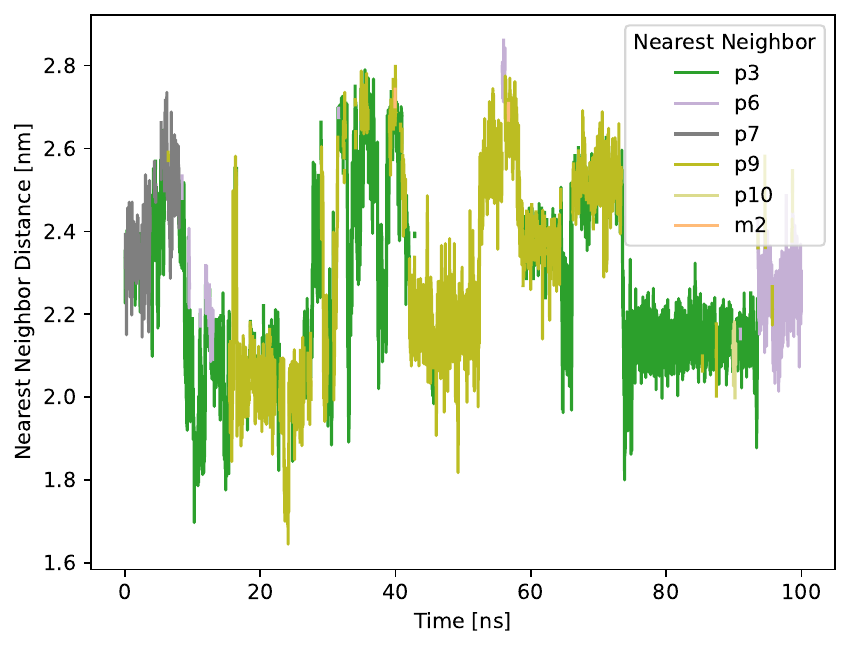}
        \caption{Nearest neighbors of m1.}
    \end{subfigure}
    \hfill
    \begin{subfigure}[b]{0.32\textwidth}
        \centering
        \includegraphics[width=\textwidth]{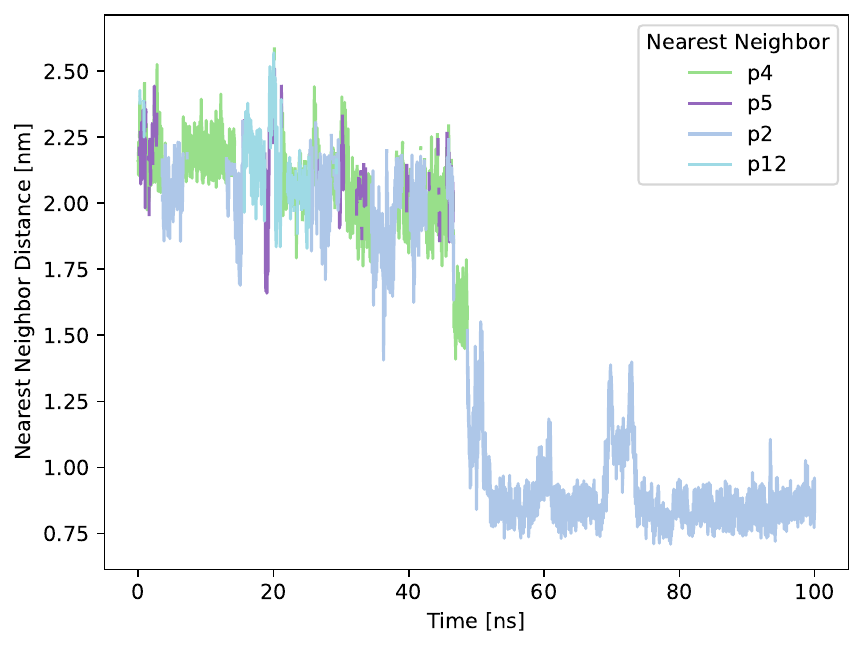}
        \caption{Nearest neighbors of m3.}
    \end{subfigure}
    \hfill
    \begin{subfigure}[b]{0.32\textwidth}
        \centering
        \includegraphics[width=\textwidth]{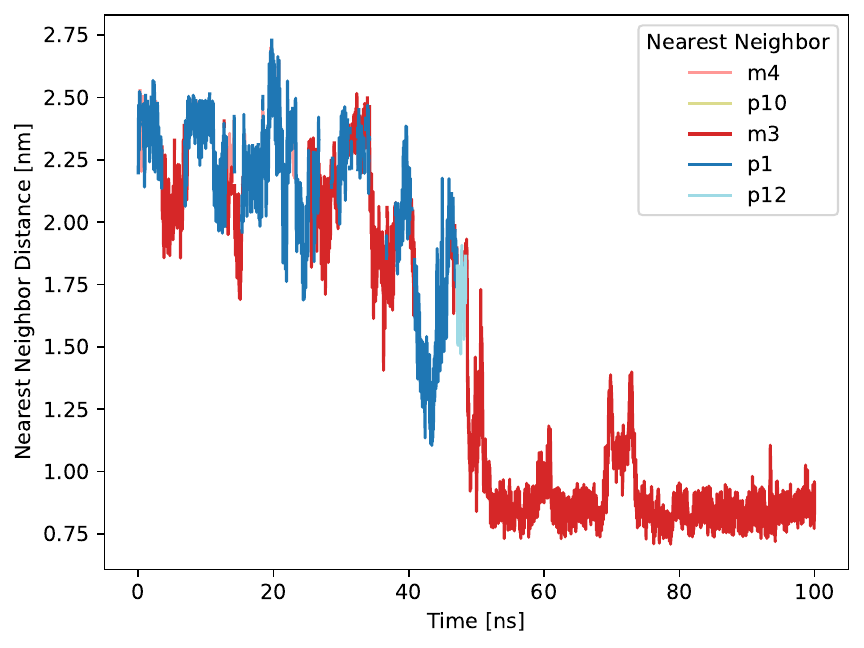}
        \caption{Nearest neighbors of p2.}
    \end{subfigure}
    
    \vspace{0.1cm} 
    
    \begin{subfigure}[b]{0.32\textwidth}
        \centering
        \includegraphics[width=\textwidth]{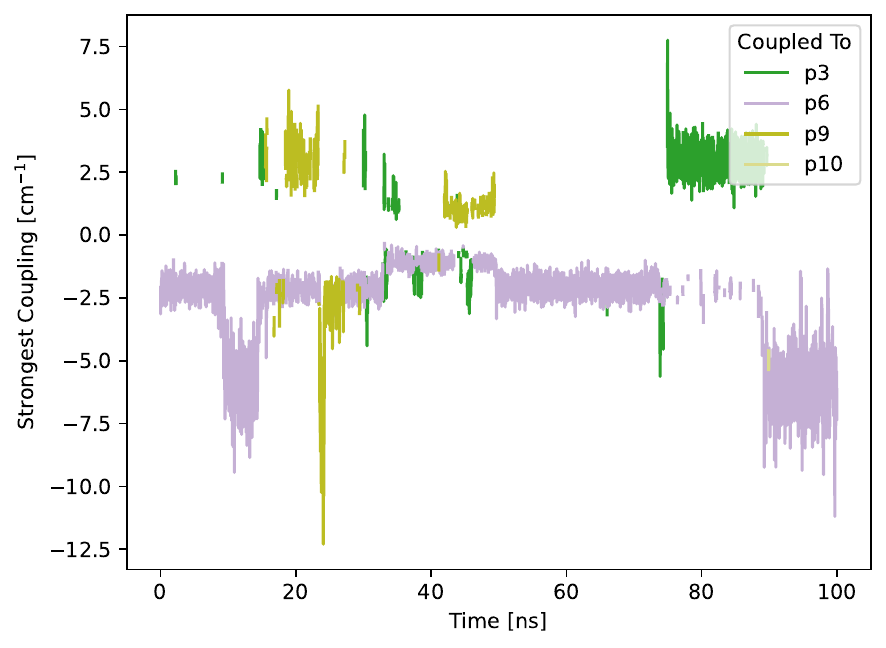}
        \caption{Strongest coupling to m1.}
    \end{subfigure}
    \hfill
    \begin{subfigure}[b]{0.32\textwidth}
        \centering
        \includegraphics[width=\textwidth]{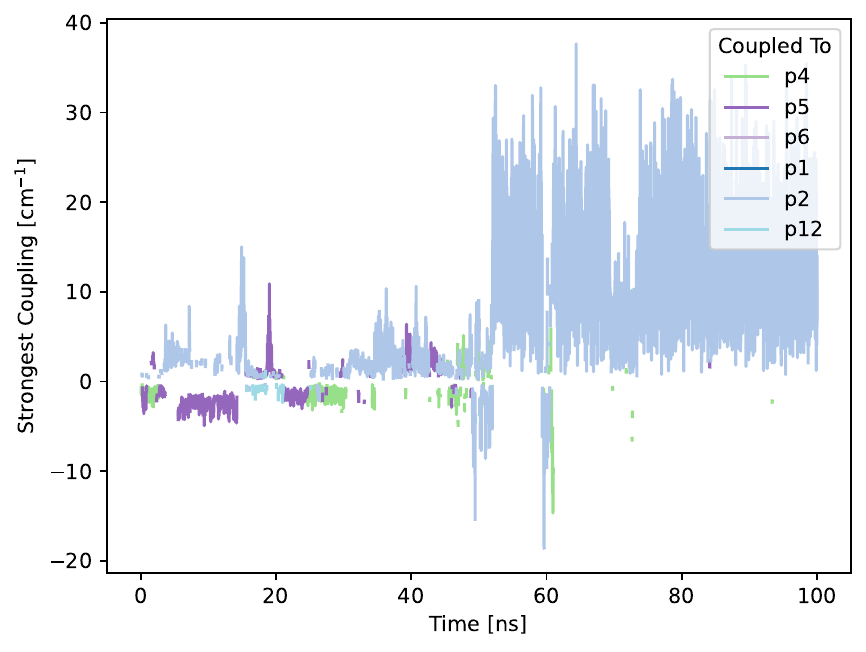}
        \caption{Strongest coupling to m3.}
    \end{subfigure}
    \hfill
    \begin{subfigure}[b]{0.32\textwidth}
        \centering
        \includegraphics[width=\textwidth]{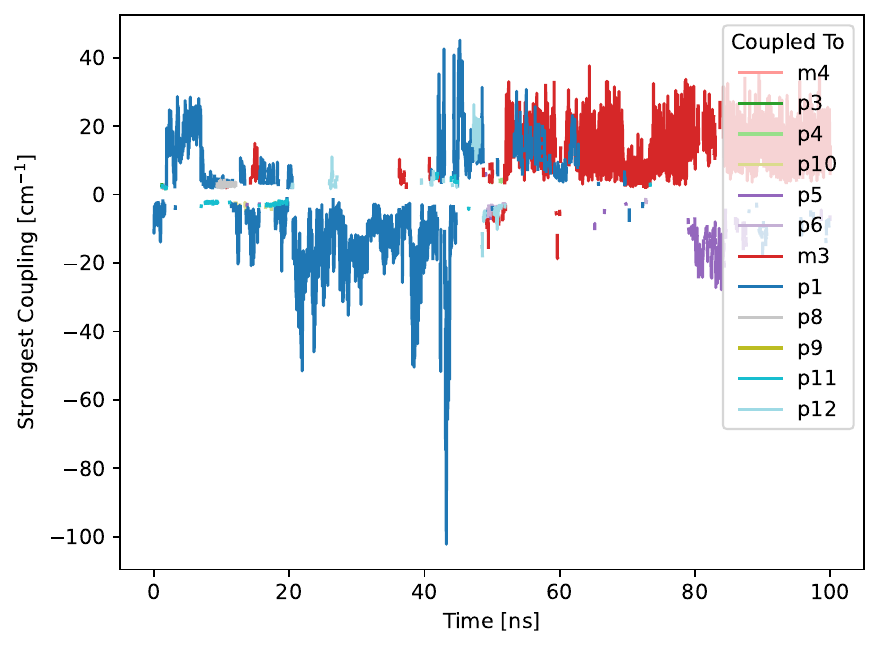}
        \caption{Strongest coupling to p2.}
    \end{subfigure}

    \caption{Nearest neighbor distances (top row) and strongest coupling (bottom row) of select pigments. The colors indicate which neighbor is the nearest (or strongest coupled) at a given time. The nearest neighbors and pigments with the largest coupling value can be across the periodic boundary.}
    \label{fig:combined_figure}
\end{figure}

In natural light-harvesting systems, the chromophores are embedded in a stable protein environment without significant changes in the distances between them. However, in this artificial porphyrin-clay surface system, not only the distance between the porphyrin molecules, but also which dye is the nearest neighbor can change continuously while the porphyrins move on the surface.  These changes in distance between pigments have a significant impact on the excitonic couplings and subsequently the exciton transfer dynamics. Considering the periodic boundary conditions, the distance between each porphyrin and its nearest neighbor was extracted from an extended box with $3\times3$ copies (see Figure \ref{fig:cm_9copy}) over the 10,000 frames. Here, we choose three porphyrins as examples and show the nearest neighbor distances in the top row of Figure \ref{fig:combined_figure}.  The different colors indicate the (frequent) change of the nearest neighbor.
The average nearest neighbor distance across all pigments and snapshots is 2.01 nm, which is of the same magnitude as the value of 2.6 nm reported in the respective experiment \cite{ishi11b}. However, in analysis of the experiment, a perfect hexagonal packing was assumed for the porphyrins on the surface.
The average distance to the six nearest neighbors is 2.92 nm in our case, suggesting that our distances on average are slightly larger than the value reported in the experiment.

The excitonic couplings were calculated based on the Coulomb interaction of the transition charges
\begin{equation}
    V_{mn}(t) = \frac{f}{4 \pi \epsilon_0} \sum_{I,J}\frac{q_{I}^{m,T} \cdot q_{J}^{n,T}}{|r_I^m(t)  - r_J^n(t) |}~,
\end{equation}
where $\epsilon_0$ denotes the vacuum permittivity, $q_{I}^{m,T}$ and $q_{J}^{n,T}$ the transition charges of atom $I$ and $J$ from pigment $m$ and $n$, while $r_I^m(t)$ 
 and $r_J^n(t)$ are the positions and $f$ is the screening factor.
The transition charges were calculated with the “transition charges from electrostatic potentials” (TrESP) method\cite{madj06a,reng13b} based on CAM-B3LYP/6-31G* level of theory and can be found in Tables S1 and S2.
The positions were based on the MD trajectory with a time step of 10~ps.
Furthermore, we chose the pairwise screening factor introduced in Ref.~\citenum{mego14a} as the screening factor $f$.
In a final step, we usually rescale the coupling to match experimental values. Since no experimental dipole moments for these types of porphyrin were available, we did not rescale the coupling.
The largest absolute coupling value was extracted for each porphyrin along the trajectory. The bottom row of Figure \ref{fig:combined_figure} shows the highest coupling values for the same pigments as in the top row. The signs of the coupling values can vary over time, as the nearest neighbor and the orientation of the porphyrin molecules with respect to each other change over time. Therefore, the coupling values indirectly also contain some information about the rotation of the molecules, i.e., mainly of the p-type molecules since the m-type molecules are rather immobile.
Usually, the closest pigment is also the one which has the largest excitonic coupling value.  In cases of similar distances to the nearest and the next nearest neighbor, however, the orientation of the pigments to each other becomes the determining factor.
The average highest absolute coupling is 8.32 cm$^{-1}$, while the median is only 4.66 cm$^{-1}$ suggesting strong outliers as can also be seen in Figure \ref{fig:combined_figure}. With such strong fluctuations and even change of sign, the usual practice of taking average couplings is likely  flawed in the present case, so that we had to take the time dependence of the couplings into account for the calculations described below.

\section{Machine Learning} \label{ML_results}
Expensive calculations of the excitation energy for the first excited state are replaced by evaluations of MFML models, one model for all 12 p-type molecules and one model for all m-type molecules. Prior to building these two MFML models on a large, computationally expensive training dataset, the characteristics of the porphyrin data of the system are being studied in the context of ML model construction for cheap single-fidelity data, namely excitation energies calculated using the TD-LC-DFTB approach. A particularly important topic is the transferability of the final models across the pigments. Also, the impact of using Active Learning is studied.

\subsection{Transferability Analysis} \label{singlefidelity_results}
\begin{figure}[bh!]
    \centering
    \begin{subfigure}[b]{0.6\textwidth}
        \centering
        \includegraphics[width=\textwidth]{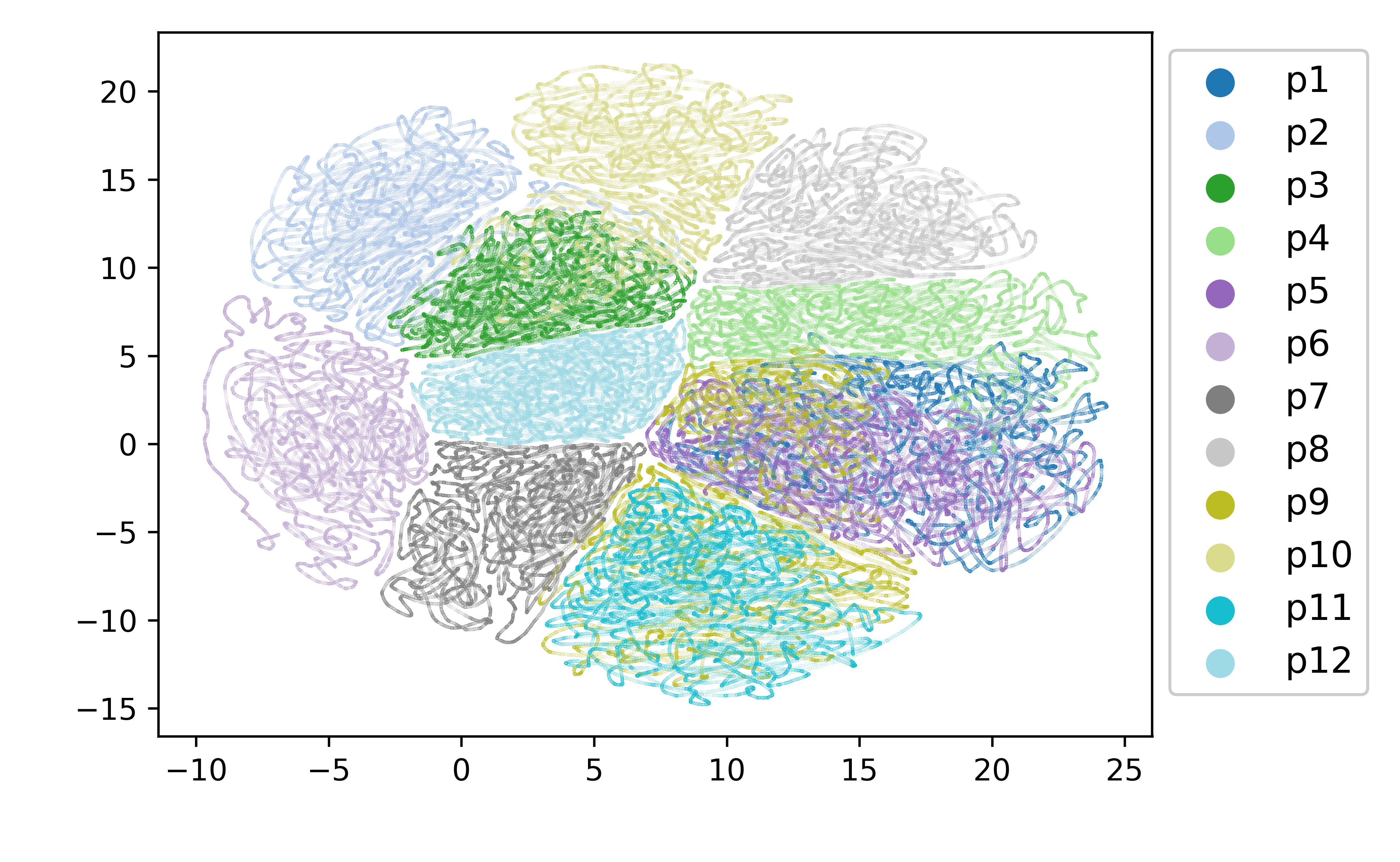}
        \caption{}
        \label{fig:single_fidelity_analysis_umap}
    \end{subfigure}
    
    \vfill
    \begin{subfigure}[b]{0.45\textwidth}
        \centering
        \includegraphics[width=\textwidth]{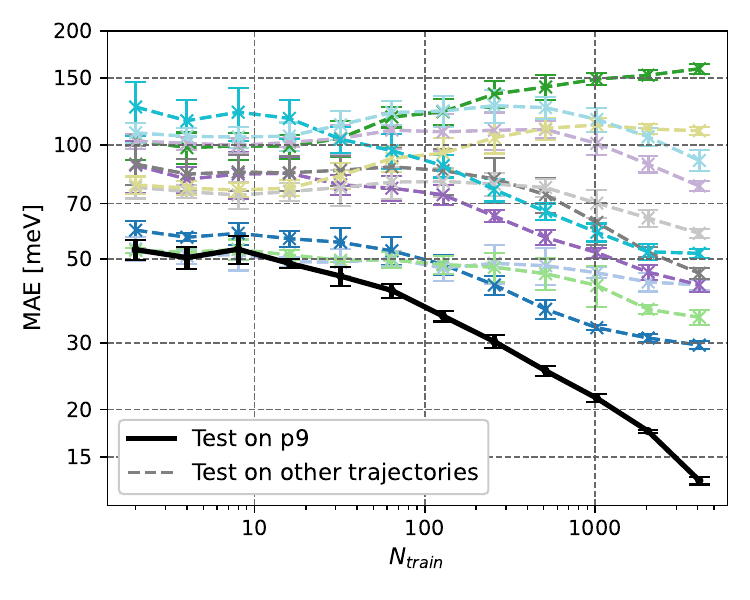}
        \caption{}
        \label{fig:single_fidelity_analysis_p_traj1}
    \end{subfigure}
    \hfill
    \begin{subfigure}[b]{0.45\textwidth}
        \centering
        \includegraphics[width=\textwidth]{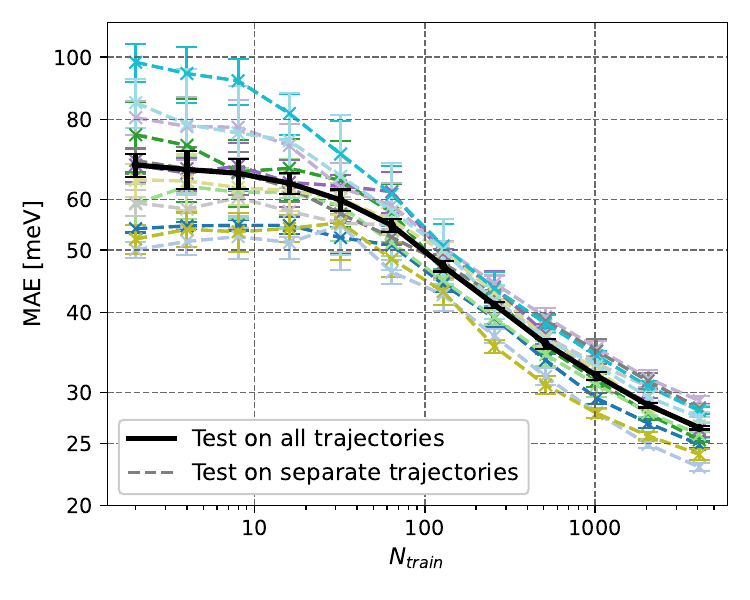}
        \caption{}
        \label{fig:single_fidelity_analysis_concatenated}
    \end{subfigure}
    
    \caption{Results of a transferability study of single-fidelity models across p-type pigments: (a) For an understanding of the variations in the porphyrin conformations, a UMAP  (Uniform Manifold Approximation and Projection) plot is provided based on the MD trajectories for the 12 p-type porphyrin molecules. (b) Learning curves for a single-fidelity model trained on p9 and separately evaluated on all p-type pigments. The model does not generalize to the other pigments.
    (c)  Learning curves for a single-fidelity model trained on the union of the trajectories of all p-type pigments. The model generalizes much better. Note that for better visibility, the vertical axis of (b) and (c) have different ranges.}
    \label{fig:single_fidelity_analysis}
\end{figure}

In Figure \ref{fig:single_fidelity_analysis} we analyze the configuration spaces of the 12 different p-type molecules, which are encountered during the MD run, and how they relate to each other. In Figure \ref{fig:single_fidelity_analysis_umap} a UMAP  (Uniform Manifold Approximation and Projection) plot \cite{mcin20a} of the Coulomb representations of the 40,000 configurations of every p-type trajectory is shown. It can be seen that, in most cases, the configuration spaces, which are covered by the different molecules, are clearly differentiated. Only for a few cases, the configuration spaces of two molecules overlap, e.g., for p9 and p1. These distinct configurations spaces are problematic, since they give a first indication towards a bad transferability of ML models being trained on a single trajectory towards model evaluations on the trajectories of the remaining pigments.

To study this transferability further, we train a single-fidelity Gaussian Process Regression (GPR) \cite{rasmussen2006gaussian} model only on p9 data, with hyperparameters being optimized on a separated, randomized validation subset with 1,000 samples of the p9 trajectory.  Figure \ref{fig:single_fidelity_analysis_p_traj1}, shows learning curves, i.e., an analysis of the prediction error for a growing number of training samples. Here, the model built on p9 data is evaluated on separate, randomized subsets of size 2,000 for each p-type molecule. Results reflect an average over five GPR models that are trained using different random selections of training data, excluding the test and validation sets.
The model trained exclusively on p9 fails to make reasonable predictions for configurations from the other trajectories. The only exception is, when testing on p1, where the error decreases with an increasing number of training samples, although the error is significantly higher than for the test set of p9. In the UMAP, see Figure~\ref{fig:single_fidelity_analysis_umap}, the configuration spaces of p1 of p9 are one of the few configuration spaces of different trajectories which are overlapping. This shows that the differences in configuration spaces among the various trajectories, as illustrated in the UMAP, are indeed substantial. A model trained solely on samples from one trajectory cannot be transferred to predict configurations from other trajectories.

To overcome this challenge, any subsequent model used in this overall study is trained over the union of all trajectories of all pigments of one type. Figure \ref{fig:single_fidelity_analysis_concatenated} depicts learning curves for a model trained on samples from all trajectories of p-type molecules, using the same construction approach outlined for the previous set of learning curves. This joint model shows a prediction error that decreases with an increasing number of training samples across the test sets of all different p-type pigments. We hence obtain a model with largely uniform prediction errors cross the individual pigments. Still, it needs to be noted that the prediction error for the joint model is higher than the error of the model trained only on p9, when predicting the p9 test configurations. This is however not surprising as we, in average, only take $1,000/12$ samples from trajectories of each pigment, compared to the $1,000$ of only p9 in the other case. At this point, we balance transferability with model accuracy.

\subsection{Assessing Active learning}

One important question is, whether we can gain from specifically choosing training samples out of a set of unlabeled candidate molecular configurations, in contrast to randomly  or uniformly sample training data from a given molecular trajectory. This is what is done in Active Learning (AL). In an exemplary study, we analyze the impact of using uncertainty sampling with the Gaussian Process Regression standard deviation as uncertainty measure to select training samples, rather than using a randomized sampling.

\begin{figure}[htb!]
    \centering
    \includegraphics[width=0.64\textwidth]{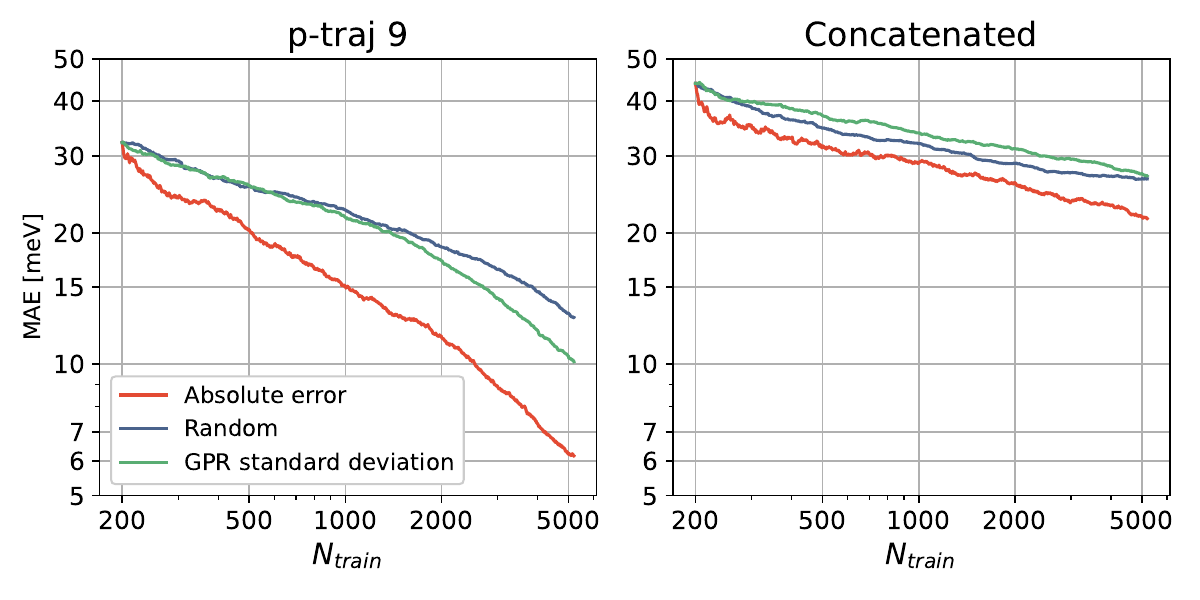}
    \caption{Comparative study between randomized training sample selection and Active Learning using GPR standard deviation for a model build for a single pigment \textit{(left)} and for a model build on all p-type pigments \textit{(right)}.}\vspace*{-1em}
    \label{fig:uncertainty_sampling_LC}
\end{figure}

Figure~\ref{fig:uncertainty_sampling_LC} shows, on the left-hand side,  learning curves comparing randomized sampling to active learning for a model that is only built on training samples of p9. Besides of the selection of the particular training set, the same data and same error evaluation approach as in the last section is used. We start the study from a model that is build on 200 randomly selected samples and consecutive add samples form the remaining full trajectory, either with active learning or in a randomized fashion. As a reference, we alternatively add samples with the highest actual absolute error, to show the optimum that we could achieve with a greedy sampling scheme such as AL. 
For p9, it can be seen that uncertainty sampling with GPR standard deviation results in a slightly improved prediction error compared to random sampling. The (red) reference result indicates, that there would be room for further improvement with a greedy sampling scheme, if a better uncertainty measure than GPR standard deviation would be available.

In a second study, for which the results are depicted in Figure~\ref{fig:uncertainty_sampling_LC} on the right-hand side, active learning is tested on the union over all trajectories for all p-type pigments, as we need it for transferability, see last section. In this case, randomized sampling gives better results than AL, in contrast to the result for p9, only. Based on this mixed picture, we decided to stick to a randomized training sample selection strategy for the remaining construction of MFML models.

\subsection{Multifidelity Machine Learning Models}
The final models for the prediction of the excitation energies of the p-type and m-type porphyrin pigments are multifidelity machine learning models (MFML). Such a MFML model involves the use of quantum chemistry training data at different \textit{fidelities}, which refer to the accuracy of the data with respect to the actual value \cite{zasp19a}. The MFML models are built with respect to some \textit{target fidelity} and a \textit{baseline fidelity}. The former, denoted hereon as $F$, refers to the most accurate (and thereby costliest) fidelity that one is interested in predicting with the ML model. The baseline fidelity, $f_b$, is the cheapest (and thereby the least accurate) fidelity data that is used in the MFML model. This MFML model is denoted by $P_{MFML}^{(F,\eta_F;f_b)}$ where $2^{\eta_F}=N^{(F)}_{train}$ is the number of training samples used at the target fidelity. The number of training samples at the subsequently cheaper fidelities are determined by the use of a \textit{scaling factor}, $\gamma$ which is conventionally set to 2 based on Ref.~\citenum{zasp19a}. That is, $N_{train}^{f}=\gamma\cdot N_{train}^{f-1}$ for all $f_b<f\leq F$. The detailed theoretical framework of MFML is further presented in the Methods Section on the \nameref{mfml_method}. In this work, the target fidelity is TD-DFT/CAM-B3LYP with the def2-SVP basis set while the cheapest baseline fidelity is the LC-DFTB approach. These will be reported using a shorthand notation, that is SVP and DFTB, respectively. The different ML models are assessed based on the MAE using learning curves which depict the MAE as a function of the number of training samples used at the target fidelity. In addition, the MAE is studied as a function of the time-cost of generating training data for a specific ML model, be it single fidelity GPR or MFML models with different values of $f_b$ (see \nameref{model_eval_methods_MAE}). These are reported alongside a recently introduced MFML approach, termed the $\Gamma$-curve which fixes the number of training samples at fidelity $F$ and varies only the value of $\gamma$ \cite{vino24b}. This is shown to be superior to the conventional approach of MFML in providing a low-cost high-accuracy ML model for the prediction of excitation energies of porphyrin.

\begin{figure}[htb!]
    \centering
    \begin{subfigure}[b]{0.32\textwidth}
        \centering
        \includegraphics[width=\textwidth]{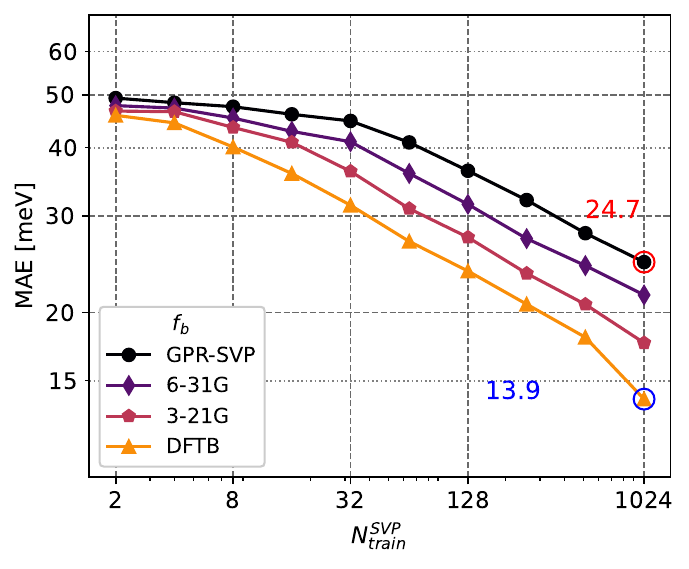}
        \caption{Based only on a trajectory of porphyrin pigment p9.}
        \label{fig_traj9_LC}
    \end{subfigure}
    \hfill
    \begin{subfigure}[b]{0.32\textwidth}
        \centering
        \includegraphics[width=\textwidth]{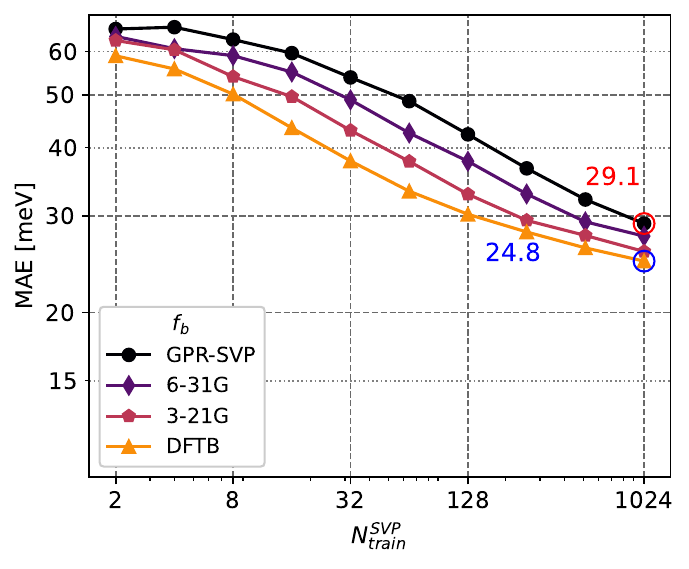}
        \caption{Concatenated trajectories of the p-type pigments.}
        \label{fig_ptype_LC}
    \end{subfigure}
    \hfill
    \begin{subfigure}[b]{0.32\textwidth}
        \centering
        \includegraphics[width=\textwidth]{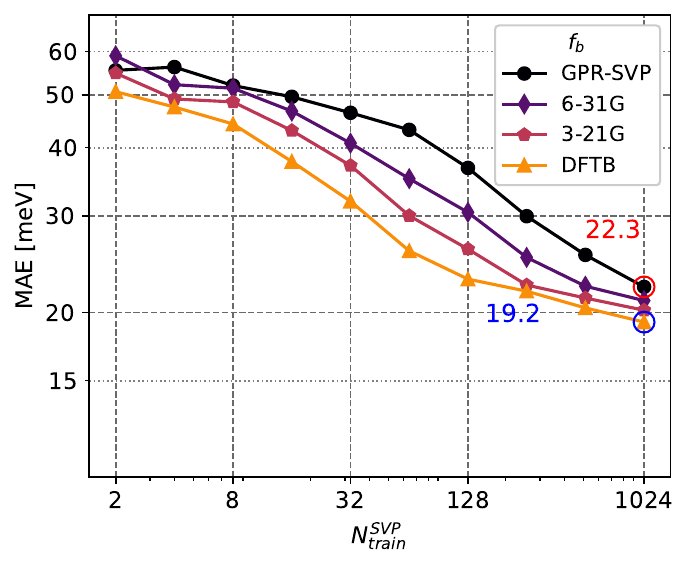}
        \caption{Concatenated trajectories of the m-type pigments.}
        \label{fig_mtype_LC}
    \end{subfigure}
    
    \caption{MFML learning curves for three cases of predicting excitation energies for porphyrin molecules. The prediction error (as MAE) of the single fidelity GPR and  of the MFML model with the DFTB baseline fidelity are explicitly stated for $N_{\rm train}^{SVP}=1024$.}
    \label{fig_combined_MFML_LC}
\end{figure}

As in the previous sections, single fidelity GPR models and MFML models were built and compared on the single pigment p9 and on the union over all p-type pigments for transferability reasons. In addition tests are conducted on the union over all m-type pigments. In the all-pigment models an even sampling of the training data is used.
The models are tested on a separated holdout test set for which the excitation energies are calculated at the target fidelity, that is SVP. For p-type pigments, 2,000 test samples are used, while for m-type pigments 500 test samples are considered, to account for the lower total amount of data.
The resulting MFML learning curves for the different cases are shown in Figure \ref{fig_combined_MFML_LC}. The shown learning curves are an average over ten learning curves created from shuffling the training data set. The different learning curves in a single study are given for a growing number of utilized fidelity levels starting from the baseline fidelity $f_b$, as indicated in the legend.

The case of training and testing on the same trajectory of p-type porphyrin molecules is shown in Figure \ref{fig_traj9_LC} for different $f_b$. With the addition of cheaper baseline fidelities, one observes that the MFML model predicts with a lower MAE in comparison to the single fidelity GPR model built with training samples only from the target fidelity. 
For instance, with $N_{train}^{SVP}=1024$, the GPR model results in an MAE of 24.7 meV while the MFML model with the baseline fidelity $f_b$  set to DFTB results in an MAE of 13.9 meV. While this is a promising result, the transferability study from above indicated that a joint model for all pigments of one type should be constructed. 
For this reason, the final MFML models that are used to predict the excitation energies for the porphyrin molecules are built with training data taken from a pool of trajectories for each type of porphyrin molecule. The MFML learning curves for the p-type porphyrin molecules are delineated in Figure \ref{fig_ptype_LC} for different baseline fidelities. While the addition of cheaper baselines does decrease the model MAE, this drop is not as significant as seen in the case for the single trajectory. The drop in error between single fidelity GPR and MFML with the DFTB baseline fidelity is about 6 meV for $N_{train}^{SVP}=1024$. However, this is anticipated since both the single fidelity GPR and the MFML models have to cover a wider region of the conformational phase space (see discussion on the UMAPs from Figure \ref{fig:single_fidelity_analysis_umap}) as opposed to a smaller region that is to be covered in the case for the single trajectory models. A similar observation is made for the MFML learning curves for m-type porphyrin molecules as seen in Figure \ref{fig_mtype_LC} with the single fidelity GPR model reporting an MAE of 22 meV and the MFML model with DFTB baseline reaching an MAE of 19 meV with 1024 training samples at the SVP fidelity. The slightly overall lower MAE for the m-type porphyrin dyes can be explained once again by the fact that the concatenated trajectories of this porphyrin type result in a lower number of geometries which in turn could span a smaller region of the conformational phase-space as opposed to the case for the p-type porphyrin molecules.
That is, the m-type set has a smaller number of total geometries when concatenated  in comparison to the total geometries of the p-TMPyP set. The larger number of total geometries for the p-type set implies that the MFML model with $N_{train}^{SVP}$ would contain a smaller amount of information about the conformation space of the molecule, in contrast to the MFML model built for m-type set. This fact is reflected in the learning curves. 

\begin{figure}[htb!]
    \centering
    \begin{subfigure}[b]{0.32\textwidth}
        \centering
        \includegraphics[width=\textwidth]{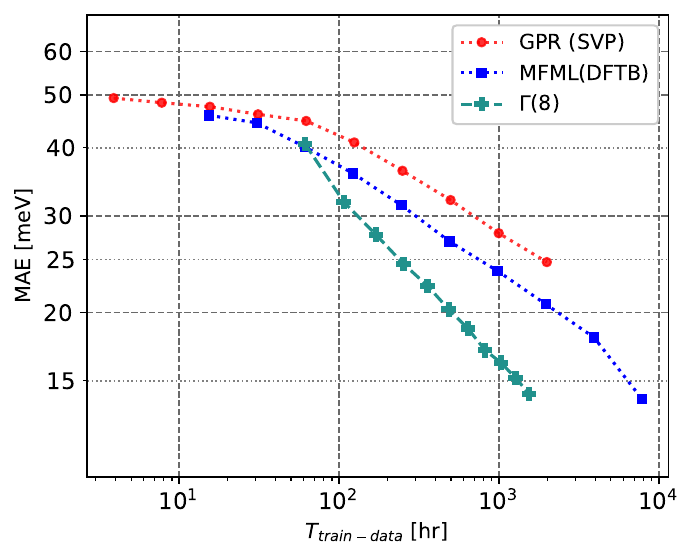}
      \caption{Based only on a trajectory of porphyrin pigment p9.}
        \label{fig_traj9_timecost}
    \end{subfigure}
    \hfill
    \begin{subfigure}[b]{0.32\textwidth}
        \centering
        \includegraphics[width=\textwidth]{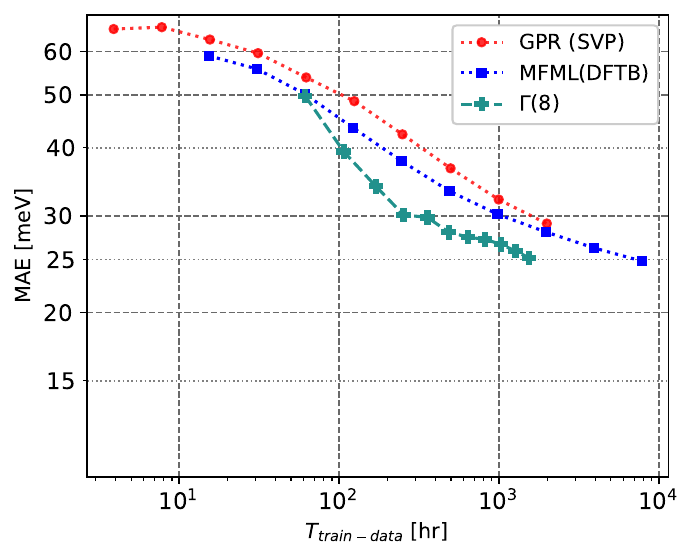}
     \caption{Concatenated trajectories of the p-type pigments.}
        \label{fig_ptype_timecost}
    \end{subfigure}
    \hfill
    \begin{subfigure}[b]{0.32\textwidth}
        \centering
        \includegraphics[width=\textwidth]{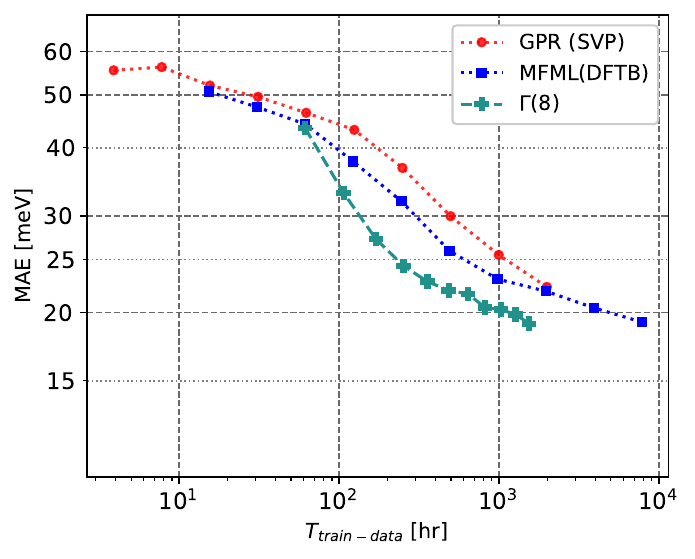}
           \caption{Concatenated trajectories of the m-type pigments.}
        \label{fig_mtype_timecost}
    \end{subfigure}
    
    \caption{Time-cost of generating training data versus MAE in meV for a single fidelity KRR contrasted with that for the MFML model built with baseline fidelity $f_b$ DFTB. The $\Gamma(8)$-curve is also depicted for increasing values of $\gamma$ as explained in \nameref{mfml_method}.}
    \label{fig_simplified_time_cost_combined}
\end{figure}
In order to better assess the computational impact of single fidelity and MFML models for porphyrin molecules, the model error is studied as a function of the cost of generating the training data used in the ML model. In this work, the QC calculation times as returned by the ORCA computing software \cite{nees20a} and DFTB+ software \cite{hour20a, soko21a} are used. For the GPR, this cost is directly related to the number of training samples. For the MFML model, this cost includes not only the training samples used at the top fidelity, but also the cost of the training samples used at the subsequent lower fidelities. These curves are shown in Figure \ref{fig_simplified_time_cost_combined}.
The time required for training the models and predictions over the holdout test set of the MFML model for the largest training set size used (that is, $N_{train}^{SVP}=1024$) was 12.97 seconds and 12.45 seconds for the p-type and m-type porphyrin molecules, respectively. Since this is such a small contribution, only the total time for generating the training data is considered in the MAE versus time-cost curves. 

In addition to the single fidelity GPR and MFML models, a recently introduced MFML approach, referred to as the $\Gamma$-curve \cite{vino24b}, is analyzed as well. In conventional MFML theory, the training samples at the various fidelities are decided by a \textit{scaling factor}, $\gamma$, that is, $N_{train}^f = \gamma\cdot N_{train}^{f-1}$ for $f\in\{2,\ldots,F\}$. Conventionally, a MFML model is built with $\gamma=2$ based on previous studies \cite{zasp19a, vinod23_MFML, vinod_2024_oMFML, Vinod_2024_nonnested}. However, the use of different values of $\gamma$ has recently been studied  resulting in a reportedly more efficient approach titled the $\Gamma$-curve. The $\Gamma$-curve is a plot of MAE versus time-cost of the MFML model with increasing values of $\gamma$. The $\Gamma$-curve is built with a fixed number of training samples at the target fidelity, SVP. Figure \ref{fig_simplified_time_cost_combined} reports the $\Gamma(8)$ curve, that is, with $NM_{train}^{SVP}=8$ with varying values of $\gamma$. Different values of $N_{train}^{SVP}$ were considered and are shown in Figure S9 in the supplementary material. 

Figure \ref{fig_traj9_timecost} depicts the MAE versus time-cost of the ML models for the case of the single trajectory of p-type porphyrin molecules. One observes that for a given time-cost on the horizontal axis, the curve for the MFML model is always below that for the single fidelity GPR curve. This implies that for a given time-cost, the MFML model results in a lower error than the single fidelity GPR model. Furthermore, the $\Gamma(8)$ curve lies lower than the conventional MFML curve. Once again, this implies that for a given time-cost, the multifidelity model built along the $\Gamma(8)$-curve results in a lower MAE. 
A similar observation is made for the case of concatenated trajectories of the p-type and m-type molecules in Figures \ref{fig_ptype_timecost} and \ref{fig_mtype_timecost}, respectively. Although for the m-type porphyrin molecules, the GPR curve does reach close to the conventional MFML curve, the $\Gamma(8)$-curve always lies beneath it. The final multifidelity models that are used in this work for the prediction of excitation energies correspond to the final data point of the $\Gamma(8)$ curve, which corresponds to $\gamma=12$. The multifidelity training structure for this model is $\{8,12\cdot8=216,12^2\cdot8=1152,12^3\cdot8=13824\}$ with decreasing fidelity.  
For the p-molecules, this model results in an MAE of $\sim 25$ meV with a time cost of about 1500 hours, while the conventional MFML model reports a similar error with a time cost of roughly 8000 hours. The single fidelity GPR model only reaches an MAE of 29 meV with a time-cost of 2000 hours. The use of the multifidelity model along the  $\Gamma(8)$ curve results in a time-cost benefit of roughly 5 over the conventional MFML model with $\gamma=2$ and $N_{train}^{SVP}=1024$. 
For the p-type porphyrin molecules, the corresponding time-benefit of using the multifidelity model along the $\Gamma(8)$ curve over the conventional MFML model is about the same with the former reporting an MAE of about 17 meV for a time-cost of roughly 1500 hours, while the latter costs as much as  8000 hours for an MAE of about 19 meV.

\section{Exciton dynamics}
In this section, we investigate the exciton dynamics in the porphyrin-clay system by constructing a time-dependent Frenkel exciton Hamiltonian, employing the Numerical Integration of the Schrödinger Equation (NISE) method, and analyzing the exciton diffusion. Our aim is to simulate exciton dynamics for nine copies of the 16 pigments, totaling  144 pigments (see Figure~\ref{fig:cm_scatter}) over 6~ns with a 1~fs time step. In that way, one experimental lifetime (5.6~ns \cite{ishi11b}) is included in the simulation

\subsection{Building the Frenkel Exciton Hamiltonian}
To model the exciton dynamics, we construct a time-dependent Frenkel exciton Hamiltonian 
\begin{equation} 
\hat{H}(t) = \sum_{i=1}^{N} \left[ E_i + \Delta E_i(t) \right] \ket{i}\bra{i} + \sum_{i \neq j} V_{ij}(t) \ket{i}\bra{j}, \label{eq:H_td} 
\end{equation}
where $E_i$ denotes the average site energies, $\Delta E_i(t)$ the site energy fluctuations, $V_{ij}(t)$  the time-dependent electronic couplings, and $N$  the total number of pigments in the system.
The average site energies $E_i$ and the site energy fluctuations $\Delta E_i(t)$ are obtained from the MFML model, which predicts the excited-state energies based on the atomic positions from the QM/MM trajectories for each pigment, as described in earlier sections.
The electronic couplings $V_{ij}(t)$ are calculated using the TrESP method \cite{madj06a,reng13b} along a 100~ns MD trajectory with a 10~ps time step, as detailed above.
In previous studies, average couplings were often used to simplify the Hamiltonian \cite{reng01,may11a, aght12a,sarn24a}. As discussed in the section on the MD and QM/MM simulations, average couplings may not be suitable here because of significant fluctuations in the couplings and the cancelation of positive and negative values. Therefore, in addition to constructing the Hamiltonian with average couplings, we also construct a Hamiltonian with time-dependent couplings $V_{ij}(t)$, where the initial couplings are chosen as a random snapshot from the MD simulation. The couplings at further times are then taken from snapshots along the MD trajectory, where coupling values between two snapshots are obtained by linear interpolation.

Our goal is to perform a 6~ns exciton dynamics simulation to investigate long-term exciton transport. However, the QM/MM trajectories used to obtain the site energies and their fluctuations are only 40~ps long. Generating 6~ns of QM/MM simulations for every pigment is computationally prohibitive.
To overcome this limitation, we employ a noise generation algorithm based on spectral densities~\cite{holt24a} to create longer site energy fluctuation trajectories $\Delta E_i(t)$. This algorithm allows us to extend the site energy fluctuations to the desired simulation length while preserving the statistical properties of the original data.
As the noise generation algorithm relies on the spectral densities of the site energy fluctuations, we first compute the spectral densities for each pigment type. As a first step, the autocorrelation function $C(t)$ of the site energy fluctuations $\Delta E_i(t)$ is obtained from the MFML model along the QM/MM trajectories using~\cite{damj02a}
\begin{equation}
    C(t_j) = \frac{1}{N-j} \sum_{i=1}^{N-j} \Delta E(t_i + t_j) \Delta E(t_i)~.
\end{equation}
 To suppress noise in the autocorrelation function, a Gaussian damping with a timescale of 5~ps is applied. The spectral density $J(\omega)$ is then obtained via the Cosine transform of the autocorrelation function
\begin{equation}
J(\omega) = \frac{\beta \omega}{\pi} \int_{0}^{\infty} C(t) \cos(\omega t) \, dt~,
\label{eq:jw}
\end{equation}
where $\beta = 1/(k_B T)$ denotes the inverse temperature with $k_B$ being the Boltzmann constant and $T$ the temperature.
Since all pigments of the same type are chemically identical, we average the spectral densities over all pigments of the same type, i.e., m-type or p-type. This averaging minimizes the contributions from specific configurations or local environments present in the relatively short 40~ps QM/MM simulations, leading to a more representative spectral density for each pigment type.
\begin{figure}[tb]
    \centering
    \begin{subfigure}[b]{0.48\textwidth}
        \centering
        \includegraphics[width=\textwidth]{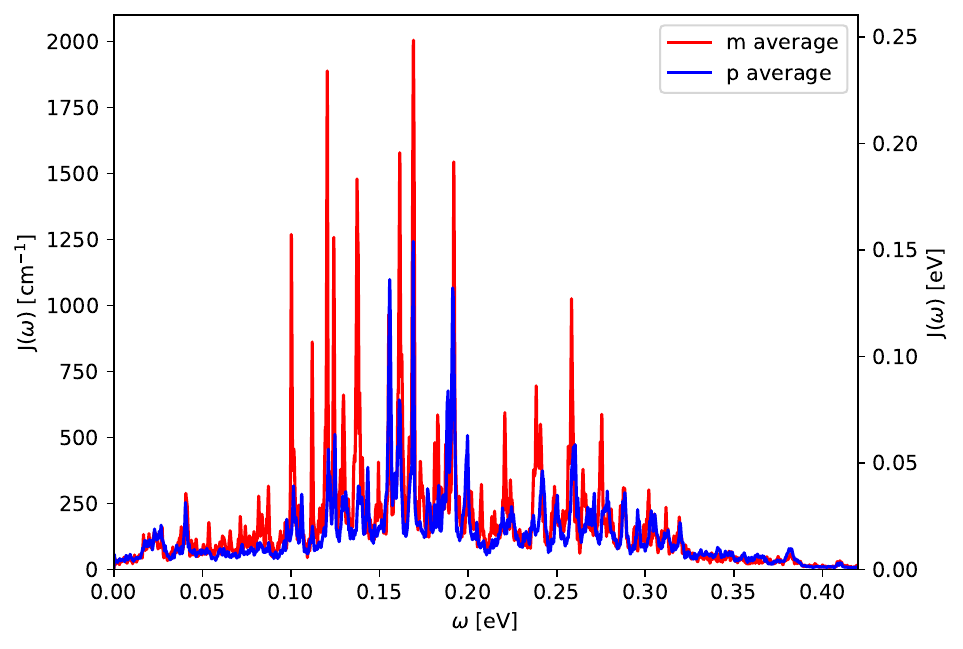}
        \caption{SD of m-type and p-type porphyrin}
        \label{fig_SD_m}
    \end{subfigure}
    \hfill
    \begin{subfigure}[b]{0.48\textwidth}
        \centering
        \includegraphics[width=\textwidth]{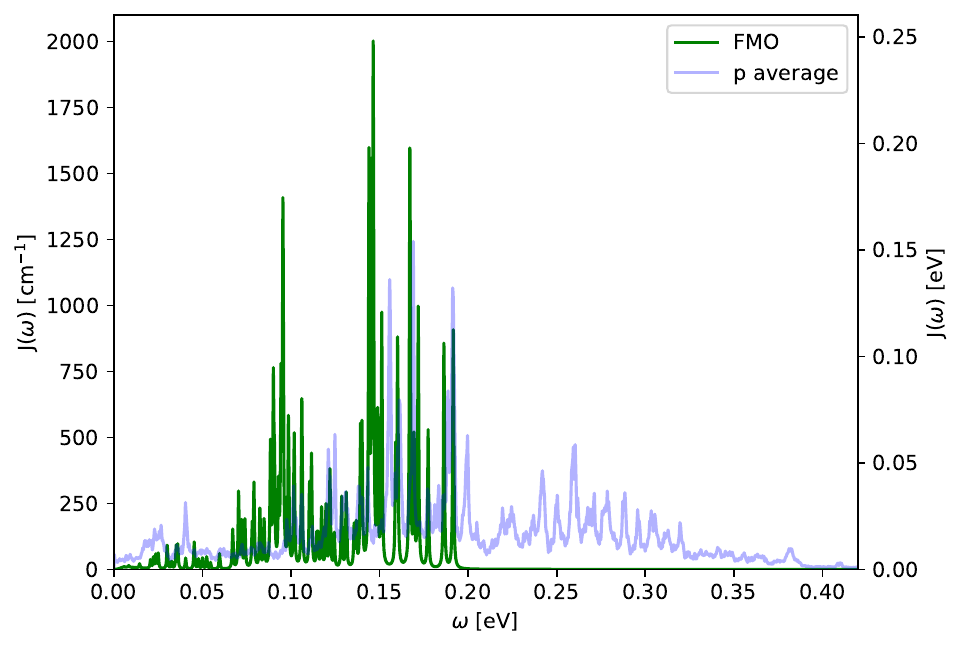}\caption{Experimental SD of FMO.}
        \label{fig_SD_p}
    \end{subfigure}
    \caption{Panel (a) shows the average spectral densities of all m-type  and p-type porphyrin molecules, while panel (b) shows an experimental spectral density of FMO as comparison \cite{raet07a,mait20a}.}
    \label{fig_SD}
\end{figure}
The spectral densities for the m-type  and p-type porphyrin molecules are shown in Figure \ref{fig_SD}, together with an experimental spectral density of the Fenna–Matthews–Olson (FMO) complex containing bacteriochlorophyll molecules (BChl) \cite{raet07a,mait20a}. Notably, the porphyrins exhibit spectral features up to approximately 0.4eV, which is significantly higher than the spectral features of BChl or chlorophyll (Chl) in plants, typically extending only up to about 0.2eV. These higher frequency components might originate from the outer rings of the porphyrin molecules, which are absent in BChl and Chl, or the central nitrogen-hydrogen (N-H) bonds, which have the highest force constant within porphyrin molecules \cite{kim81a}.

To generate the noise $\Delta E_i(t)$ following the spectral densities, we first create white noise $\eta(t)$ with zero mean and unit variance. The target noise $\Delta E_i(t)$ is then obtained by multiplying the Fourier transform of the white noise $\tilde{\eta}(\omega)$ with the square root of the power spectrum $\tilde{C}(\omega) = J(\omega) 2\pi / (\beta\omega)$  followed by an inverse Fourier transform~\cite{holt24a}
\begin{equation}
    \Delta E_i(t) = \Re (\textbf{IFFT}(\tilde{\eta}(\omega) \tilde{C}(\omega)) )~.
\end{equation}
This process yields noise $\Delta E_i(t)$ that follows the desired spectral density and can be used for the desired 6~ns simulations.
Combining all components, the final Hamiltonian $\hat{H}(t)$ is constructed with average site energies $E_i$ from the MFML model along the QM/MM trajectories, time-dependent fluctuations $\Delta E_i(t)$ generated from the noise algorithm, and the time-(in)dependent electronic couplings $V_{ij}(t)$ obtained from average linearly interpolated TrESP couplings along the MD simulation.

\subsection{Numerical integration of Schrödinger equation}
To enable exciton calculations over 6~ns for 144 pigments, i.e., 9 copies of 16 pigments, a numerically efficient algorithm is needed. To this end, we choose the Numerical integration of Schrödinger equation (NISE) as it is fast, easily parallelizable and works with the time dependent Hamiltonian we just constructed \cite{jans06a,aght12a, holt24a}. The NISE approach treats only the system quantum mechanically and the coupling to the environment classically. This treatment includes an implicit high temperature limit, so that detailed balance is not fulfilled, i.e., the long term distribution is an even distribution rather than the expected Boltzmann distribution \cite{para06a}. Modifications to the NISE that address this limitation exist \cite{jans18a,holt23a}, but have shortcomings in systems with low couplings $V_{ij}$\cite{holt23a} and high frequency contributions in the noise\cite{holt24a}, both of which are present in the porphyrin-clay system. Therefore, we proceed without the modifications, but must keep the limitations of the NISE approach in mind when interpreting the results.

\begin{figure}[tb]
    \centering
    \begin{subfigure}[b]{0.48\textwidth}
        \centering
        \includegraphics[width=\textwidth]{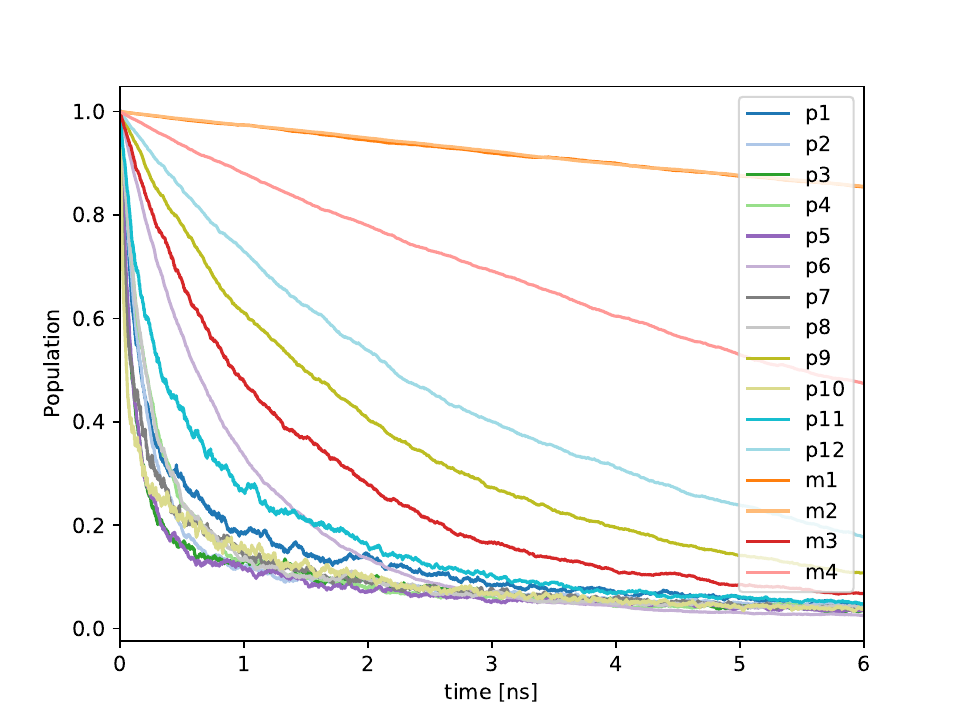}
        \caption{Average coupling values.}
    \end{subfigure}
    \hfill
    \begin{subfigure}[b]{0.48\textwidth}
        \centering
        \includegraphics[width=\textwidth]{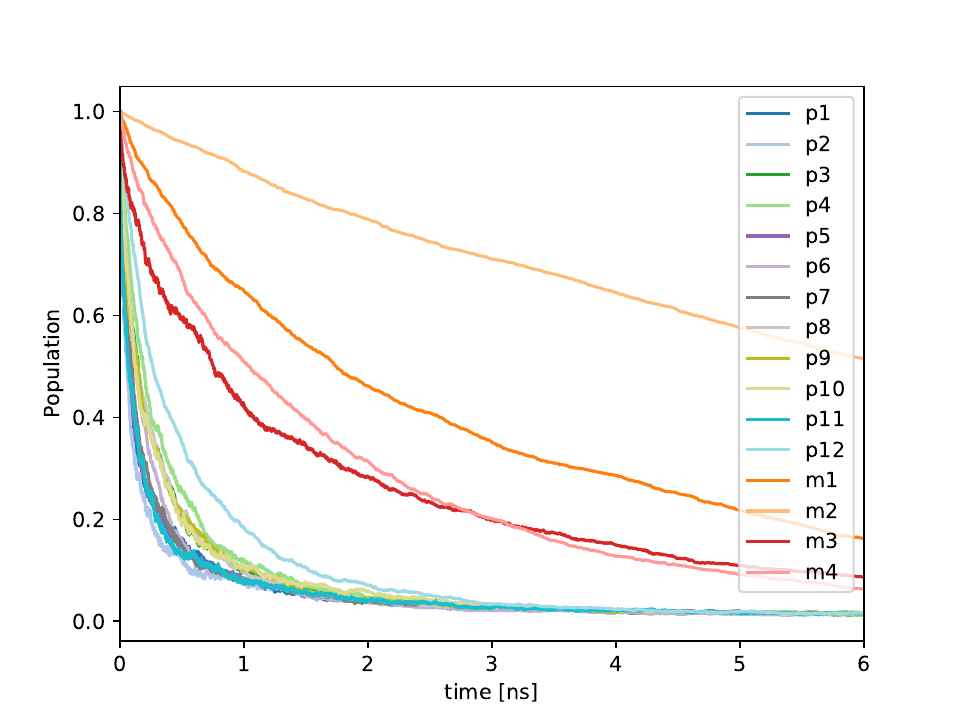}
        \caption{Time-dependent coupling values.}
    \end{subfigure}   
    \caption{Population dynamics of an exciton initially placed on the respective pigment in the central copy. The populations were calculated using an average over 200 realizations. Panel (a) shows the simulation for average coupling values, and panel (b) for time-dependent coupling values.}
    \label{fig_lifetimes}
\end{figure}

In the NISE method, we solve the time-dependent Schrödinger equation for the excitonic wave function $\ket{\psi_s(t)}$
\begin{equation} 
i\hbar \frac{\partial}{\partial t} \ket{\psi_s(t)} = \hat{H}(t) \ket{\psi_s(t)}~, \label{eq:TDSE} 
\end{equation}
where $\hat{H}(t)$ is the time-dependent Hamiltonian we constructed earlier. The excitonic state can be expressed in terms of site basis states $\ket{m}$
\begin{equation} \ket{\psi_s(t)} = \sum_{m} c_m(t) \ket{m}~, \end{equation}
with time-dependent coefficients $c_m(t)$. These coefficients can be obtained by solving Eq.~\eqref{eq:TDSE} numerically by assuming $H(t)$ is constant for small time steps $\Delta$t, where we use a time step of 1~fs.  The population of an exciton on site $m$ is then given by
\begin{equation} P_m(t) = |c_m(t)|^2. \end{equation}
Finally this has to be repeated for many realizations to get reasonable results.
The software code TorchNISE which is available on GitHub is used to run the calculations \cite{holt24b}.
The exciton dynamics are simulated by initially exciting a single porphyrin molecule in the central copy and observing the population decay over time as depicted in  Figure \ref{fig_lifetimes}. Two versions are shown for the Hamiltonian constructed based on the average and on the time-dependent coupling values. 
The results indicate significant differences in the decay of the exciton population between the two cases. With average couplings, certain pigments (most notably p9, p12 and m1) retain higher populations over longer times, which is not observed with the time-dependent coupling values. We suspect that averaging the couplings, which can be both positive and negative due to the relative orientations of the transition dipole moments (see Figure \ref{fig:combined_figure}), leads to an underestimation of the effective coupling strengths as positive and negative values cancel out each other.
Moreover, the transfer from pigment m3 shows notably different behavior between the two cases. In simulations with average couplings, the exciton population on m3 shows a continuous exponential decay, while the simulations with time-dependent couplings are better described by a double exponential decay; a fast decay at short times is followed by a slower decay at longer times. This behavior can be explained by examining the coupling patterns in the system: pigment m3 is strongly coupled to p2 during certain periods (as shown in Figure \ref{fig:combined_figure}), and p2 becomes strongly coupled to p1 at other times. With average couplings, these interactions are always present in the Hamiltonian, allowing the exciton to continuously transfer from m3 to p2, then to p1, and further into the system. In contrast, time-dependent couplings capture these strong interactions only during specific intervals. The exciton initially equilibrates quickly between m3 and p2 due to their strong coupling at that time, but transfer to other pigments is slower because they are not coupled strongly at the same time.

Despite the improved realism provided by time-dependent couplings, our simulations show that the exciton lifetimes for the m-type porphyrins are much longer than experimental observations. In the experimental study by Ishida et al.~\cite{ishi11b}, the exciton lifetime of m-type porphyrin was determined to be approximately 0.4~ns, whereas in our simulations, the lifetimes of p-type porphyrins are on a similar timescale, but the lifetimes of m-type porphyrins are clearly much longer (see Figure~\ref{fig_lifetimes}). There are several potential sources for this discrepancy. Firstly, we have included only the Q\textsubscript{x} state in our exciton dynamics simulations. Incorporating the Q\textsubscript{y} state might increase the transfer rates significantly when weak couplings are the result of misaligned transition dipoles. Whenever the transition dipole of the Q\textsubscript{x} states of neighboring pigments is not aligned, the dipole of the Q\textsubscript{y} state would be very well aligned with the Q\textsubscript{x} state, thereby increasing overall transfer. Additionally, a higher time-resolution of the couplings might impact the transfer rates. While fluctuating couplings are usually considered less important in exciton calculations in light-harvesting systems, they clearly have an impact on the present system. It is possible that variations of the couplings on faster timescales would further affect the lifetimes. Furthermore, in the experiment, a hexagonal packing of the porphyrins on the clay surface is assumed, with an inter-pigment distance of approximately 2.6~nm~\cite{ishi11b}, whereas in the present simulations, the average distance to the six nearest neighbors is about 2.9~nm. Since dipole-dipole coupling decreases with the 6th power of the distance, we might underestimate the couplings about 50\% due to the larger distances, resulting in slower exciton transfer and longer lifetimes in the simulations. Another possibility is the absence of thermalization effects in the NISE as p-type porphyrin molecules generally have lower energy levels. Finally, the coupling values are usually scaled so that the calculated transition dipole moments match experimental ones \cite{mait20a}. Because no experimental dipole moments for the porphyrins were available, this step was skipped.

\subsection{Diffusion}
To quantify the exciton transport, we analyze the diffusion of the exciton over time. The diffusion is defined as the mean squared displacement (MSD) of the exciton and can be determined using the positions of the porphyrin molecules and the exciton populations
\begin{equation} 
    \langle d(t)^2 \rangle = \sum_{i} |(x_i(t)-x_0(0))|^2 P_i(t)~, 
\label{eq:diffusion} 
\end{equation}
where $P_i(t)$ denotes the exciton population on site $i$ obtained from NISE, $x_i(t)$  the center-of-mass position of porphyrin $i$, and $x_0(0)$  the initial center-of-mass position of the initially excited porphyrin.

As a comparison with experimental values, we employ a simple classical model to describe the exciton transport on a two-dimensional hexagonal lattice with nearest-neighbor spacing of 2.6~nm, as reported in Ref.~\citenum{ishi11b}.
Each site on the grid represents a porphyrin molecule, and the exciton population on site $i$ at time $t$ is denoted $P_i(t)$. We assume that an exciton can transfer from site $i$ to each of its six neighbors with a constant transfer rate $t_r$, leading to the following update equation for the population
\begin{equation} 
\label{eq:hex_mastereq} P_i(t+\Delta t) = P_i(t) - \sum_{j\in \langle i \rangle} t_r P_i(t) \Delta t + \sum_{j\in \langle i \rangle} t_r P_j(t)  \Delta t~, 
\end{equation} where $\langle i \rangle$ denotes the set of six nearest neighbors of site $i$, and $\Delta t$ is a small time step.
In Ref.~\citenum{ishi11b}, the transfer rate from m-type to p-type porphyrins was measured to be 2.4~ns$^{-1}$, with each m-type porphyrin having on average 4.5 p-type neighbors. Because the measured transfer rate describes the total transfer from m-type to p-type porphyrins, we estimate the site-to-site transfer rate to be $t_r=$~2.4~ns$^{-1}$/4.5 $\approx$~0.53~ns$^{-1}$. As no other transfer rates were determined experimentally, we assume this same value for all transfers (m,$\to$,m, p,$\to$,p, m,$\to$,p, p,$\to$,m). This is, of course, a significant assumption, since in the present calculations the transfer from p-type porphyrins is much faster than from m-type. Furthermore, we used the initial condition that the exciton is fully localized on one site, surrounded by 30 hexagonal rings. 30 rings were chosen to ensure that edge effects do not play a role in the simulated time frame of 6~ns, as the exciton does not significantly reach the outermost ring in this timescale. The time step $\Delta t$ was chosen to be 0.1~ps. 
We then propagate the population over time using Eq.~\eqref{eq:hex_mastereq} and calculate the diffusion based on the MSD, i.e., Eq.~\eqref{eq:diffusion}.

\begin{figure}[htb!]
    \centering
    \begin{subfigure}[b]{0.48\textwidth}
        \centering
        \includegraphics[width=\textwidth]{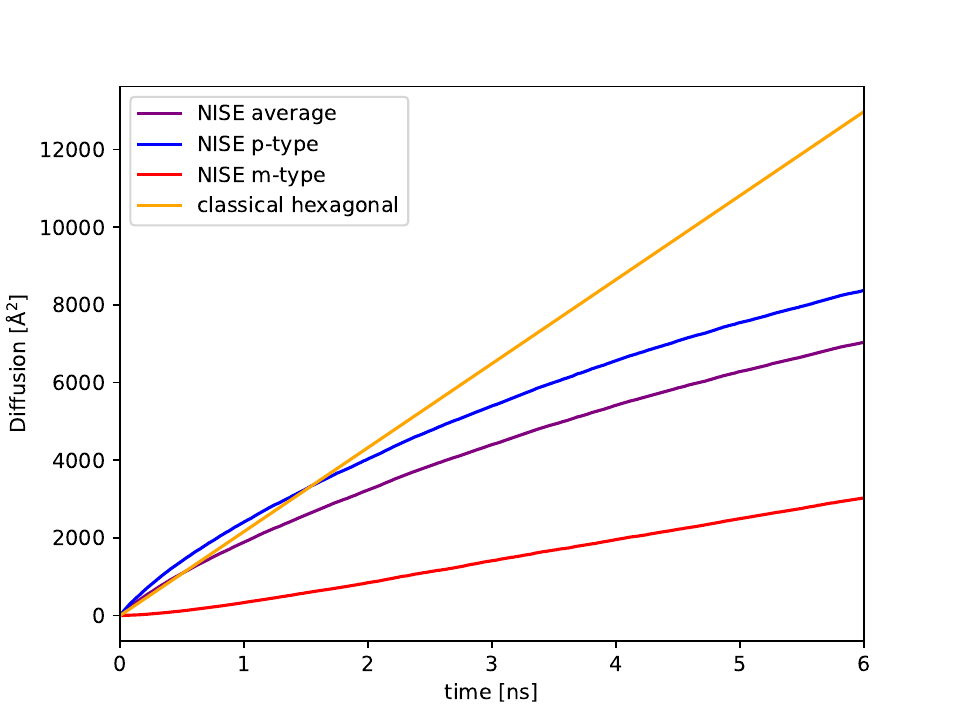}
        \caption{Average Coupling}
        \label{fig_Diffusion_t}
    \end{subfigure}
    \hfill
    \begin{subfigure}[b]{0.48\textwidth}
        \centering
        \includegraphics[width=\textwidth]{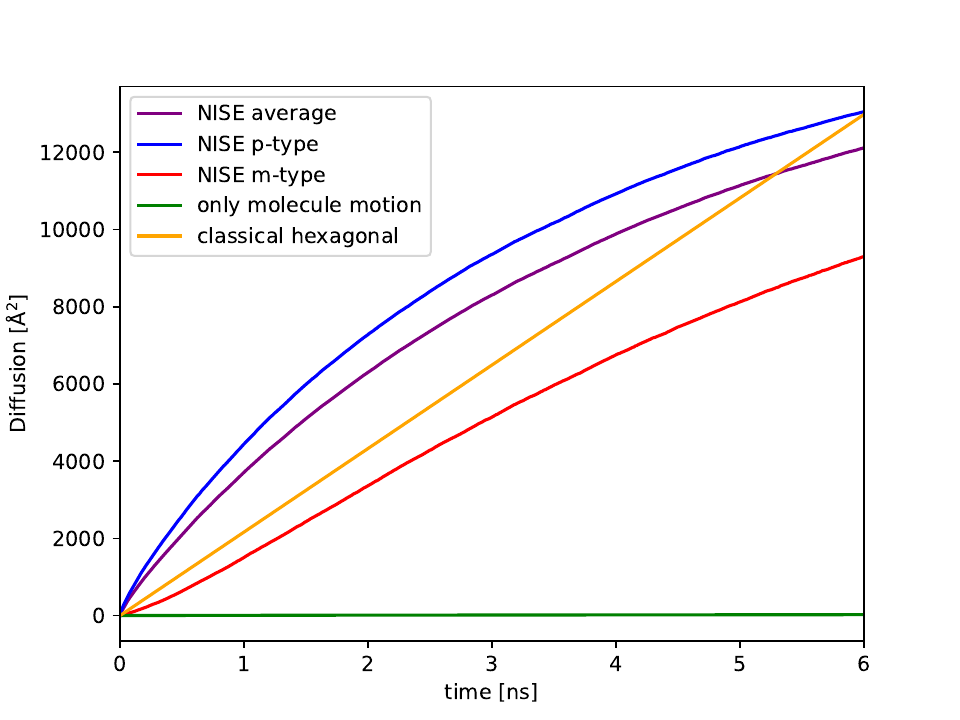}
        \caption{Time-dependent Coupling}
        \label{fig_Diffusion_C}
    \end{subfigure}   
    \caption{Diffusion with average coupling (panel (a)) and time-dependent couplings and positions (panel (b)) calculated from the same NISE simulation as Figure \ref{fig_lifetimes}. In addition, the diffusion based on the classical hexagonal model is shown in both panels and, furthermore, the diffusion based only on the molecular motion is shown in panel (b)}    \label{fig_Diffusion}
\end{figure}
Figure \ref{fig_Diffusion} displays the exciton diffusion averaged over the four NISE calculations with the exciton placed on an m-type porphyrin in the central copy, the 12 calculations with the exciton placed on a p-type porphyrin, and averaged over all 16 calculations. Additionally, the classical diffusion based on the hexagonal grid is shown.  Panel (a) shows the simulation for average couplings, and panel (b) for time-dependent couplings. Panel (b) also includes the diffusion of the porphyrin molecules themselves, i.e., $ \langle d_{i,mol}(t)^2\rangle  = |x_i(t)-x_i(0)| ^2$, where $x_i$ denotes the position of  porphyrin $i$. The average over all porphyrins is finally included in the graph. 

The results show that time-dependent couplings significantly enhance the exciton diffusion rate compared to average couplings, which aligns with the generally faster decay of initial populations in Figure \ref{fig_lifetimes}. This also brings the NISE diffusion results closer to the classical diffusion. The movement of the pigments themselves contributes negligibly to the diffusion, indicating that exciton transfer is the main driver of exciton diffusion. It can also be seen that, diffusion is slower for m-type porphyrins compared to p-type porphyrins, which aligns again with Figure \ref{fig_lifetimes}, where the exciton decay for m-type porphyrins is slower.
\begin{figure}[htb!]
    \centering
    \begin{subfigure}[b]{0.48\textwidth}
        \centering
        \includegraphics[width=\textwidth]{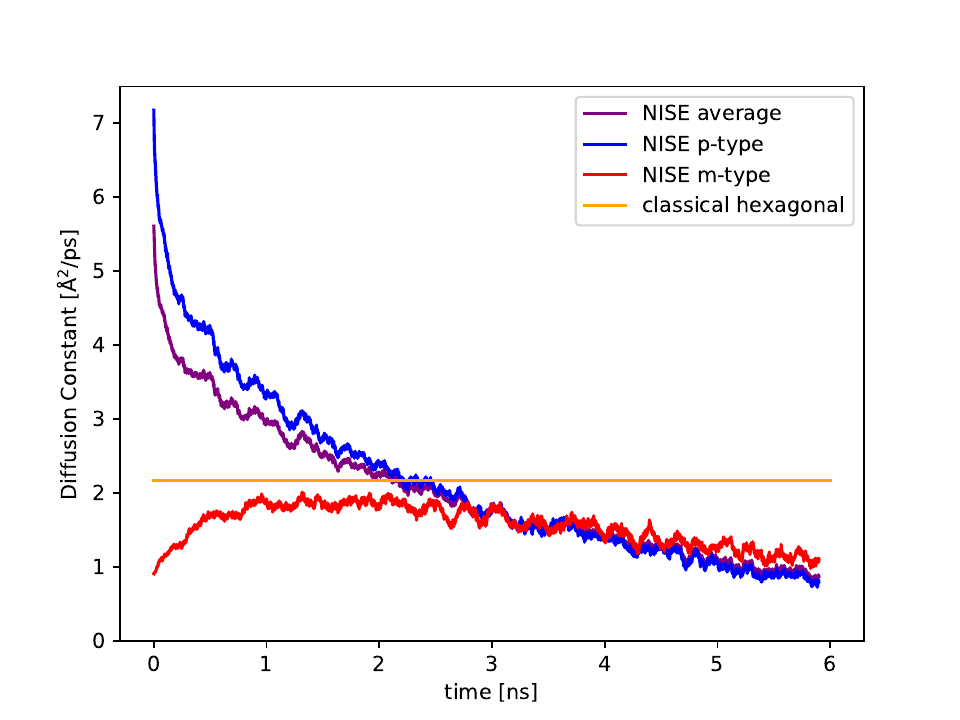}
        \caption{Long time (6ns)}
        \label{fig_Diffusion_constant_t}
    \end{subfigure}
    \hfill
    \begin{subfigure}[b]{0.48\textwidth}
        \centering
        \includegraphics[width=\textwidth]{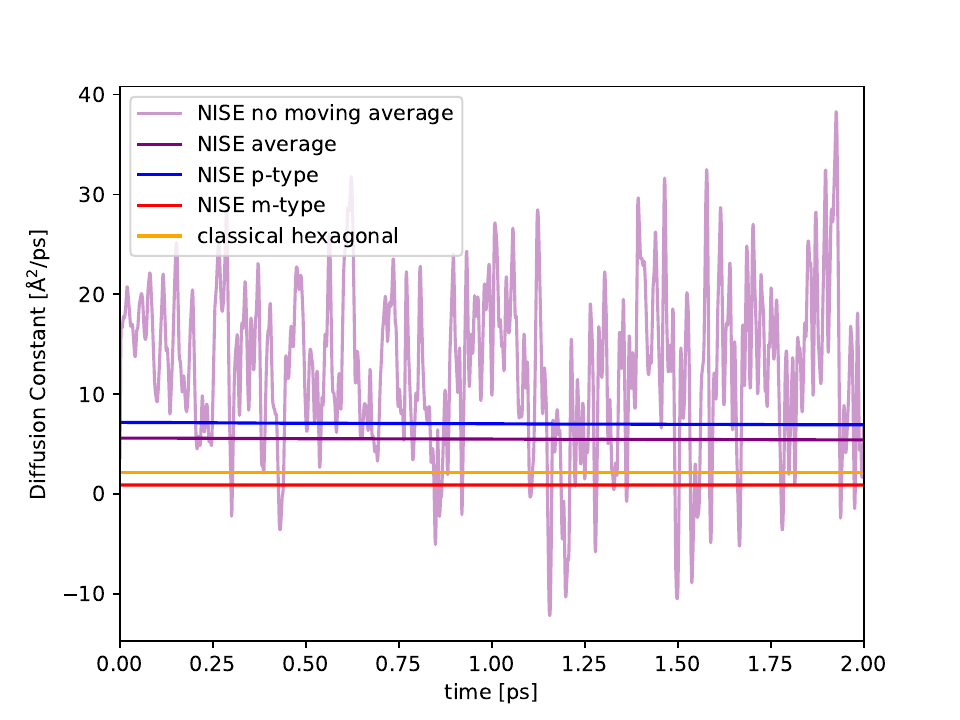}
        \caption{Short time (2ps)}
        \label{fig_Diffusion_constant_short}
    \end{subfigure}   
    \caption{Diffusion “constant” over time, shown as (a) a moving average over 100ps and (b) compared to instantaneous value for short times. The NISE calculations are taken from the simulation with the time-dependent couplings.}    \label{fig_Diffusion_constant}
\end{figure}
The diffusion “constant” $D(t)$ is defined as the time derivative of the diffusion
\begin{equation} 
D(t) = \frac{d}{dt} < d(t)^2 >. 
\label{eq:Diffusion_Constant} 
\end{equation}
In the classical model, the diffusion constant remains constant. However, in the NISE-based calculations, we observe that the diffusion “constant” fluctuates on very short timescales, comparable to the Rabi frequencies between porphyrin pairs
which can be calculated to be on the order of 30-150~fs based on the average energy difference and coupling (see Figure \ref{fig_Diffusion_constant}(b)) \cite{merl21a}. While these very fast fluctuations are a result of the quantum description of diffusion, they have little impact on the long term behavior.
Hence, we compute a moving average of the diffusion constant over 100~ps to avoid the fluctuations dominating the graph and to enable a better comparison with the classical diffusion (see Figure \ref{fig_Diffusion_constant}(a)). Overall, the diffusion constants calculated with the NISE approach are on a scale similar to that of the classical diffusion based on the experimental values. The moving average reveals that the diffusion constant is initially larger for p-type porphyrins but becomes similar to that of m-type porphyrins at a timescale comparable to the lifetime of excitons placed on the m-type porphyrins. This suggests that while p-type porphyrins facilitate faster exciton transport initially, the diffusion rates become equal once the exciton population is distributed between m- and p-type porphyrins.
The initial drop in the diffusion constant can be attributed to an initial equilibration of the exciton between strongly coupled nearest neighbors. Such strongly coupled nearest neighbors are absent on an orderly hexagonal grid. The reduction of the diffusion constant over longer times might also be related to some  population reaching the edge of our 3×3 copy simulation. Even small populations at the edges will have a large influence on diffusion as it grows with the square of the distance. 

A property often referred to in experimental papers is the diffusion length, which, for classical diffusion, can be calculated from the diffusion constant $D$ and the lifetime of the exciton $\tau$ as\cite{scul06a}
\begin{equation} 
L = \sqrt{ D \tau }~. 
\label{eq:Diffusion_lengths} 
\end{equation}
This formula equation is usually used because $D$ is often calculated directly without calculating $< d(t)^2 >$ first and $D \tau$ is equal to $< d(\tau)^2 >$ for classical diffusion.  In our case, we have calculated $< d(t)^2 >$ directly. Hence, determining the diffusion length as $L=\sqrt{(< d(\tau)^2 >)}$ is more sensible as it avoids complications with the fluctuations in $D$ for the quantum cases. We chose the experimental lifetime of $\tau=5.6$~ns of the combined porphyrin-clay complex for all our calculations \cite{ishi11b}. For the time-dependent coupling calculations, we obtain a diffusion length of 9.4~nm for excitations starting on an m-type porphyrin molecules, 11.2~nm for excitations starting on a p-type dye, and 11.0~nm for the classical hexagon-based diffusion.
These values are generally in line with experimental values of organic semiconductors, which are typically between 5 and 20~nm, although longer diffusion lengths have also been observed\cite{fird20a}. For porphyrin-based systems, diffusion lengths between 7.5 and 40~nm have been observed \cite{huij05a,huij08a,kaus16a,gu21a}. Interestingly, most of these have higher diffusion constants, up to 900Å$^2$/ps for porphyrin-based metal–organic frameworks~\cite{gu21a}, but shorter diffusion times, which ultimately lead to comparable diffusion lengths.

Since the NISE simulations underestimate the transfer from m- to p-type porphyrins, we expect that our calculations also underestimate the diffusion length for the porphyrin-clay complex. If that underestimation is of similar magnitude as for the population decay of the m-type, the porphyrin-clay complex might have one of the largest diffusion lengths among comparable systems. This would be an important finding because a large diffusion lengths is an important property for organic solar cells.
To investigate this further, one could determine the other pigment-pigment transfer rates or directly measure the diffusion length experimentally. Alternatively, the previously mentioned improvements to the exciton dynamics (inclusion of the Q\textsubscript{y} state, better-resolved couplings, and closer aligned nearest-neighbor distances) could align the simulations closer to reality and predict a better diffusion length.

\section{Conclusions and Outlook}

Inspired by the energy transfer networks present in the light-harvesting complexes of plants, bacteria, and algae, a significant amount of work has been done over the past years to replicate these natural systems artificially. Experimental evidence \cite{ishi11b} has shown that porphyrin molecules attached to inorganic clay surfaces can achieve remarkable energy transfer efficiencies close to 100\%. In this study, we tried to emulate the experimental setup in Ref.~\citenum{ishi11b}, i.e., cationic free-base porphyrin molecules adsorbed on an anionic inorganic clay surface. The surface model was created using the CHARMM-GUI webserver. Two types of porphyrins, m-TMPyP and p-TMPyP, were selected due to their positive charge, in a ration of 1:3 and 
with a total of 16 porphyrin molecules.

Classical MD simulations were conducted to equilibrate the system, allowing each porphyrin molecule to find its optimal binding location on the clay surface. The equilibrated structures were subsequently used as starting structures  for QM/MM MD simulations, followed by excitation energy calculations using the TC-LC-DFTB method. This workflow established the computational protocol for investigating energy transfer processes among the pigment molecules. In addition, a novel multifidelity machine learning-based approach, MFML, was used to compute excitation energies at the computationally demanding TD-DFT/Def2-SVP level for all porphyrin molecules, each consisting of 90 atoms, across 640,000 geometries. The extensive sampling of high-level quantum excitation energies was then used to construct a time-dependent exciton Hamiltonian, enabling simulations of exciton dynamics and diffusion using the NISE approach in a periodic system. 
The results demonstrate the feasibility of simulating energy transfer in artificial systems by utilizing porphyrin-clay hybrid materials. The high efficiency observed in the energy transfer process underscores the potential for porphyrin-based materials in designing light-harvesting devices. The computational protocol in the present study effectively bridges classical and quantum simulations, enabling an in-depth exploration of energy dynamics at the molecular level. The successful implementation of the MFML technique for large-scale quantum calculations also highlights the potential for machine learning to significantly accelerate complex computational tasks in quantum chemistry.

As an outlook, the computational framework developed in this study can be extended to design more sophisticated artificial light-harvesting networks using various dye molecules, paving the way for further innovations in solar energy harvesting. By providing a detailed theoretical understanding of the mechanisms underlying efficient energy transfer, this work contributes to the development of sustainable energy solutions. Furthermore, the results serve as a valuable benchmark for experimental investigations, offering a basis for future experimental studies to validate and refine the proposed models. Subsequent studies might investigate more intricate combinations of clay and dye, varying dye proportions, and external influences like controlling the temperature to enhance the computational design of synthetic light-harvesting systems to adhere to the “size-matching” principle.

\section{Methods}\label{Methods}

\subsection{Simulation and Calculation Details}


Here we give some more details for modelling the  montmorillonite nanosheet. The aluminum atoms in the octahedral layer of the system, i.e., the central layer in Figure \ref{fig:montmorillonite surface}, are partially replaced by magnesium, which leads to a charge difference and forms a negatively charged surface. 
The parameter $x$, which can vary between 0 and 0.95, indicates the percentage of magnesium defects in the system. The cations in the formula imply that the system is in a water environment with dissolved cations to achieve neutrality.
It has to be noted that in the experimental setup with the saponite system, the silicon atoms in the tetrahedral layer, which is shown in yellow in Figure \ref{fig:montmorillonite surface}, were replaced by aluminum. Although the layer of atoms exchanged in our simulations were different from the experiment, the distances between replacement sites were consistent, so the “size-matching rule” still applies. 
The clay surface with a size of 10.38 nm $\times$ 10.82 nm $\times$ 1 nm in vacuum was generated using the online tool CHARMM-GUI \cite{jo08a,choi22a} where the ratio of defects and ion options can be customized and modified. Four different values of $x$ (0.13, 0.2, 0.45 and 0.94) were selected to generate the topology and perform individual MD simulations. The size of the initial simulation box was chosen to fit the material size, with an height of 4 nm. Periodic boundary conditions (PBC) were first only  applied to the  $x$ and $y$ directions.
The structures of the porphyrin molecules  were obtained from ChemSpider (CSID: 133612 and 4086), and the ACPYPE (AnteChamber PYthon Parser interfacE) tool \cite{dasi12a} was then employed to generate the topology and coordinates in GROMACS format. After randomly placing four m-TMPyP and twelve p-TMPyP molecules on top of the clay surface, a short preliminary simulation was performed in vacuum for a maximum of 5~ns, allowing the porphyrins to be attracted to the clay surface solely through electrostatic interactions. The number of the porphyrins was chosen as it was proved to be one of the ratios with the highest energy transfer efficiency \cite{ishi11b}. The last frame of the simulation was extracted and then used as the initial configuration for the further MD simulations.

Due to technical reasons, 3D-PBC was required in the subsequent MD simulations. The height of the simulation box was extended to 25~nm, which is about 2.5 times the width, to minimize the effect of the mirror structure on the adsorption behavior of porphyrins. Afterward, the system was solvated using TIP3P water molecules, and neutralized with sodium and chloride ions, followed by energy minimization, a 5~ns NVT equilibration at 300~K and a 5~ns NPT equilibration. Thereafter, the classical MD simulations for 30~ns were performed with a time step of 1~fs employing GROMACS 5.1.4\cite{abra15a} using the general AMBER force field (GAFF) \cite{wang04a} and the interface force field (IFF) \cite{hein13a} for porphyrins and clay surface, respectively. The Nose-Hoover thermostat\cite{evan85a} and Berendsen barostat\cite{bere84a} were employed to control the temperature and pressure, respectively. The cutoff for short-range non-bonded interaction was set to 1.2 nm and the long-range electrostatics was treated using the Particle Mesh Ewald (PME) method\cite{essm95a}. In addition, the LINCS algorithm was used for  bond constrains \cite{hess97a}.

The montmorillonite surface with a $x$ value of 0.45 was chosen for the subsequent quantum mechanics/molecular mechanics dynamics (QM/MM) since all sixteen porphyrins were still adsorbed on the surface after the 30~ns MD simulation, while some porphyrins detached in the simulations using the other $x$ values. Each of the porphyrin molecules was assigned to the QM region and treated separately using the DFTB3 approach with the 3OB-f parameter set \cite{gaus13a}. Classical force fields were employed for the remainder of the system. A 50~ps NPT equilibration was performed in GROMACS including the DFTB$+$ interface \cite{bold20a}, followed by another 40~ps QM/MM MD simulation with a smaller integrator time step, which was set to 0.5~fs. The atomic coordinates were stored every two steps with 40,000 frames, which were subjected to excited state calculations employing TD-LC-DFTB \cite{kran17a} and TD-DFT with the CAM-B3LYP functional, as implemented in DFTB$+$ 21.1\cite{hour20a} and the ORCA 5.0.3 package \cite{nees22a}, respectively. Four different basis sets, STO-3G, 3-21G, 6-31G and def2-SVP, were used for the TD-DFT calculations according to their quantum chemical hierarchy, with a time stride of 8, 16, 32 and 64~fs, respectively.

Furthermore, another 100~ns classical MD simulation was performed for 10,000 snapshots after the equilibration, in order to evaluate the excitonic couplings between the porphyrin molecules the transition charges from electrostatic potentials (TrESP) method was applied \cite{madj06a,reng13b}.

\subsection{Gaussian Process Regression}

In GPR\cite{williams2006gaussian}, a machine learning model is found for a training set $\mathcal{T}=\{(\mathbf{x}_i,y_i)\}_{i=1}^N$, with training inputs $\mathbf{x}_i$ and corresponding output measurements $y_i$. Then, starting from a Gaussian process prior $\mathcal{GP}(0,k(\mathbf{x},\mathbf{x}^\prime))$, with covariance (kernel) function $k\left(\mathbf{x}, \mathbf{x'}\right)$,  a posterior predictive distribution is found by conditioning on the training set $\mathcal{T}$. The posterior distribution has a the mean
\begin{align}\label{eq:GPRmodel}
m\left(\mathbf{x}\right) = \mathbf{K}\left(\mathbf{x}, \mathbf{X}\right) \bigl(\mathbf{K}\left(\mathbf{X}, \mathbf{X}\right) + \sigma_n^2\mathbf{I}\bigr)^{-1}\mathbf{y}
\end{align}
that gives predictions for unknown inputs $\textbf{x}$, i.e., it is the constructed ML model. 
By $\mathbf{X}$, we denote the set of training inputs and $\mathbf{y}$ is a vector with the corresponding outputs. In the model, $k(\mathbf{x},\mathbf{X})$ denotes the vector of pair-wise kernel evaluations between the query input $\mathbf{x}$ and the set of training inputs $\mathbf{X}$, whereas $\mathbf{K}\left(\mathbf{X}, \mathbf{X}\right)$ is the matrix of all pair-wise kernel evaluations between all training inputs $\mathbf{X}$. In addition, $\sigma_n^2$ is a hyperparameter that models the (unknown) variance of the noise that is assumed on the outputs of the training data.

For our tests we chose a Gaussian covariance function
\begin{align}
    k\left(\mathbf{x}, \mathbf{x'}\right) = \sigma_f^2\text{exp}\left(-\frac{1}{2}\left(\mathbf{x} - \mathbf{x}'\right)^T\mathbf{M}\left(\mathbf{x} - \mathbf{x}'\right)\right).
    \label{eq:gaussian_cov}
\end{align}
Here, $\mathbf{M}=\text{diag}\left(\frac{1}{l_1^2},...,\frac{1}{l_D^2}\right)$ is a diagonal matrix, where $D$ is the number of features. This results in a different length-scale for every feature, known as automatic relevance determination\cite{williams2006gaussian}. Moreover, $\sigma_f^2$ is a hyperparameter, the output-scale. All hyperparamters are found by finding by maximising the marginal log-likelihood, with details in the supplementary material. We used the GPR implementation of GPyTorch \cite{gpytorch}.

In this work, the inputs $\textbf{x}$ are \textit{Coulomb matrices}\cite{rupp2012fast}, representing the molecular configurations, and the outputs $y$ are the corresponding excitation energies. For Coulomb matrices, to ensure invariance against atom permutation, often a sorting is carried out, which results in discontinuities and makes them unfavorable for training ML models \cite{langer2022representations}.
In our tests, all we  have fixed atoms given in a fixed order. Therefore, we can safely ignore invariance against atom permutation and hence use unsorted Coulomb matrices.

\subsection{Active learning}
Active learning (AL) algorithms try to choose or create new training samples, which are maximally informative to an ML model. In our setup, we have a large amount of unlabeled molecular configurations and generating the respective labels (excitation energies) via quantum chemical calculations is expensive. For this setting, uncertainty sampling \cite{uncertainty_sampling} can be a well suited AL scheme: First, for a small random subset of the data, excitation energies are calculated. Then, this data is used to train an initial model. In the iterative process to add favorable new training samples, the standard deviation of the predictive distribution of the GPR model 
\begin{align}\label{eq:sigma}
\sigma\left(\mathbf{x}\right) = \sqrt{k\left(\mathbf{x}, \mathbf{x}\right) - \mathbf{K}\left(\mathbf{x}, \mathbf{X}\right) \bigl(\mathbf{K}\left(\mathbf{X}, \mathbf{X}\right) + \sigma_n^2\mathbf{I}\bigr)^{-1}\mathbf{K}\left(\mathbf{X}, \mathbf{x}\right)}
\end{align}
is used as uncertainty measure. Those samples are added that have the highest uncertainty. Once a samples is selected from the remaining pool of unlabeled samples, its label, i.e.~the excitation energy for the given molecule, is calculated. The input-output pair is then added as additional training sample. The full procedure is summarized in Algorithm~\ref{alg:uncertainty_sampling}.


\begin{algorithm}
\caption{Active Learning by Uncertainty Sampling}
\begin{algorithmic}[1]
\REQUIRE Set of unlabeled inputs $U$, initial sample size $n_\text{init}$, number of iterations $n_\text{iter}$
\STATE Randomly select $n_\text{init}$ samples from $U$ and store them in set $L$
\STATE Obtain labels $y(L)$ for all inputs in $L$
\STATE $U$ = $U$ \textbackslash $L$
\STATE Train initial model $m$, following eq.~\eqref{eq:GPRmodel}, using labeled data $(L,y(L))$
\FOR{$i = 1$ to $n_\text{iter}$}
    \STATE Compute uncertainty $\sigma(x)$, from eq.~\eqref{eq:sigma}, for each sample in $U$ using model $m$
    \STATE Select sample $x^*$ from $U$ with highest uncertainty
    \STATE Obtain label $y^*$ for selected sample $x^*$
    \STATE $L = L \,\cup\, \{x^*\}$
    \STATE $U = U$ \textbackslash $\{x^*\}$
    \STATE Retrain model $m$, following eq.~\eqref{eq:GPRmodel}, using updated labeled dataset $(L,y(L))$
\ENDFOR
\STATE \textbf{return} Trained model $m$
\end{algorithmic}
\label{alg:uncertainty_sampling}
\end{algorithm}

\subsection{Multifidelity Machine Learning Approach} \label{mfml_method}
Multifidelity machine learning (MFML) systematically combines ML \textit{sub-models} trained on multiple fidelities, $f$, to produce a low-cost high accuracy ML model interested in predicting the excitation energies for a target fidelity
$F$ \cite{zasp19a, vinod_2024_oMFML}. 
A composite index, $\boldsymbol{s}=(f,\eta_f)$, is used to identify the sub-models of MFML. In this index, we have $2^{\eta_f}=N_{\rm train}^{(f)}$. The MFML model is built with a target fidelity, $F$, for a given baseline fidelity $f_b$, which refers to the lowest QC fidelity used in the model. One can write the MFML model for a query molecular descriptor, $\boldsymbol{X}_q$ as
\begin{equation}
    P_{\rm MFML}^{(F,\eta_F;f_b)}\left(\boldsymbol{X}_q\right) := \sum_{\boldsymbol{s}\in\mathcal{S}^{(F,\eta_F;f_b)}} \beta_{\boldsymbol{s}} P^{(\boldsymbol{s})}_{\rm GPR}\left(\boldsymbol{X}_q\right)~,
    \label{eq_MFML_linearsum}
\end{equation}
where, the summation is made for the set of selected sub-models of MFML, $\mathcal{S}^{(F,\eta_F;f_b)}$. This selection is decided by the choice of $f_b$, and $N_{\rm train}^{(F)}=2^{\eta_F}$, the number of training samples at the target fidelity, as presented in Ref.~\citenum{vinod_2024_oMFML}.
The $\beta_{\boldsymbol{s}}$ from Eq.~\eqref{eq_MFML_linearsum} are referred to as the coefficients of the linear combination of these selected sub-models. 
For MFML, the $\beta_{\boldsymbol{s}}$ are chosen to be
\begin{equation}
    \beta_{\boldsymbol{s}}^{\rm MFML} = \begin{cases}
        +1, & \text{if } f+\eta_f = F+\eta_F\\
        -1, & \text{otherwise}
    \end{cases}~,
    \label{eq_MFML_beta_i}
\end{equation}
based on previous work from Ref.~\citenum{zasp19a}.

In the MFML method, the number of training samples at the subsequent fidelities are set by a \textit{scaling factor}, $\gamma$. That is, $N_{train}^{f} = \gamma\cdot N_{train}^{f-1}$ for $f_b<f\leq F$. The value of $\gamma=2$ is conventionally used in MFML based on previous work \cite{zasp19a, vinod23_MFML, vinod_2024_oMFML, Vinod_2024_nonnested}. In a recent work, the effect of different values of $\gamma$ in the model error of MFML has been studied \cite{vinod2024_gamma_curve_error_contours}.
Ref.~\citenum{vinod2024_gamma_curve_error_contours} reports that the use of very little training data at the target fidelity combined with increasing values of $\gamma$, results in a more data efficient model. These models are studied using the model error and time-cost of generating the training data required for the model. This specific curve is referred to as the $\Gamma(N_{train}^{F})$ curve with each data point essentially being the MFML model built for a specific value of $\gamma$. Herein, the $\Gamma(8)$ curve is built for $\gamma\in\{2,\ldots,12\}$. The supplementary information associated with this work also reports $\Gamma(2)$, $\Gamma(4)$, and $\Gamma(16)$ curves in Figure~S9. The difference between the $\Gamma(8)$ and $\Gamma(16)$ curves were not seen to be significantly different. The final model used in this work in the prediction of def2-SVP energies is the MFML model along the $\Gamma(8)$ curve for $\gamma=12$ as discussed in Figure \ref{fig_simplified_time_cost_combined}.

\subsection{Model Evaluation} \label{model_eval_methods_MAE}
The different ML models that are built in this work, both single fidelity and MFML, are evaluated using mean absolute error (MAE) over holdout test sets. Given a test set at target fidelity, $F$, denoted as $\mathcal{Q}^F:=\{(\boldsymbol{X_q},y_q^{ref})\}_{q=1}^{N_{eval}}$, the MAE for an ML model is calculated as
\begin{equation}
    MAE:= \frac{1}{N_{eval}}\sum_{q=1}^{N_{eval}} \left\lvert y^{ML}_q-y^{ref}_q\right\rvert_1~.
\end{equation}
The model error is initially reported as a function of number of training samples used at the target fidelity in the form of learning curves\cite{li2013learning, muller1996numerical_LearningCurve, cortes1993learning} for both single fidelity and MFML models since these are a common metric of model evaluation for the ML for QC quantities workflow \cite{Westermayr2020review, Westermayr2020NNKRR, dral2020quantum}. 

Since the main aim of multifidelity models is to reduce the time-cost of generating training data, this work also presents a recently introduced analysis of MAE versus time-cost of generating training data for a certain ML model \cite{vinod23_MFML}. For single fidelity models, this is simply the QC calculation cost of the number of training samples used for the ML model. For MFML models, this cost takes into account the QC computation cost of not just the training samples used at the target fidelity, but also the cost of training samples used in the entire multifidelity data structure. 


\begin{acknowledgement}
The authors are grateful to the developers of the webtool CHARMM-GUI and in particular Prof.\ Wonpil Im for assistance in modelling and modifying the  montmorillonite nanosheet.
Furthermore, the authors acknowledge support by the Deutsche Forschungsgemeinschaft (DFG) through the 
Priority Program SPP 2363 on “Utilization and Development of Machine Learning for Molecular Applications – Molecular Machine Learning” through the project ZA 1175/4-1 and  KL 1299/25-1 and further funding through projects ZA 1175/3-1 and  KL 1299/24-1.
VV and PZ would also like to acknowledge the support of the `Interdisciplinary Center for Machine Learning and Data Analytics (IZMD)' at the University of Wuppertal. Furthermore, part of the simulations were performed on a compute cluster in Bremen funded through the DFG project INST 676/7-1 FUGG, while part of the machine learning training was carried out on the {PLEIADES cluster} at the University of Wuppertal, which is supported by the DFG through grant INST 218/78-1 FUGG and the Bundesministerium für Bildung und Forschung (BMBF). 
\end{acknowledgement}

\begin{suppinfo}
Additional figures, tables, and explanations on excitation energies, transition charges, spectral densities and the MFML approach.
\end{suppinfo}

\bibliography{bibvivinod,ukleine,holzenkamp,holtkamp,lyu} 
%
\end{document}





\section{Distributions of Excitation energies}
\begin{figure}[H]
    \centering
    \begin{subfigure}[b]{0.494\textwidth}
        \centering
        \includegraphics[width=\textwidth]{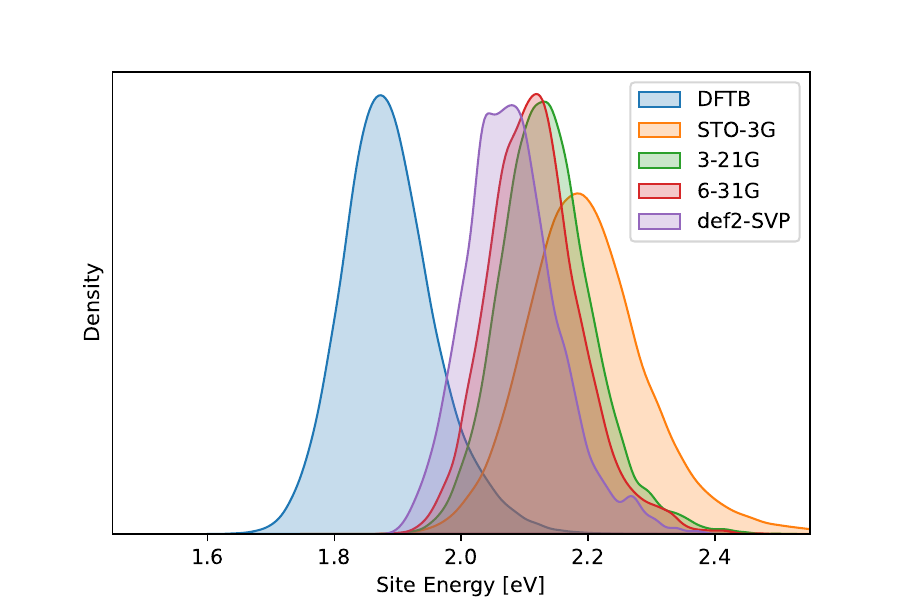}
        \caption{m-type porphyrin molecules.}
    \end{subfigure}
    \hfill
    \begin{subfigure}[b]{0.494\textwidth}
        \centering
        \includegraphics[width=\textwidth]{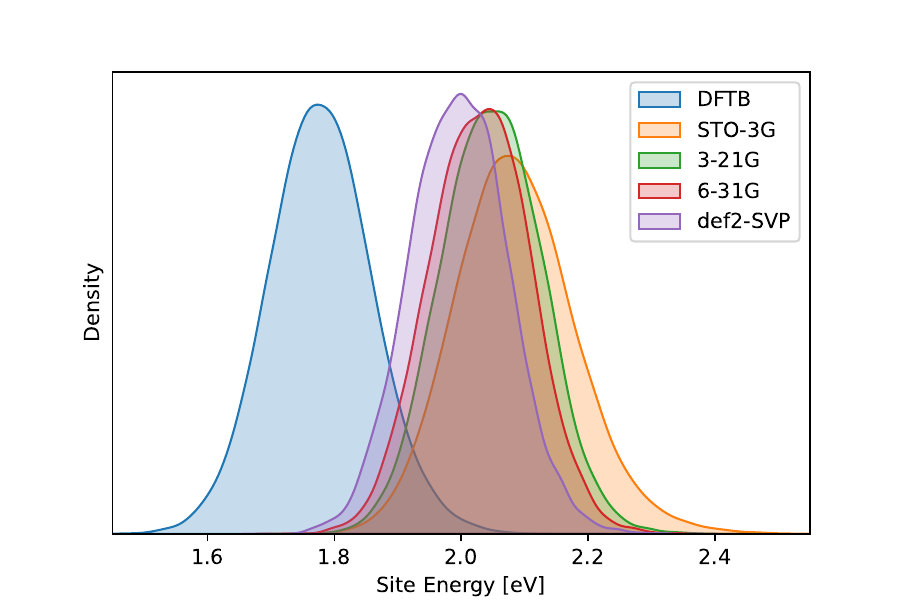}
        \caption{p-type porphyrin molecules.}
    \end{subfigure}    
    \caption{Distributions of excitation energies for different levels of quantum chemistry averaged over all pigments of the same type.}
    \label{fig:dos_m_p}
\end{figure}

\section{Transition charges from electrostatic potentials}

\begin{longtable}[H]{|c|c|c|c|c|}
        \hline
        Atom & x (Å) & y (Å) & z (Å) & Charge (e) \\
        \hline
        N & 0.421597 & -2.132765 & -0.040233 & 0.004221 \\
        N & -1.983771 & -0.368911 & -0.035220 & 0.134251 \\
        N & -3.605662 & -4.502220 & -1.031529 & 0.005530 \\
        N & 2.050558 & 0.310653 & -0.195799 & -0.137842 \\
        N & -0.340619 & 2.060808 & -0.211058 & 0.004226 \\
        N & 5.008518 & -2.994448 & 1.040074 & -0.003230 \\
        N & -4.820361 & 3.055132 & 1.091452 & -0.000599 \\
        N & 3.012745 & 4.886841 & 0.779343 & -0.018144 \\
        C & -1.936494 & -2.908660 & 0.017158 & 0.227779 \\
        C & -0.592297 & -3.057533 & -0.006731 & -0.161439 \\
        C & -2.572244 & -1.533003 & 0.043459 & -0.153517 \\
        C & 1.689852 & -2.671401 & -0.041360 & 0.148796 \\
        C & -2.985299 & 0.594980 & -0.016338 & -0.175660 \\
        C & 2.910429 & -2.092278 & -0.071507 & -0.218195 \\
        C & -2.857323 & 1.947617 & -0.081107 & 0.192928 \\
        C & -2.821890 & -4.088093 & 0.085790 & -0.057952 \\
        C & 0.120829 & -4.341481 & -0.001371 & 0.046537 \\
        C & -4.043513 & -1.414690 & 0.156894 & -0.044387 \\
        C & 3.016401 & -0.576386 & -0.173204 & 0.145679 \\
        C & -1.602042 & 2.649152 & -0.219007 & -0.166818 \\
        C & 1.442074 & -4.115945 & -0.008222 & -0.046181 \\
        C & -4.240539 & -0.120133 & 0.091302 & 0.006181 \\
        C & 2.053042 & 2.787755 & -0.289399 & -0.213202 \\
        C & 4.144607 & -2.906612 & -0.087775 & 0.056347 \\
        C & -4.077582 & 2.804998 & -0.094049 & -0.040800 \\
        C & 2.645689 & 1.566310 & -0.252764 & 0.189595 \\
        C & 0.630520 & 3.043435 & -0.323455 & 0.172590 \\
        C & 2.909964 & 4.009287 & -0.347216 & 0.047661 \\
        C & 4.364091 & 0.032361 & -0.242907 & 0.041203 \\
        C & -1.396375 & 4.010535 & -0.353687 & 0.082123 \\
        C & 4.082246 & 1.325772 & -0.268444 & -0.007633 \\
        C & -0.025732 & 4.244872 & -0.440427 & -0.076580 \\
        C & -2.902804 & -4.747067 & 1.263714 & 0.018754 \\
        C & 4.443901 & -3.576882 & -1.230780 & -0.019234 \\
        C & -4.455729 & 3.341823 & -1.283890 & 0.012739 \\
        C & -4.441843 & -5.605528 & -0.861493 & 0.014622 \\
        C & -3.595555 & -3.841328 & -2.329381 & 0.003907 \\
        C & -3.767104 & -5.892177 & 1.392982 & 0.003225 \\
        C & 3.515863 & 4.302707 & -1.512417 & -0.008985 \\
        C & 6.150667 & -3.800456 & 0.934022 & -0.015111 \\
        C & -5.976699 & 3.836917 & 0.977024 & 0.011786 \\
        C & -4.550884 & -6.306116 & 0.265665 & 0.002901 \\
        C & 4.780040 & -2.326283 & 2.313005 & -0.005672 \\
        C & -4.470466 & 2.530065 & 2.401559 & 0.003310 \\
        C & 5.621544 & -4.404871 & -1.287265 & -0.002870 \\
        C & -5.647546 & 4.157855 & -1.347639 & 0.003229 \\
        C & 3.806318 & 6.031995 & 0.630129 & 0.002088 \\
        C & 6.492501 & -4.485792 & -0.149290 & -0.004012 \\
        C & -6.411432 & 4.386493 & -0.152522 & 0.000575 \\
        C & 2.371226 & 4.668773 & 2.066005 & 0.002163 \\
        C & 4.323421 & 5.489409 & -1.610854 & -0.003148 \\
        C & 4.454485 & 6.367095 & -0.484315 & -0.003731 \\
        H & 0.261394 & -1.143165 & -0.051560 & 0.000000 \\
        H & -0.348384 & -5.312816 & 0.010812 & -0.000000 \\
        H & -4.775179 & -2.197865 & 0.252831 & 0.000000 \\
        H & 2.195243 & -4.890485 & -0.003348 & 0.000000 \\
        H & -5.225600 & 0.348567 & 0.120893 & 0.000000 \\
        H & 5.327980 & -0.443695 & -0.262616 & -0.000000 \\
        H & -2.149577 & 4.785825 & -0.398553 & 0.000000 \\
        H & 4.842302 & 2.094002 & -0.314675 & 0.000000 \\
        H & 0.436296 & 5.225117 & -0.552454 & 0.000000 \\
        H & -2.344832 & -4.471084 & 2.148587 & 0.000000 \\
        H & 3.835005 & -3.548910 & -2.119916 & 0.000000 \\
        H & -3.932434 & 3.191837 & -2.207290 & 0.000000 \\
        H & -5.020987 & -5.888868 & -1.736485 & -0.000000 \\
        H & -2.913668 & -2.995132 & -2.404093 & 0.000000 \\
        H & -4.601480 & -3.488041 & -2.561983 & 0.000000 \\
        H & -3.295689 & -4.575277 & -3.093514 & 0.000000 \\
        H & -3.834607 & -6.433284 & 2.323061 & 0.000000 \\
        H & 3.440235 & 3.703766 & -2.403736 & 0.000000 \\
        H & 6.780809 & -3.835399 & 1.823010 & -0.000000 \\
        H & -6.528279 & 3.986771 & 1.900444 & -0.000000 \\
        H & -5.197358 & -7.159778 & 0.370213 & 0.000000 \\
        H & 3.865095 & -1.729997 & 2.344930 & -0.000000 \\
        H & 5.625547 & -1.647177 & 2.515078 & 0.000000 \\
        H & 4.727504 & -3.063028 & 3.119384 & 0.000000 \\
        H & -3.555538 & 1.946794 & 2.429889 & -0.000000 \\
        H & -5.301639 & 1.897090 & 2.760437 & -0.000000 \\
        H & -4.346592 & 3.366288 & 3.097250 & 0.000000 \\
        H & 5.863872 & -4.952648 & -2.187886 & -0.000000 \\
        H & -5.968453 & 4.591096 & -2.279423 & -0.000000 \\
        H & 3.867598 & 6.673476 & 1.512256 & -0.000000 \\
        H & 7.372166 & -5.102891 & -0.211581 & 0.000000 \\
        H & -7.291128 & 4.991701 & -0.210475 & 0.000000 \\
        H & 1.781549 & 3.759567 & 2.129459 & 0.000000 \\
        H & 1.709796 & 5.517388 & 2.282484 & 0.000000 \\
        H & 3.142764 & 4.622414 & 2.844645 & -0.000000 \\
        H & 4.837824 & 5.728414 & -2.536957 & -0.000000 \\
        H & 5.058418 & 7.248463 & -0.568306 & 0.000000 \\
        H & -0.130919 & 1.090401 & -0.079390 & 0.000000 \\
        \hline
    \caption{Transition charges and Atomic coordinates (x, y, z) for m-TMPyP}
\end{longtable}

\begin{longtable}[H]{|c|c|c|c|c|}
        \hline
        Atom & x (Å) & y (Å) & z (Å) & Charge (e) \\
        \hline
        C & 5.005311 & 1.315763 & -0.062397 & -0.063429 \\
        C & 6.822858 & 2.793393 & 0.644527 & 0.016456 \\
        C & 5.452531 & 2.475338 & 0.575636 & 0.031097 \\
        C & 6.008654 & 0.519655 & -0.631443 & 0.030054 \\
        C & 7.373325 & 0.807816 & -0.442659 & 0.010262 \\
        N & 7.754891 & 1.940274 & 0.188017 & 0.001928 \\
        C & 9.188915 & 2.200493 & 0.418145 & 0.013398 \\
        C & -5.052217 & -1.423867 & -0.169090 & 0.051190 \\
        C & -6.992483 & -2.902767 & 0.223869 & -0.011712 \\
        C & -5.617956 & -2.671033 & 0.111290 & -0.041290 \\
        C & -6.044986 & -0.459596 & -0.490947 & -0.020794 \\
        C & -7.423356 & -0.678102 & -0.297683 & -0.014965 \\
        N & -7.865256 & -1.882258 & 0.091848 & 0.000800 \\
        C & -9.287730 & -2.097999 & 0.423846 & -0.014479 \\
        C & 2.323920 & -6.683852 & -1.045090 & -0.001502 \\
        C & 1.939680 & -5.337467 & -1.137863 & -0.008750 \\
        C & 1.059098 & -5.468637 & 1.087439 & -0.009720 \\
        C & 1.439782 & -6.810833 & 1.128542 & -0.000243 \\
        N & 2.071177 & -7.382541 & 0.074935 & -0.005793 \\
        C & 2.504820 & -8.798642 & 0.187687 & 0.001520 \\
        C & 1.313582 & -4.709116 & -0.054869 & 0.043909 \\
        C & -2.118989 & 6.731555 & -1.097417 & -0.000216 \\
        C & -1.701528 & 5.401750 & -1.199121 & 0.007361 \\
        C & -1.069160 & 5.470925 & 1.112486 & -0.000117 \\
        C & -1.502945 & 6.798853 & 1.162907 & 0.001945 \\
        N & -2.026010 & 7.401639 & 0.074949 & 0.003539 \\
        C & -2.514783 & 8.782573 & 0.176029 & -0.002770 \\
        C & -1.172951 & 4.756131 & -0.084051 & -0.032454 \\
        C & 0.927242 & -3.257775 & -0.121570 & -0.216674 \\
        C & 1.900035 & -2.332055 & -0.022505 & 0.170788 \\
        C & 3.298715 & -2.654680 & 0.238283 & -0.040914 \\
        C & 4.015500 & -1.521009 & 0.319571 & 0.048500 \\
        C & 3.151940 & -0.382099 & 0.022333 & -0.150854 \\
        C & 3.534724 & 0.912782 & -0.119660 & 0.235659 \\
        C & 2.386349 & 1.918228 & -0.278449 & -0.139459 \\
        C & 2.716319 & 3.313392 & -0.590601 & -0.041075 \\
        C & 1.533552 & 3.914371 & -0.474028 & 0.007256 \\
        C & 0.477991 & 2.953611 & -0.235422 & -0.150672 \\
        C & -0.839282 & 3.276963 & -0.124360 & 0.185411 \\
        C & -1.936775 & 2.350795 & 0.048328 & -0.178131 \\
        C & -3.243524 & 2.666098 & 0.380580 & 0.063236 \\
        C & -3.946086 & 1.469323 & 0.554469 & -0.072974 \\
        C & -3.177941 & 0.416909 & 0.093080 & 0.143864 \\
        C & -3.519603 & -1.018215 & -0.090718 & -0.211386 \\
        C & -2.502465 & -1.941113 & -0.125223 & 0.181780 \\
        C & -2.618071 & -3.364187 & -0.155226 & 0.008662 \\
        C & -1.444374 & -3.979942 & -0.285920 & 0.038749 \\
        C & -0.506687 & -2.849032 & -0.194898 & 0.145077 \\
        N & 1.895430 & -0.958442 & -0.083072 & 0.002560 \\
        N & 1.090244 & 1.704229 & -0.142735 & 0.115325 \\
        N & -1.906131 & 0.973502 & -0.123502 & 0.004150 \\
        N & -1.119084 & -1.687825 & -0.124413 & -0.134106 \\
        H & 7.172348 & 3.718628 & 1.101911 & 0.000000 \\
        H & 4.761783 & 3.168529 & 1.058701 & 0.000000 \\
        H & 5.776178 & -0.378675 & -1.199207 & -0.000000 \\
        H & 8.155051 & 0.129758 & -0.787542 & 0.000000 \\
        H & 9.354566 & 3.286331 & 0.428975 & 0.000000 \\
        H & 9.772077 & 1.746511 & -0.380679 & 0.000000 \\
        H & 9.441675 & 1.762876 & 1.379397 & 0.000000 \\
        H & -7.392522 & -3.892071 & 0.453755 & -0.000000 \\
        H & -5.065105 & -3.573468 & 0.325220 & 0.000000 \\
        H & -5.761459 & 0.517887 & -0.874408 & -0.000000 \\
        H & -8.156848 & 0.098515 & -0.433902 & 0.000000 \\
        H & -9.481020 & -3.168262 & 0.520690 & 0.000000 \\
        H & -9.887691 & -1.677071 & -0.367777 & 0.000000 \\
        H & -9.471451 & -1.601390 & 1.374755 & 0.000000 \\
        H & 2.832358 & -7.179728 & -1.865393 & 0.000000 \\
        H & 2.161299 & -4.791212 & -2.057151 & 0.000000 \\
        H & 0.565313 & -5.033971 & 1.948220 & 0.000000 \\
        H & 1.240358 & -7.440234 & 1.993114 & 0.000000 \\
        H & 3.085954 & -9.072683 & -0.698984 & -0.000000 \\
        H & 1.607704 & -9.411396 & 0.257808 & 0.000000 \\
        H & 3.122131 & -8.887079 & 1.080551 & 0.000000 \\
        H & -2.551078 & 7.256783 & -1.950198 & 0.000000 \\
        H & -1.822645 & 4.871201 & -2.145350 & 0.000000 \\
        H & -0.682918 & 4.995776 & 2.013124 & 0.000000 \\
        H & -1.440300 & 7.384339 & 2.075035 & 0.000000 \\
        H & -2.928017 & 9.113924 & -0.769246 & 0.000000 \\
        H & -1.673227 & 9.420067 & 0.469777 & 0.000000 \\
        H & -3.299635 & 8.802407 & 0.949666 & 0.000000 \\
        H & 3.709373 & -3.638351 & 0.430252 & -0.000000 \\
        H & 5.050082 & -1.504249 & 0.607771 & 0.000000 \\
        H & 3.650642 & 3.770489 & -0.840408 & 0.000000 \\
        H & 1.401908 & 4.975229 & -0.586045 & 0.000000 \\
        H & -3.636689 & 3.653312 & 0.563152 & 0.000000 \\
        H & -4.965491 & 1.428800 & 0.911641 & 0.000000 \\
        H & -3.453088 & -4.022879 & -0.048589 & 0.000000 \\
        H & -1.277767 & -5.040636 & -0.309835 & 0.000000 \\
        H & 1.054709 & -0.401934 & -0.231692 & 0.000000 \\
        H & -1.081090 & 0.425675 & -0.356299 & 0.000000 \\
        \hline
    \caption{Transition charges and Atomic coordinates (x, y, z) for p-TMPyP}
\end{longtable}

\section{Hyperparameter optimization}
In the GPR ML models, we have the set of hyperparamters $\sigma_n^2, \sigma_f^2, l_1^2, \ldots, l_D^2$, which needs to be optimized as part of the training process. To this end, we use the marginal log likelihood (MLL) of the model. GPyTorch\cite{gpytorch} provides the ADAM optimizer, which we run with a learning rate of $lr=0.1$ for a total of 200 steps on a separated, randomly chosen validation set of 1,000 samples. Since optimizers like ADAM are sensible to initial value selection, we first select an improved set of initial values by grid search with 5-times repeated 5-fold cross validation on the MLL optimization task over the grid given in Table~\ref{tab:hyper_inits}.
The best initial values are then used in a final optimization procedure over the full validation set.

\begin{table}[H]
    \centering
    \renewcommand{\arraystretch}{1.2}
    \begin{tabular}{c|c}
        Hyperparameter & Possible values\\ \hline
         lengthscales $l_1,...,l_D$& \{2, $10^{0.75}$, $10^{1.5}$\}\\
         output scale $\sigma_f$& \{1\}\\
         noise $\sigma^2_n$& \{$10^{-4}$, $10^{-6}$, $10^{-8}$\} \\
    \end{tabular}
    \caption{Grid of possible initial values for the ADAM-based hyperparameter optimization.}
    \label{tab:hyper_inits}
\end{table}

\section{Preliminary Multifidelity Data Analysis}
\begin{figure}[H]
    \centering
    \includegraphics[width=0.8\linewidth]{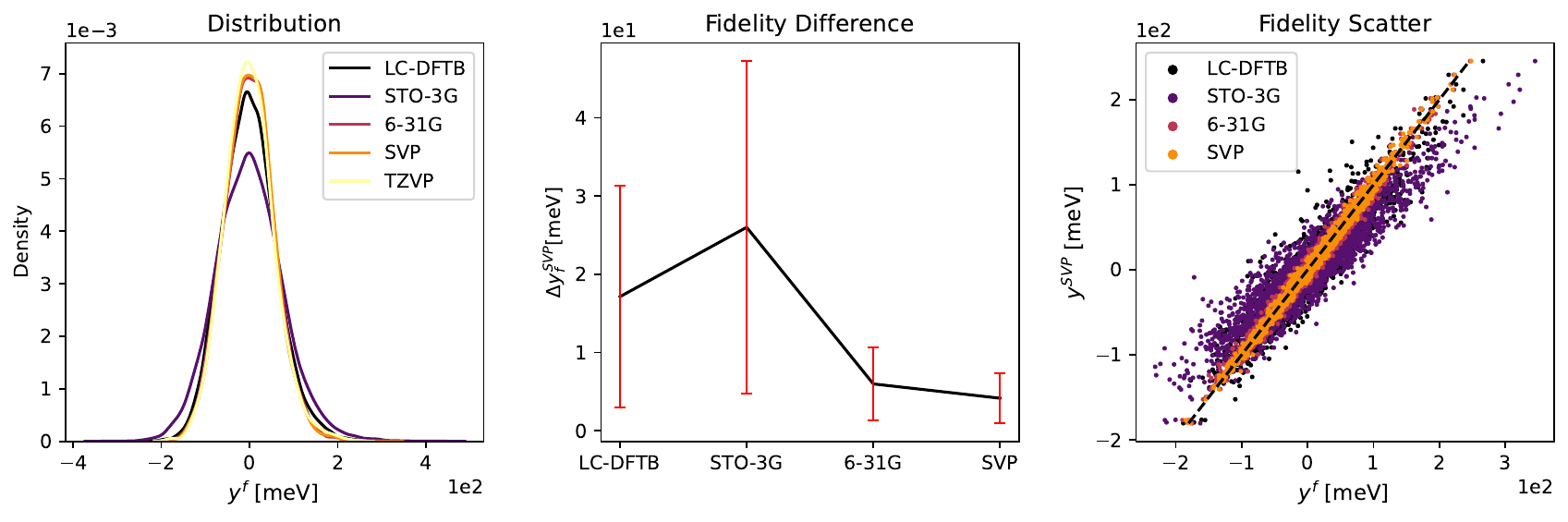}
    \caption{Preliminary multifidelity data analysis as recommended in Ref.~\cite{vinod23_MFML} for porphyrin molecule p9. The STO3G fidelity shows an unfavorable distribution with respect to the target fidelity of SVP.}
    \label{fig_traj9_prelim}
\end{figure}

\begin{figure}[H]
    \centering
    \includegraphics[width=0.8\linewidth]{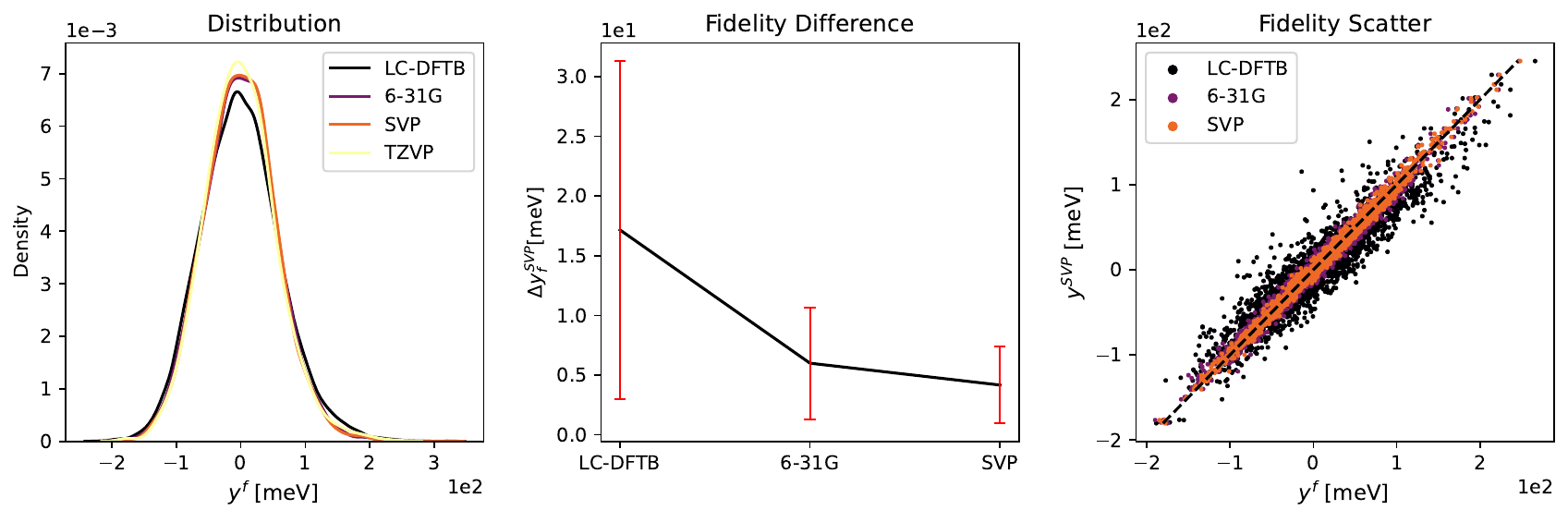}
    \caption{Same as Fig.~S2  but 
    without the STO3G energies. }
    \label{fig_traj9_prelim_nosto3g}
\end{figure}

\begin{figure}[H]
    \centering
    \includegraphics[width=0.8\linewidth]{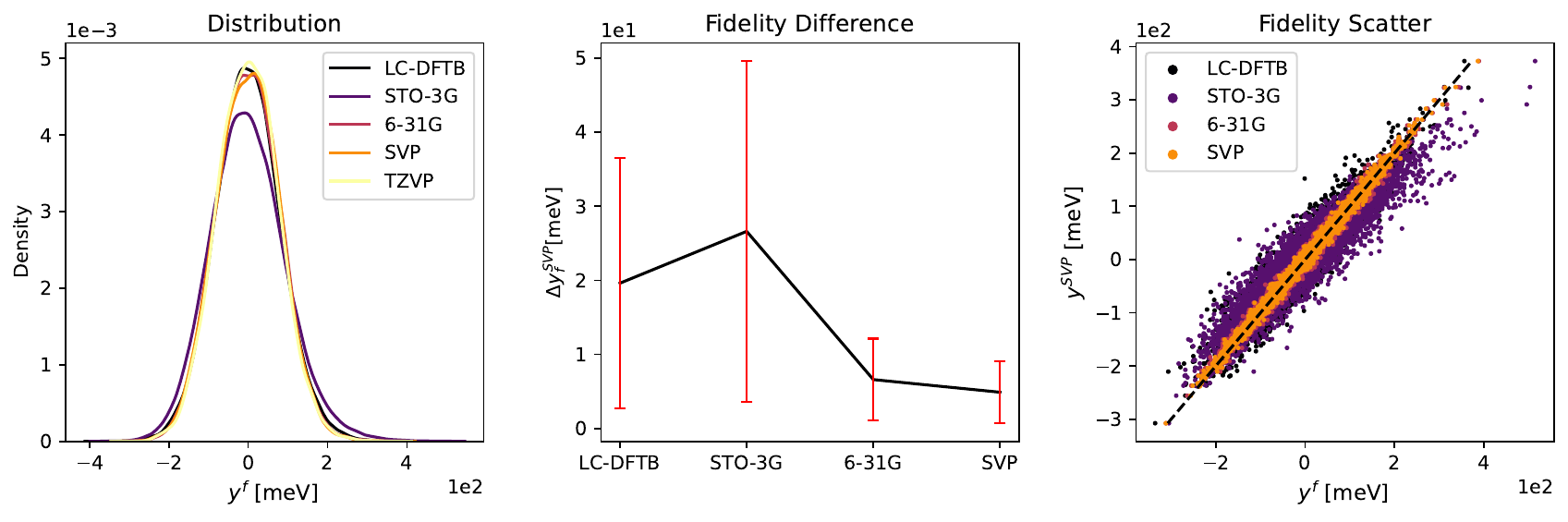}
    \caption{Preliminary multifidelity data analysis as recommended in Ref.~\cite{vinod23_MFML} for the concatenated trajectories of the p-type porphyrin molecules. The STO3G fidelity shows unfavorable distribution with respect to the target fidelity of SVP.}
    \label{fig_ptype_prelim}
\end{figure}

\begin{figure}[H]
    \centering
    \includegraphics[width=0.8\linewidth]{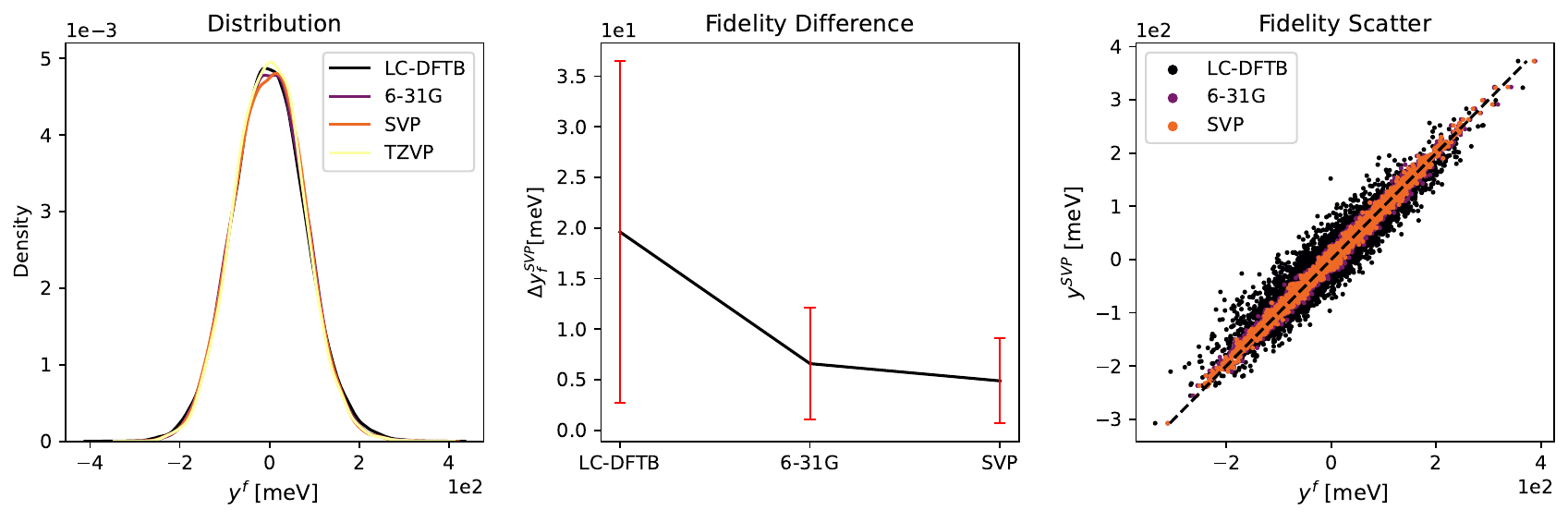}
    \caption{Same as Fig.~S4  but 
    without the STO3G energies. }
    \label{fig_ptype_prelim_nosto3g}
\end{figure}

\begin{figure}[H]
    \centering
    \includegraphics[width=0.8\linewidth]{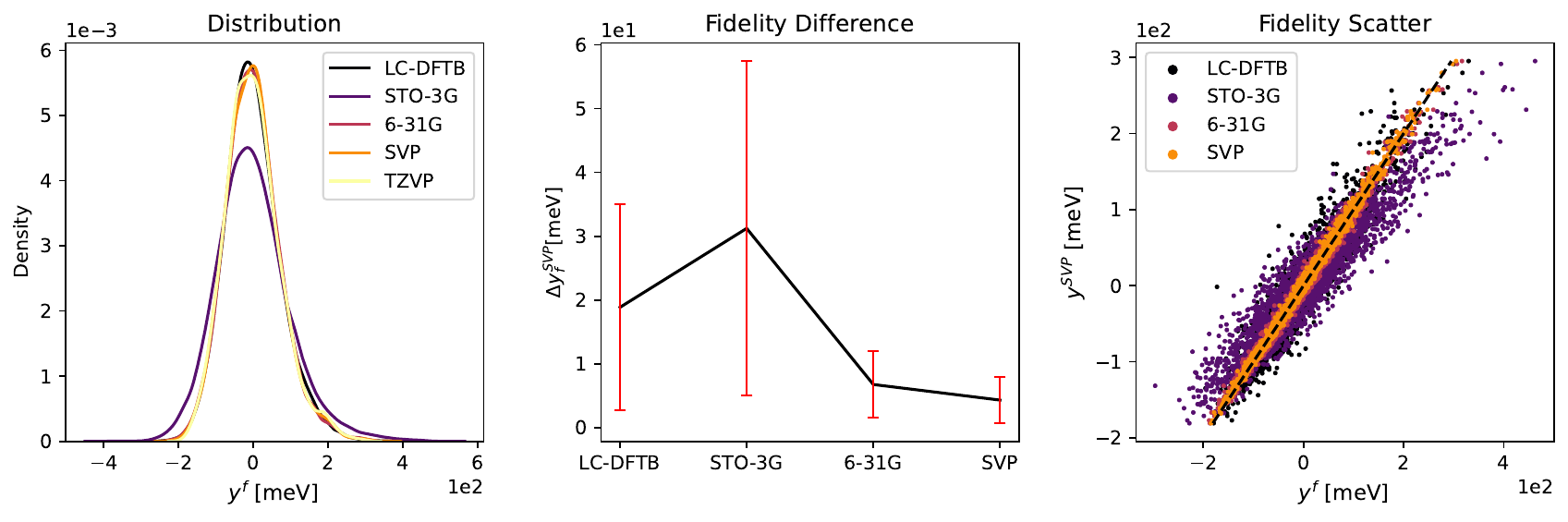}
    \caption{Preliminary multifidelity data analysis as recommended in Ref.~\cite{vinod23_MFML} for the concatenated trajectories of the m-type porphyrin molecules. The STO3G fidelity shows unfavorable distribution with respect to the target fidelity of SVP.}
    \label{fig_mtype_prelim}
\end{figure}

\begin{figure}[H]
    \centering
    \includegraphics[width=0.8\linewidth]{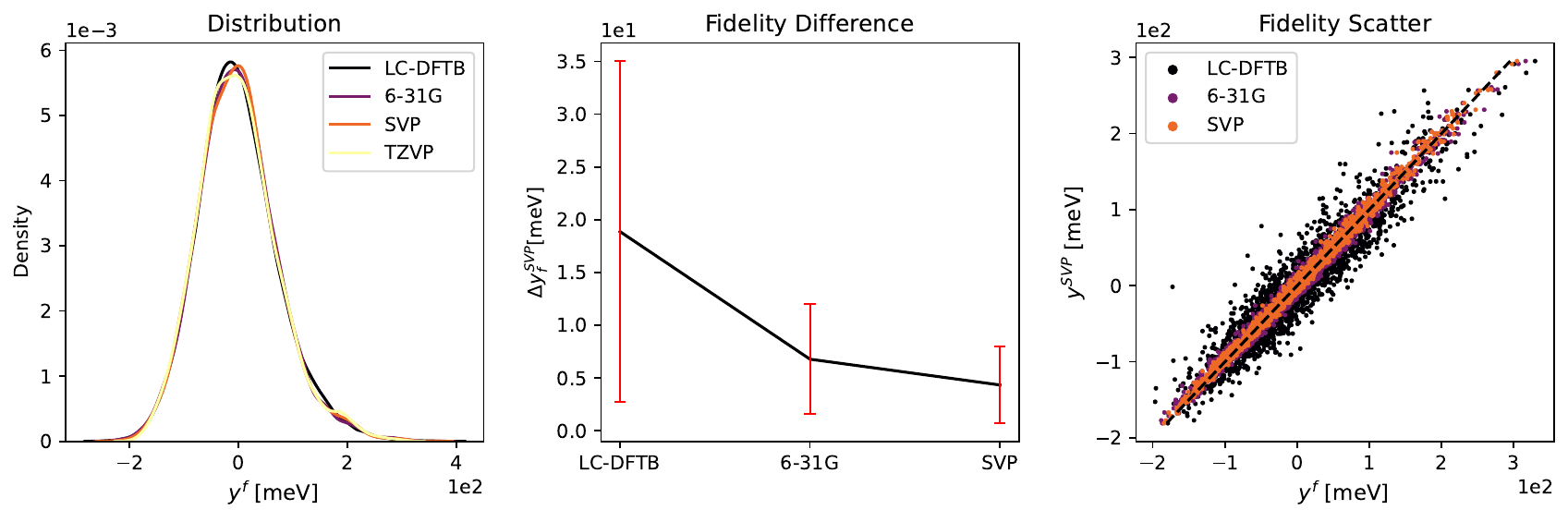}
    \caption{Same as Fig.~S6  but 
    without the STO3G energies. }
    \label{fig_mtype_prelim_nosto3g}
\end{figure}

\section{Additional Machine Learning Results}

\begin{figure}[H]
    \centering
    \includegraphics[width=0.9\linewidth]{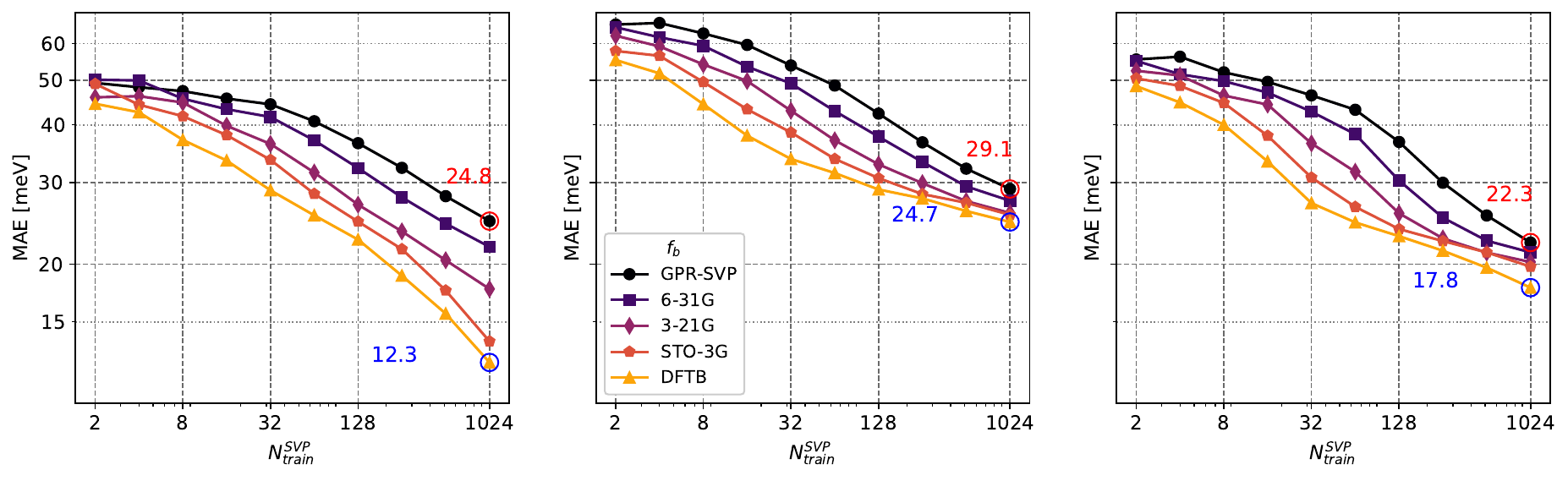}
    \caption{MFML learning curves with STO3G fidelity included.}
    \label{fig_LC_withsto}
\end{figure}

\begin{figure}[H]
    \centering
    \includegraphics[width=0.9\linewidth]{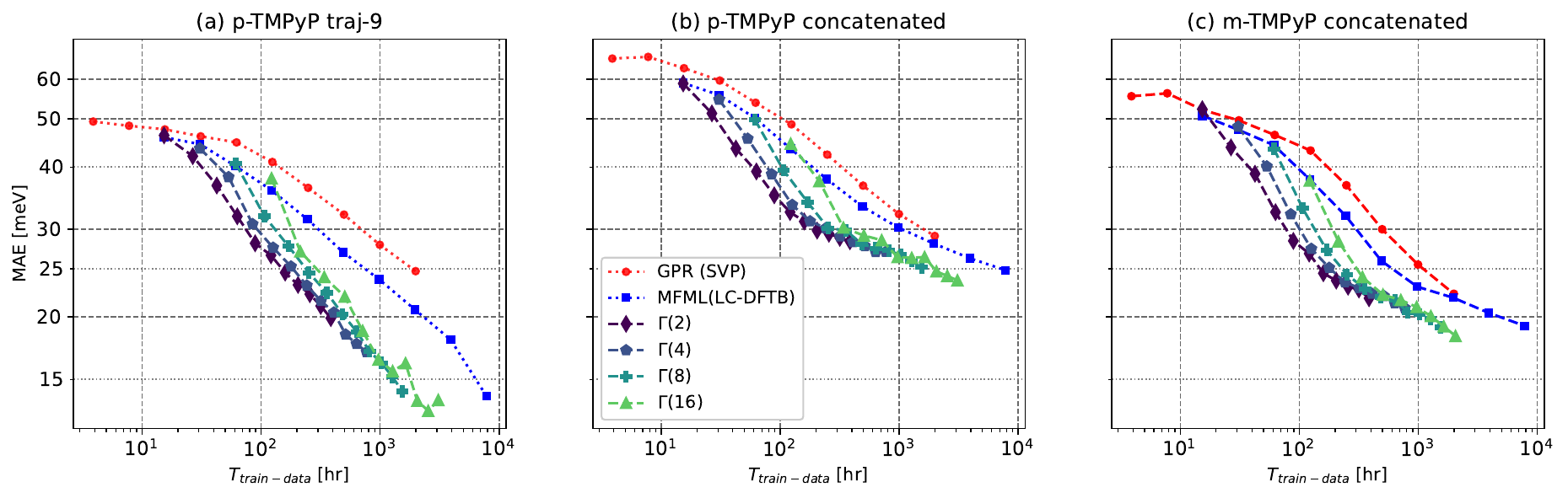}
    \caption{Full time-cost plots.}
    \label{fig_full_time_cost}
\end{figure}
\section{Spectral Densities}

\begin{figure}[H]
    \centering
    \begin{subfigure}[b]{0.24\textwidth}
        \centering
        \includegraphics[width=\textwidth]{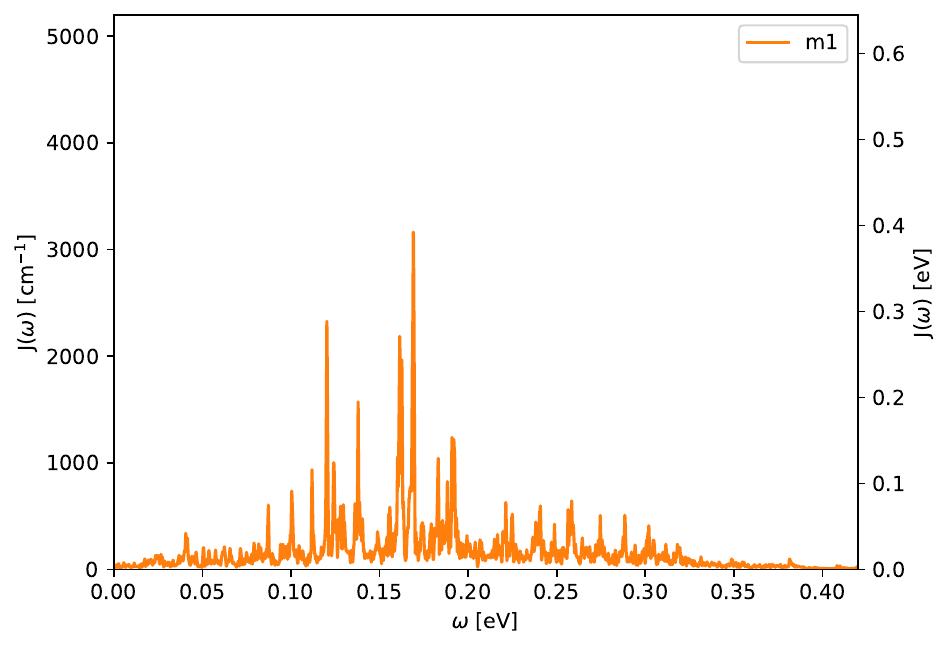}
        \caption{m1}
    \end{subfigure}
    \hfill
    \begin{subfigure}[b]{0.24\textwidth}
        \centering
        \includegraphics[width=\textwidth]{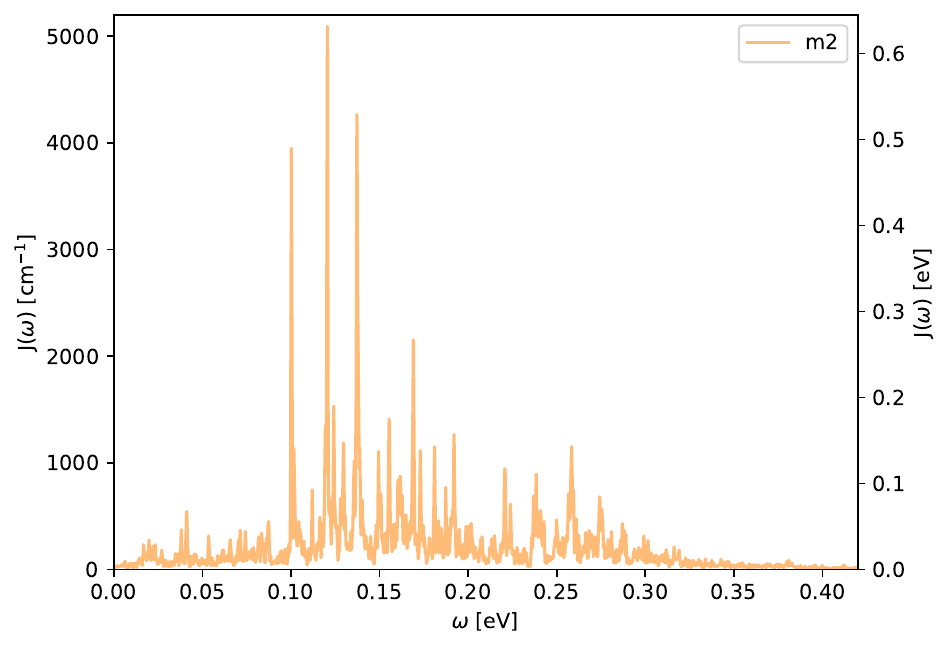}
        \caption{m2}
    \end{subfigure}
    \hfill
    \begin{subfigure}[b]{0.24\textwidth}
        \centering
        \includegraphics[width=\textwidth]{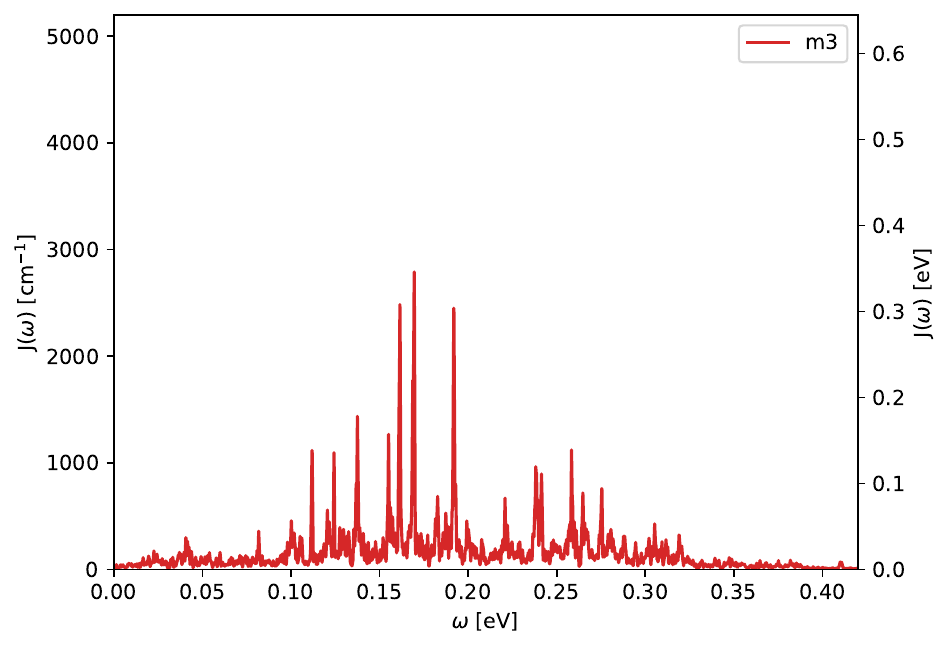}
        \caption{m3}
    \end{subfigure}
    \hfill
    \begin{subfigure}[b]{0.24\textwidth}
        \centering
        \includegraphics[width=\textwidth]{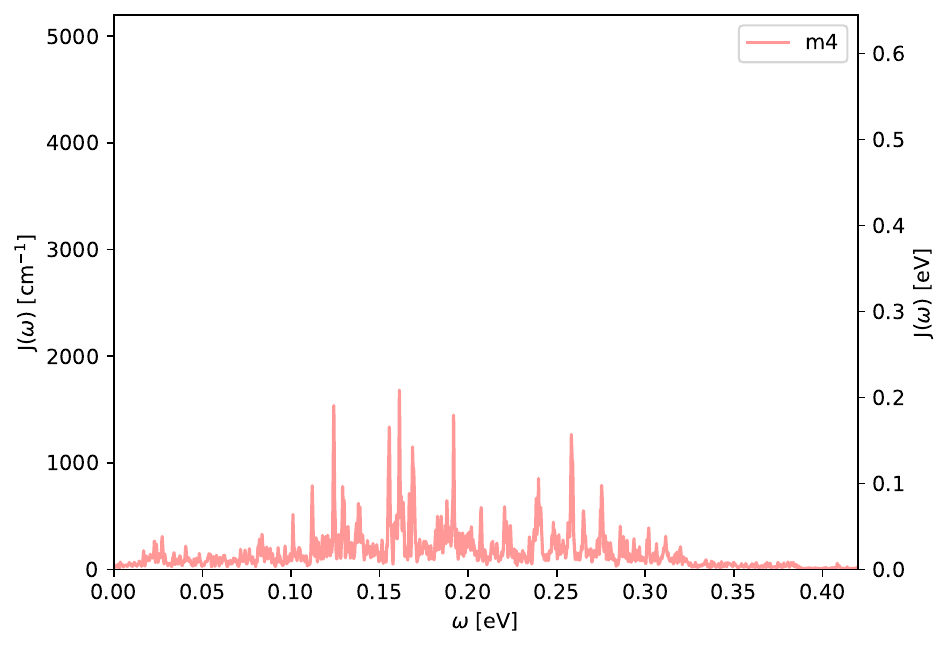}
        \caption{m4}
    \end{subfigure}
    
    \vspace{0.2cm} 
    
    \begin{subfigure}[b]{0.24\textwidth}
        \centering
        \includegraphics[width=\textwidth]{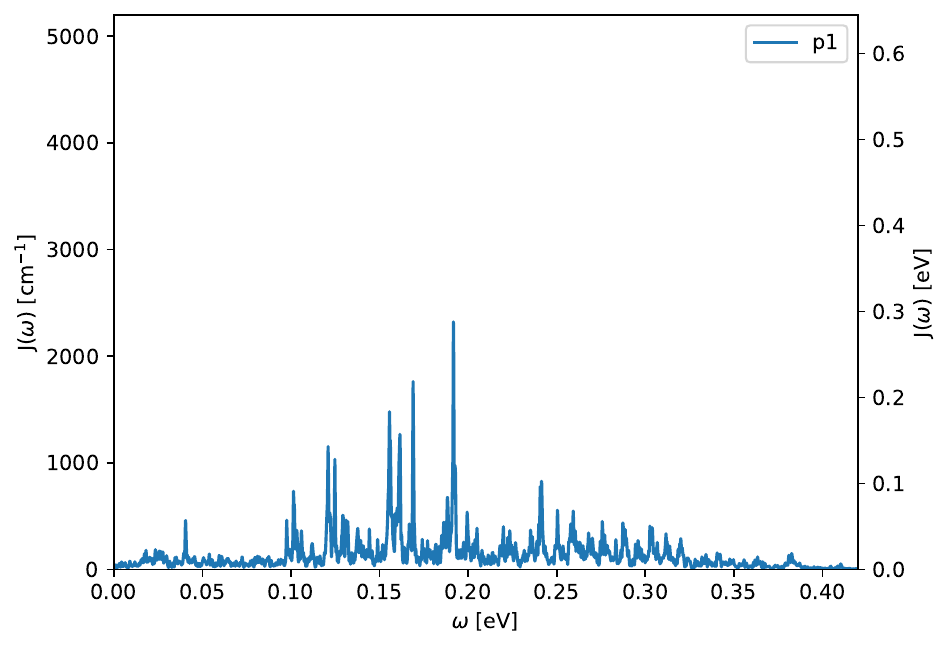}
        \caption{p1}
    \end{subfigure}
    \hfill
    \begin{subfigure}[b]{0.24\textwidth}
        \centering
        \includegraphics[width=\textwidth]{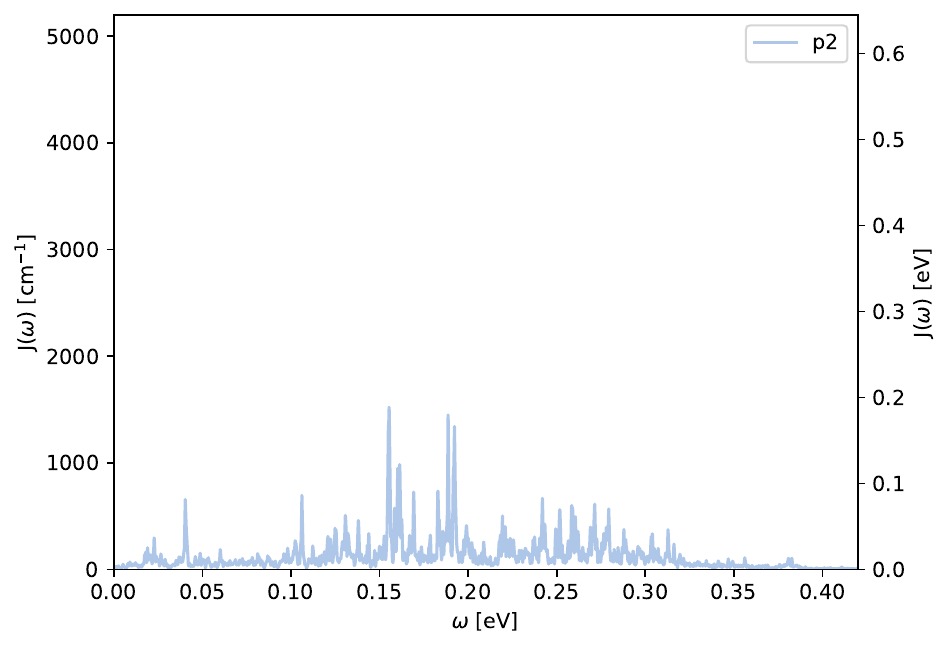}
        \caption{p2}
    \end{subfigure}
    \hfill
    \begin{subfigure}[b]{0.24\textwidth}
        \centering
        \includegraphics[width=\textwidth]{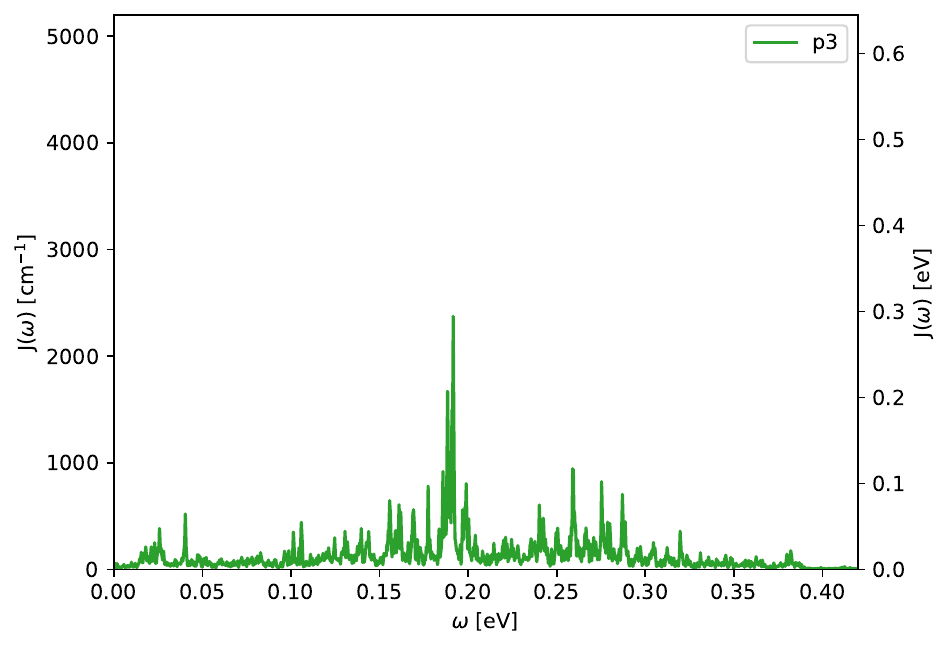}
        \caption{p3}
    \end{subfigure}
    \hfill
    \begin{subfigure}[b]{0.24\textwidth}
        \centering
        \includegraphics[width=\textwidth]{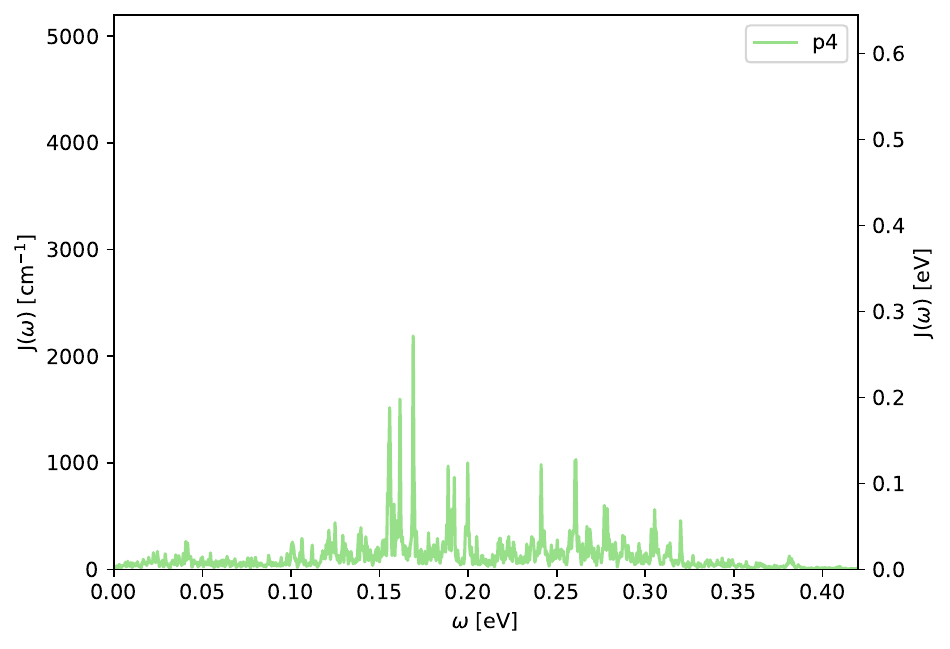}
        \caption{p4}
    \end{subfigure}

    \vspace{0.2cm} 
    
    \begin{subfigure}[b]{0.24\textwidth}
        \centering
        \includegraphics[width=\textwidth]{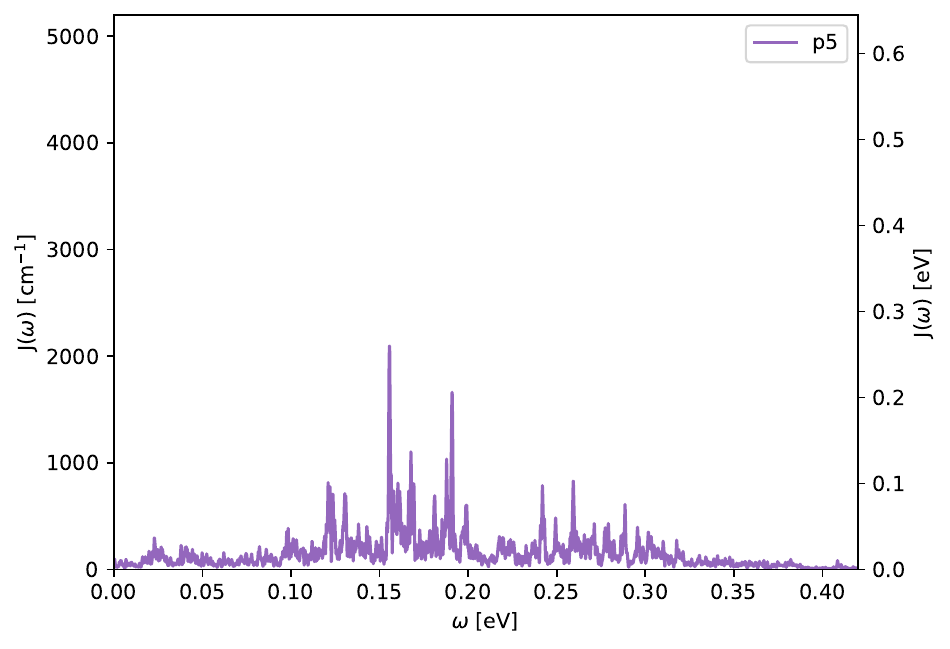}
        \caption{p5}
    \end{subfigure}
    \hfill
    \begin{subfigure}[b]{0.24\textwidth}
        \centering
        \includegraphics[width=\textwidth]{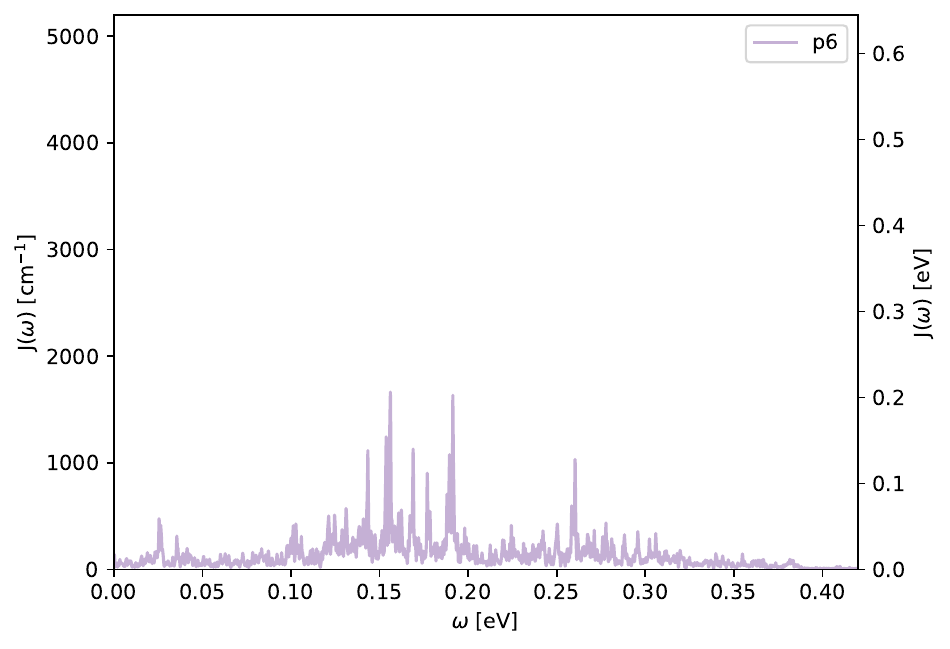}
        \caption{p6}
    \end{subfigure}
    \hfill
    \begin{subfigure}[b]{0.24\textwidth}
        \centering
        \includegraphics[width=\textwidth]{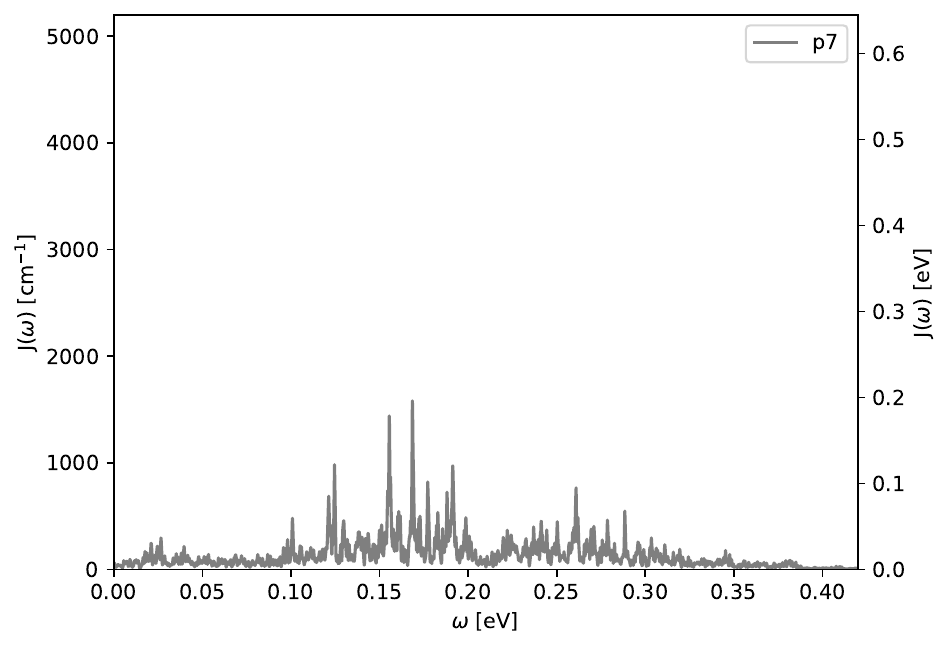}
        \caption{p7}
    \end{subfigure}
    \hfill
    \begin{subfigure}[b]{0.24\textwidth}
        \centering
        \includegraphics[width=\textwidth]{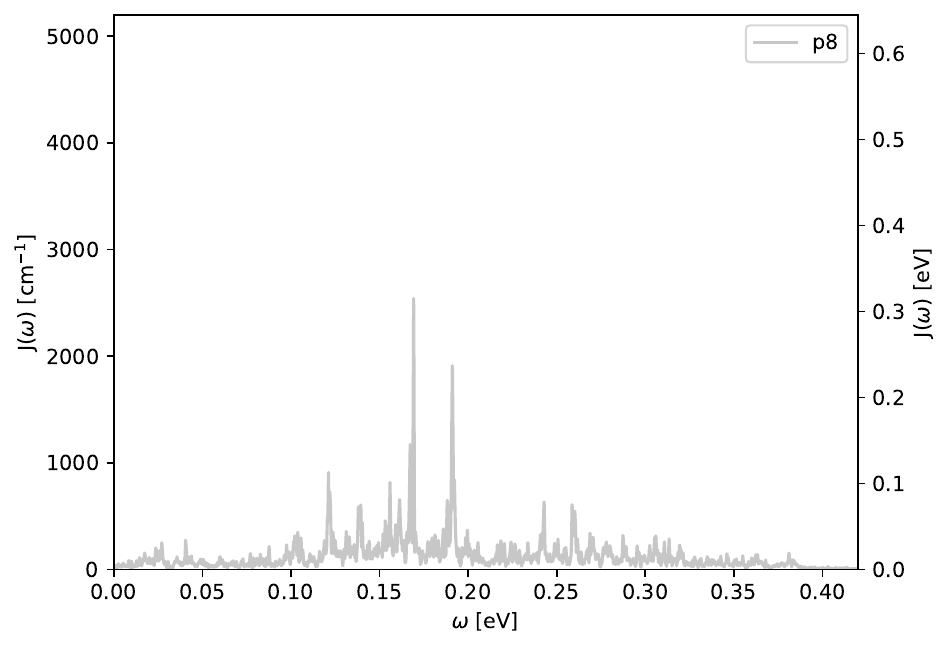}
        \caption{p8}
    \end{subfigure}

    \vspace{0.2cm} 
    
    \begin{subfigure}[b]{0.24\textwidth}
        \centering
        \includegraphics[width=\textwidth]{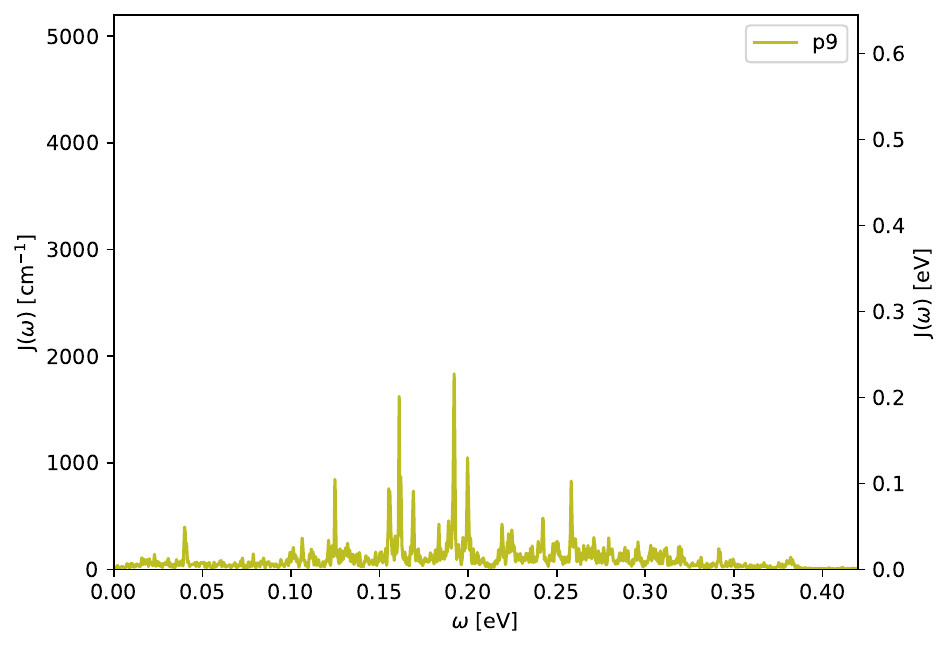}
        \caption{p9}
    \end{subfigure}
    \hfill
    \begin{subfigure}[b]{0.24\textwidth}
        \centering
        \includegraphics[width=\textwidth]{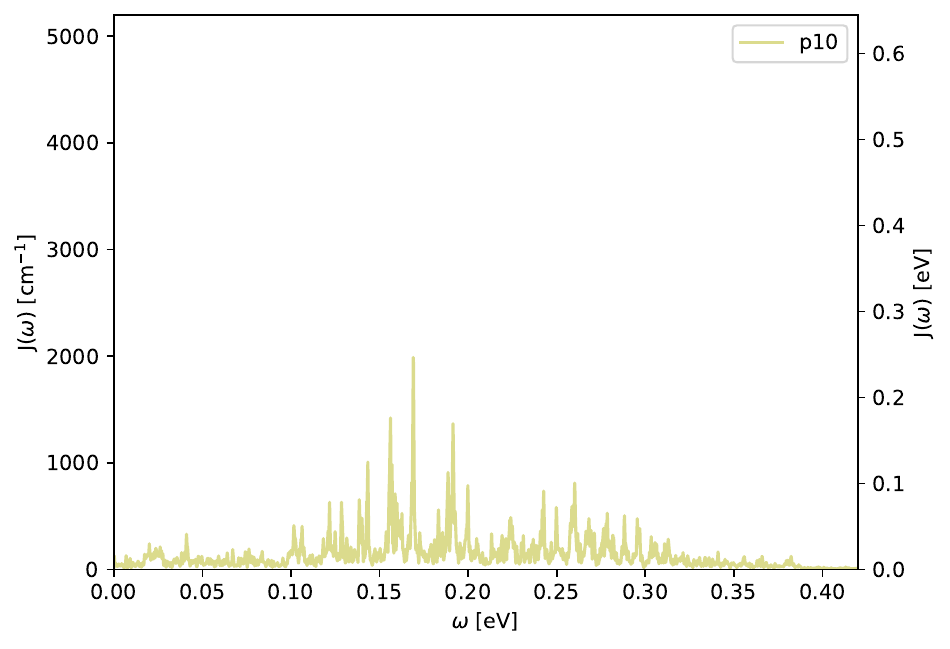}
        \caption{p10}
    \end{subfigure}
    \hfill
    \begin{subfigure}[b]{0.24\textwidth}
        \centering
        \includegraphics[width=\textwidth]{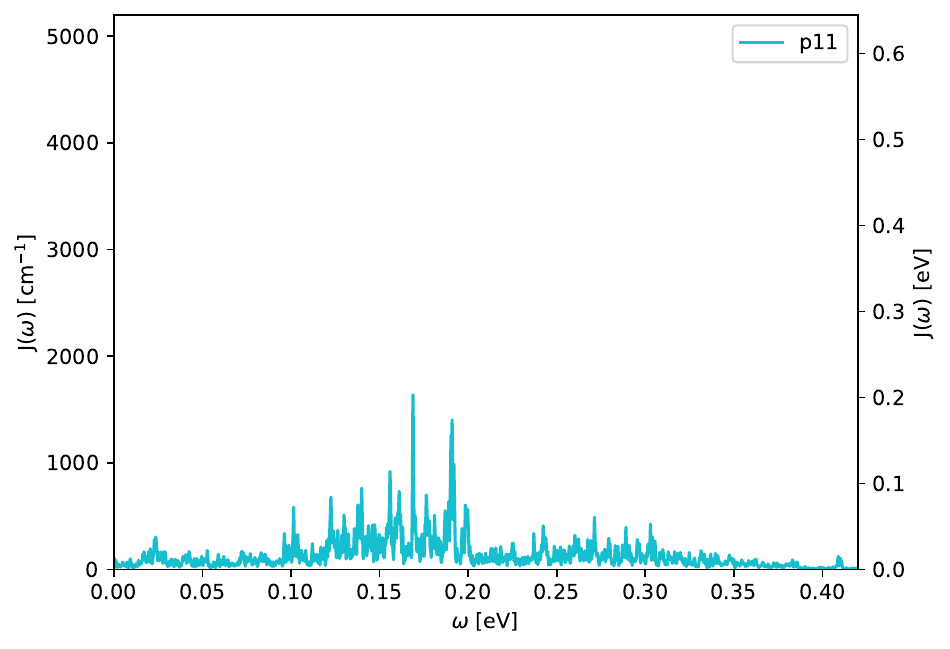}
        \caption{p11}
    \end{subfigure}
    \hfill
    \begin{subfigure}[b]{0.24\textwidth}
        \centering
        \includegraphics[width=\textwidth]{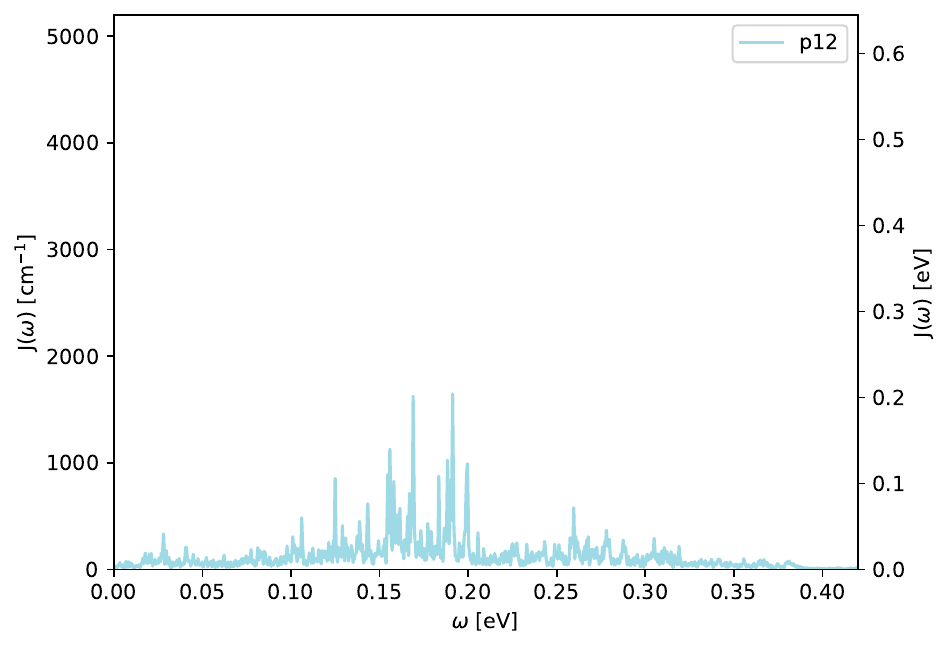}
        \caption{p12}
    \end{subfigure}

    \caption{Spectral densities of the individual pigments.}
    \label{fig:SI spectral densities}
\end{figure}

\begin{figure}[H]
    \centering
    \begin{subfigure}[b]{0.48\textwidth}
        \centering
        \includegraphics[width=\textwidth]{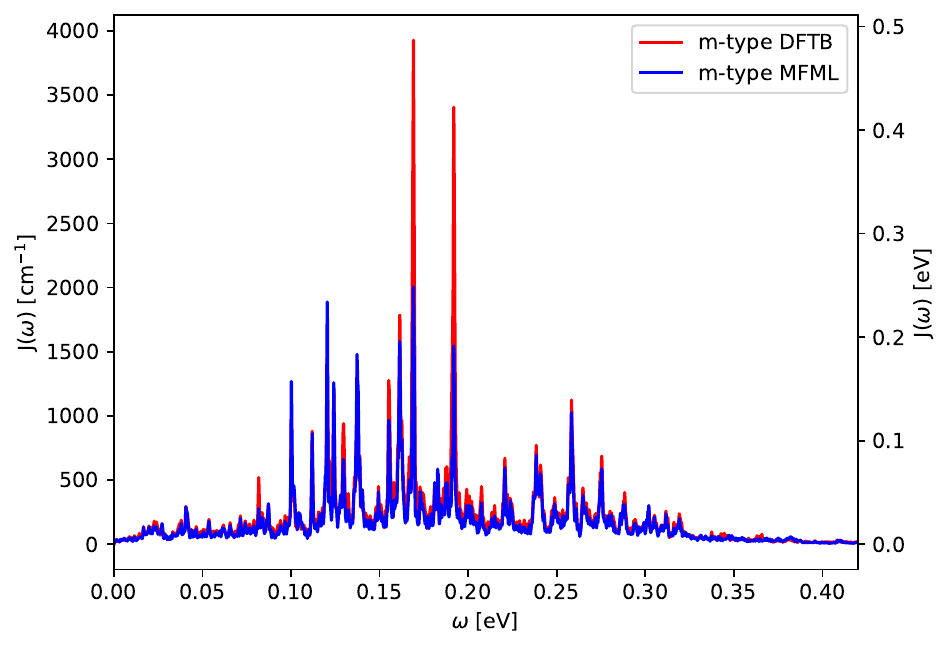}
        \caption{m-type porphyrins.}
        \label{fig_SD_m}
    \end{subfigure}
    \hfill
    \begin{subfigure}[b]{0.48\textwidth}
        \centering
        \includegraphics[width=\textwidth]{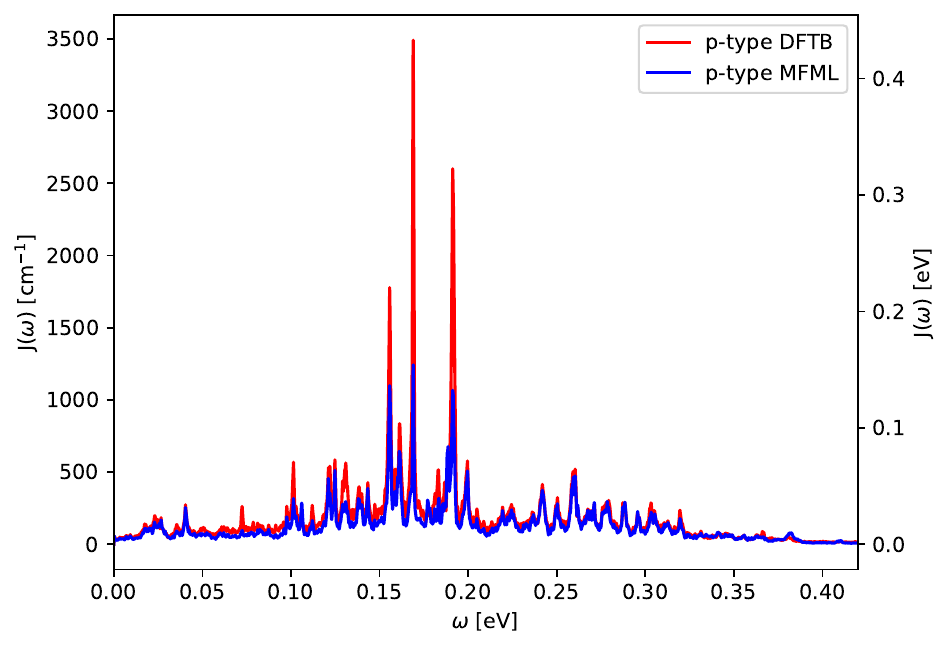}\caption{p-type porphyrins.}
        \label{fig_SD_p}
    \end{subfigure}
    \caption{Comparison of the average spectral densities based on the MFML and DFTB approaches.}
    \label{fig_SD_ml_dftb}
\end{figure}

\bibliography{bibvivinod,ukleine,holzenkamp,holtkamp,lyu}